%% file: compiler_paper.tex
\documentclass[journal]{IEEEtran}

\usepackage[english]{babel}

\usepackage{ifpdf}

\usepackage{cite} % Orders citations.
\usepackage{url}
\usepackage{hyperref}

\ifCLASSINFOpdf
	\usepackage[pdftex]{graphicx}
	\graphicspath{{./figures/}}
\else
	\usepackage[dvips]{graphicx}
	\graphicspath{./figures/}
\fi
\usepackage{color}
\usepackage{amssymb}
\usepackage{pgf}
\usepackage{tikz-cd}
\usepackage{pgf, tikz, pgfplots}
\usepackage{tikz-3dplot}
\usepackage{tikz-qtree}
\usetikzlibrary{shapes, arrows, automata, plotmarks}
\usetikzlibrary{calc,hobby,decorations,math}
\usepackage{tcolorbox}

\usepackage{amsmath}
\usepackage{amsfonts, amssymb, amsthm,mathtools}
\usepackage{mathrsfs}

\usepackage{algorithm,algpseudocode}
	\algnewcommand{\LeftComment}[1]{\Statex \(\triangleright\) #1}

\usepackage{multirow}
\usepackage{rotating}
\usepackage{subcaption}
\usepackage[shortlabels]{enumitem}

\input{./latex_resources/mySymbol.sty}

\input{./latex_resources/asp_colors.sty}

\newtheorem{lemma}{\hspace{0pt}\bf Lemma}
\newtheorem{proposition}{\hspace{0pt}\bf Proposition}

\newtheorem{theorem}{\hspace{0pt}\bf Theorem}
\newtheorem{corollary}{\hspace{0pt}\bf Corollary}

\newtheorem{definition}{\hspace{0pt}\bf Definition}

\def\pathfigs{./figures}

\begin{document}
\title{Fundamental Limits of Transferability and Equivariance in Algebraic Signal Models I: Finite Dimensions}
\author{Alejandro Parada-Mayorga
        % <-this % stops a space
\thanks{The author is with the Department of Electrical Engineering, University of Colorado (Denver). Email: alejandro.paradamayorga@ucdenver.edu}}

\markboth{IEEE Transactions on Signal Processing (submitted)}%
{Parada-Mayorga: Transferability and Equivariance in Algebraic Signal Models}
\maketitle

%%%%%%%%%%%%%%%%%%%%%%%%%%%%%%%%%%%%%%%%%%%%%%%%%%%%
%%%%%%%%%%%%%%%%%%%% ABSTRACT %%%%%%%%%%%%%%%%%%%%%%
%%%%%%%%%%%%%%%%%%%%%%%%%%%%%%%%%%%%%%%%%%%%%%%%%%%%

\input{\pathsections/sec_abstract.tex}


\IEEEpeerreviewmaketitle

%%%%%%%%%%%%%%%%%%%%%%%%%%%%%%%%%%%%%%%%%%%%%%%%%%
%%%%%               S  E  C  T  I  O  N :   I  N  T  R  O  D  U  C  T  I  O  N              %%%%%%%%%
%%%%%%%%%%%%%%%%%%%%%%%%%%%%%%%%%%%%%%%%%%%%%%%%%%

\input{\pathsections/sec_introduction}

%%%%%%%%%%%%%%%%%%%%%%%%%%%%%%%%%%%%%%%%%%%%%%%%%%%%%%%%%%%%
%%%%%               S  E  C  T  I  O  N :   H  O  M  O  M  O  R  P  H  I  S  M  S   I N    A S P             %%%%%%%%%
%%%%%%%%%%%%%%%%%%%%%%%%%%%%%%%%%%%%%%%%%%%%%%%%%%%%%%%%%%%%

\input{\pathsections/sec_hom_in_asm}

%%%%%%%%%%%%%%%%%%%%%%%%%%%%%%%%%%%%%%%%%%%%%%%%%%
%%%%%%%%%%%%%%% SECTION: ASP MOdel  %%%%%%%%%%%%%%%%
%%%%%%%%%%%%%%%%%%%%%%%%%%%%%%%%%%%%%%%%%%%%%%%%%%

\input{\pathsections/sec_hom_trans_equiv_fsm.tex}

%%%%%%%%%%%%%%%%%%%%%%%%%%%%%%%%%%%%%%%%%%%%%%%%%%%%%%%%%
%%%%%%%%%%%%%%% SECTION: GENERALIZED CONV LIESP  %%%%%%%%%%%%%%%%
%%%%%%%%%%%%%%%%%%%%%%%%%%%%%%%%%%%%%%%%%%%%%%%%%%%%%%%%%

\section{Applications}
In this section, we discuss the implications of previous results in sampling, dimensionality reduction, compressed sensing, and equivariance. 
%
\input{\pathsections/sec_implications_in_samp.tex}
\input{\pathsections/sec_implications_in_cs.tex}
\input{\pathsections/sec_implications_in_eqvml.tex}

%%%%%%%%%%%%%%%%%%%%%%%%%%%%%%%%%%%%%%%%%%%%%%%
%%%%%%%%%%%% SECTION: NUMERICAL RESULTS  %%%%%%%%%%%%
%%%%%%%%%%%%%%%%%%%%%%%%%%%%%%%%%%%%%%%%%%%%%%%

%\input{\pathsections/numerical_results}

%%%%%%%%%%%%%%%%%%%%%%%%%%%%%%%%%%%%%%%%%%%%%%%
%%%%%%%%%%%%%%% SECTION: DISCUSSION  %%%%%%%%%%%%%%%
%%%%%%%%%%%%%%%%%%%%%%%%%%%%%%%%%%%%%%%%%%%%%%%

\input{\pathsections/sec_discussion}

\bibliography{bibliography}
\bibliographystyle{unsrt}

\ifCLASSOPTIONcaptionsoff
  \newpage
\fi

%%%%%%%%%%%%%%%%%%%%%%%%%%%%%%%%%%%%%%%%%%%%%%%%%%%%%%%%%%
%%%% SUPPLEMENTARY MATERIAL SECTION %%%%
%%%%%%%%%%%%%%%%%%%%%%%%%%%%%%%%%%%%%%%%%%%%%%%%%%%%%%%%%%

\clearpage
\newpage
\begin{center}
{\LARGE \textbf{Supplementary Material}}\\[1em]
{\large Proofs and Derivations}
\end{center}
\vspace{1em}

%%%%%%%%%%%%%%%%%%%%%%%%%%%%%%%%%%%%%%%%%%%%%%%%%
%%%%%%%%%%%%%% SECTION: APPENDICES  %%%%%%%%%%%%%
%%%%%%%%%%%%%%%%%%%%%%%%%%%%%%%%%%%%%%%%%%%%%%%%%

 \appendices
%%%%%%%%%%%%%%%%%%%%%%%%%%%%%%%%%%%%%%%%%%%%%%%%%%
%%%%%%%%%%%%%%% SECTION: BACKGROUND  %%%%%%%%%%%%%%%%
%%%%%%%%%%%%%%%%%%%%%%%%%%%%%%%%%%%%%%%%%%%%%%%%%%

\input{\pathsections/sec_appendix.tex}

%\newpage
%\input{\pathsections/sec_background.tex}

\end{document}

%% file: v09/sec_abstract.tex
%!TEX root =../compiler_paper.tex

%%%%%%%%%%%%%%%%%%%%%%%%%%%%%%%%%%%%%%%%%%%%%%
%%%%%%%%%%%%%%%%%% ABSTRACT %%%%%%%%%%%%%%%%%%
%%%%%%%%%%%%%%%%%%%%%%%%%%%%%%%%%%%%%%%%%%%%%%

\begin{abstract}
We study the fundamental limits of transferability in
algebraic signal processing through homomorphisms between algebraic signal
models. Homomorphisms are linear maps between the signal spaces of two
models that commute with filtering, so filtering a signal and transferring
it across domains can be done in either order. The existence of such maps
is governed entirely by coincidences among the filtered eigenvalues of the
two models' shift operators, but existence alone is insufficient: the space
of homomorphisms always contains trivial elements that destroy all
information. We introduce the spectral transfer efficiency
$\eta(\theta)\in[0,1]$ to quantify information-preserving quality, prove
that every homomorphism decomposes into unconstrained blocks over
coincidence classes, derive the dimension of the homomorphism space, and
characterize exactly when lossless transfer is achievable. Beyond normal
shift operators, we quantify a departure-from-normality penalty and show
how filter derivatives can repair spectral defectiveness. The theory yields
concrete consequences in three settings: for sampling, eigenvalue
interlacing converts transferability under subsampling into an explicit
filter design constraint; for compressed sensing, $\eta(\theta)$ controls
the restricted isometry constant and the coherence of the resulting
measurements, and recovery decouples across coincidence classes; and for
machine learning, spectral aliasing emerges as the controlled symmetry
breaking that makes transfer between mismatched domains possible at all.
\end{abstract}

%% file: v09/sec_introduction.tex
%!TEX root =../compiler_paper.tex

%%%%%%%%%%%%%%%%%%%%%%%%%%%%%%%%%%%%%%%%%%%%%%
%%%%%%%%%%%%%%%%%% SECTION %%%%%%%%%%%%%%%%%%%
%%%%%%%%%%%%%%%%%%%%%%%%%%%%%%%%%%%%%%%%%%%%%%

\section{Introduction}
\label{sec_introduction}

\IEEEPARstart{S}{ignal} processing and machine learning models are
increasingly designed on one domain and deployed on another: a graph filter
is designed on one network and executed on a subsampled version of it, or a
neural network is trained at one scale and deployed at another. In each case
the question is the same: under which conditions can such processing be
transferred without retraining or redesign? A companion question, central to
modern learning architectures, is \emph{equivariance}: which maps commute
with the symmetries of the domain on which the data is
processed~\cite{villar2021scalars}? This paper shows that both questions admit a common
formulation, and a common answer, at the level of the algebras that define
filtering.

We argue that algebraic signal processing (ASP)~\cite{algSP0,algSP1,algSP2}
provides the natural language for answering these questions. A signal model
is a triplet $(\ccalA,\ccalH,\rho)$, with the filters forming an abstract
algebra $\ccalA$, the signals living in a vector space $\ccalH$, and a
homomorphism $\rho$ realizing the filters as operators on $\ccalH$. This
abstract formulation is what gives ASP its unifying power: discrete-time,
graph, graphon, multigraph, and Lie group signal
processing~\cite{algSP1,gphon_pooling_j,msp_j,lga_j} are all
instances of one construction. These examples also point to transferability
being intrinsically algebraic. Two domains that process signals with the
\emph{same} filters are two signal models sharing the \emph{same} algebra
but carrying different signal spaces $\ccalH_{1}$ and $\ccalH_{2}$, and
transferring information between them is the role of a
\emph{homomorphism}: a linear map $\theta:\ccalH_{1}\to\ccalH_{2}$ that
commutes with filtering. When $\ccalH_{1}=\ccalH_{2}$ and $\rho_{1}=\rho_{2}$, the same condition
says that $\theta$ commutes with every filter of the model, which is
precisely equivariance. Transferring between two domains and filtering
within one are therefore the same requirement, stated for two
representations or for a single one.

The existence of such maps is a classical subject: homomorphisms between
signal models are the morphisms of representation theory, and their
existence is governed by the Sylvester equation~\cite{horn1991topics}.
Existence, however, is not the central question. The space of homomorphisms
always contains the zero map, and more generally admits rank-deficient
elements that satisfy the intertwining condition exactly while destroying
all information. Algebraic validity and informational richness are thus
\emph{independent} requirements that a theory of transferability must
quantify jointly. Existing analyses, most notably in the graphon
literature~\cite{gphon_pooling_j,gphon_samp,gphon_leus}, are
asymptotic or perturbative, bounding the discrepancy when a filter designed
on one domain is executed on a similar one. What is missing is a
non-asymptotic, structural theory that characterizes when \emph{exact}
transfer is possible, counts the degrees of freedom available to construct
it, measures how much information a given map preserves, and identifies the
fundamental limits that no design can overcome. Providing such a theory is
the purpose of this paper. Our contributions are the following:
\begin{itemize}%[leftmargin=*]
	\item[\textbf{(C1)}] \textbf{Transferability as a homomorphism problem.}
	We formalize transferability of a filter subset
	$\ccalA_{0}\subseteq\ccalA$ as the existence of a nonzero homomorphism,
	and characterize it in finite dimensions through a linear operator whose
	zero eigenvalues are indexed by the coincidences
	$p(\lambda_{i})=p(\widetilde{\lambda}_{j})$ between the filtered spectra
	of the two shift operators (Theorem~\ref{thm_trans_asp_finite}). We show that the
	freedom available to build a transfer map trades directly against
	spectral separation and filter variability across the two domains.

	\item[\textbf{(C2)}] \textbf{Degree of transferability.} We prove that
	every homomorphism decomposes into a direct sum of unconstrained linear
	maps over the \emph{coincidence classes} of the filtered spectra
	(Lemma~\ref{lem:support}), derive the dimension of the space of
	homomorphisms (Proposition~\ref{prop:dimension}), and introduce the
	\emph{spectral transfer efficiency} $\eta(\theta)\in[0,1]$, an intrinsic
	measure of information-preserving quality that detects trivial
	homomorphisms and characterizes lossless transfer. We show that an isometric
	homomorphism exists if and only if the filtered source spectrum is
	contained in the target's with dominated multiplicities
	(Proposition~\ref{prop:achievability}). We also extend the analysis
	beyond normal shift operators, showing that the algebraic theory survives
	via oblique projections. We quantify a departure-from-normality penalty
	and establish the effect of the filter derivatives on the Jordan
	structure of the filtered shift operators (Lemma~\ref{lem:pprime}).

	\item[\textbf{(C3)}] \textbf{Consequences for sampling, compressed
	sensing, and equivariant learning.} For sampling, we show that eigenvalue
	interlacing converts the coincidence requirement into an explicit design
	constraint localizing where the filter must be flat. For compressed sensing, richness
	controls the restricted isometry constant of a homomorphism seen as a
	measurement operator (Proposition~\ref{prop:ric}). The coherence is a
	purely intra-class quantity (Lemma~\ref{lem:gram}) subject to a
	Welch-type lower bound (Proposition~\ref{prop:mudesign}), and sparse
	recovery decouples across coincidence classes. For equivariant learning, spectral
	aliasing emerges, within a hierarchy of symmetry constraints, as the
	\emph{controlled symmetry breaking} that makes transfer between
	mismatched domains possible at all
	(Proposition~\ref{prop_equivariance}).
\end{itemize}

\noindent\textbf{Related work.} Transferability across
domains of different sizes has been analyzed mainly through graphon
limits~\cite{gphon_pooling_j,gphon_samp,gphon_leus,diao2016model,9053882}, with transfer error
controlled by filter variability. We complement these perturbative bounds
with an exact, non-asymptotic theory explaining the spectral restrictions
such analyses impose on filters. Equivariant maps are
a cornerstone of geometric deep learning~\cite{villar2021scalars,blumsmith2023invariant}, typically for a group
acting on a single domain. Our notion, with the same algebra acting on two
domains, contains the single-domain case and connects it to
transferability. Our compressed sensing results relate to the classical
restricted isometry property and coherence-based
recovery, imported at the level of individual coincidence
blocks.

\noindent\textbf{Organization.} Section~\ref{sec_asp} reviews ASP and
introduces homomorphisms between signal models. Section~\ref{sec_finite_asm}
defines transferability of filter subsets and the existence theory in
finite dimensions. Section~\ref{sec_deg_transf} develops the degree of
transferability: coincidence structure, block decomposition, the richness
measure $\eta(\theta)$, achievability, and extensions beyond normality.
Section~\ref{sec_sampling} through Section~\ref{sec_eqvml} develop the
applications to sampling, compressed sensing, and equivariant learning, and
we close with a discussion in Section~\ref{sec_discussion}. Proofs are collected in the appendices of the supplementary materials.

%% file: v09/sec_hom_in_asm.tex
%!TEX root =../compiler_paper.tex

%%%%%%%%%%%%%%%%%%%%%%%%%%%%%%%%%%%%%%%%%%%
%%%%%%%%%%%%% SECTION %%%%%%%%%%%%%%%%%%%%%
%%%%%%%%%%%%%%%%%%%%%%%%%%%%%%%%%%%%%%%%%%%

\section{ASP and Homomorphisms between Signal Models}
\label{sec_asp}
Filtering, polynomial diffusions, and signal models in general admit a common
description within \textit{algebraic signal processing (ASP)}, a framework
that has established itself as a consistent foundation for an extensive
collection of convolutional signal frameworks. These include discrete-time
signal processing~\cite{algSP1}, discrete space models governed by symmetric
shift operators~\cite{algSP2}, models defined on 2D hexagonal
lattices~\cite{algSP6}, models on general lattices~\cite{puschel_asplattice},
signal processing over sets~\cite{puschel_aspsets}, Lie group signal
processing~\cite{lga_j,lga_icassp}, graphon signal
processing~\cite{gphon_pooling_c,gphon_pooling_j,gphon_samp,gphon_leus},
multigraph signal processing~\cite{msp_j,msp_icassp2023}, models on
digraphs~\cite{puschel_digraphs1,puschel_digraphs2}, and several
others~\cite{algSP7,algSP8,algnn_nc_j,parada_quiversp}. What underlies this breadth is a deliberate separation of concerns: the
filters are abstracted into an algebra, formulated independently of any
specific data domain, and are realized on concrete signals through a
representation whose operators carry the structure of the domain at hand.

The fundamental object in ASP is the algebraic signal model (ASM), which is
specified by a triplet
\begin{equation}
	\left(
	\ccalA
	,
	\ccalH
	,
	\rho
	\right)
	,
\end{equation}
in which $\ccalA$ denotes an associative algebra with a unit, $\ccalH$ is a
vector space, and $\rho:\ccalA\to\text{End}(\ccalH)$ is a homomorphism
assigning to each element of $\ccalA$ a linear endomorphism of $\ccalH$, that
is, an element of $\text{End}(\ccalH)$. The three components carry
complementary roles: $\ccalA$ is the home of the \emph{filters} as abstract
algebraic entities, $\ccalH$ stores the \emph{signals} to be processed, and
$\rho$ is the link that converts the former into operators acting on the
latter. We examine each in turn.

By an algebra $\ccalA$ we mean a vector space endowed with an additional
product that is internal and associative. A canonical illustration is
$\mbC[t]$, the polynomials in one variable with complex coefficients. Seen
purely as a vector space, $\mbC[t]$ comes with the usual addition and scaling
of polynomials. The extra ingredient that elevates it to an algebra is
polynomial multiplication, which sends any two polynomials to a third one and
is hence internal. The constant polynomial $p(t)=1$ behaves as a
multiplicative identity, so $\mbC[t]$ is, in addition, unital. Throughout, we
designate the elements of $\ccalA$ as the \textit{filters} of the ASM. The
algebraic structure is considerably richer than this single example
suggests~\cite{algSP0,parada_algnn,algnn_nc_j,parada_algnnconf}: $\ccalA$ may be  non commutative, and an ASM can be built upon operator algebras, group
algebras, or path algebras, among other choices. It has likewise been shown
that reproducing kernel Hilbert spaces give rise to algebras tied to the
domain of the reproducing kernel~\cite{rkhs_conv,int_rkhs_conv}.

It is often convenient to describe the algebra through its generators. For the
polynomial model, every element of $\mbC[t]$ arises from the single
indeterminate $t$ via sums and products, so $t$ alone generates $\mbC[t]$.
Regarding $t$ as an abstract object, a filter $a\in\mbC[t]$ becomes the
polynomial $a=\sum_{k} h_k\, t^{k}$ in that object, with coefficients
$h_k\in\mbC$. The general case admits a finite set of generators
$g_1,\dots,g_m$, in which a filter is a---possibly
non-commutative---polynomial in $g_1,\dots,g_m$. This perspective is what
connects the algebra to the operational meaning of filtering.

The vector space $\ccalH$ holds the data that the model is meant to process,
and it is precisely through $\ccalH$ that this data is equipped with an
algebraic structure. We call its elements the \textit{signals}. On its own,
$\ccalH$ carries no signal-processing content. That content emerges only once
the filters are made to act on it, which is the function of the homomorphism.

The map $\rho$ is linear, multiplicative, and it carries $\ccalA$
into $\text{End}(\ccalH)$ in a way that respects the product of $\ccalA$. In
explicit terms, for every pair $a,b\in\ccalA$ we have
\begin{equation}\label{eq_hom_def}
	\rho\left(
	ab
	\right)
	=
	\rho(a)\,
	\rho(b)
	,
\end{equation}
with $\rho(a),\rho(b)\in\text{End}(\ccalH)$ acting as linear operators on
$\ccalH$. We additionally take $\rho$ to be unital, so that
$\rho(1)=\mathrm{Id}_{\ccalH}$. Relation~\eqref{eq_hom_def} is what allows us
to interpret $\rho$ as the device that materializes the abstract filters of
$\ccalA$ as concrete operators transforming the signals in $\ccalH$. In the context of representation theory of algebras~\cite{repthysmbook,repthybigbook,barot2014introduction,folland2016course}, the pair
$(\ccalH,\rho)$ is exactly a representation of $\ccalA$, and it is the
representation---not the algebra by itself---that pins down how filtering is
performed on a particular domain. Returning to the polynomial model, the
representation is fixed by the image of the generator,
$\mathbf{S}=\rho(t)\in\text{End}(\ccalH)$, which we refer to as the
\emph{shift operator}. Then, a filter $a=\sum_{k} h_k t^{k}$ is correspondingly
implemented as
\begin{equation}\label{eq_filter_rep}
	\rho(a)
	=
	\sum_{k}
	h_k\,
	\mathbf{S}^{k}
	,
\end{equation}
which restores the familiar picture of filters as polynomials in a shift
operator. The same construction carries over directly to several generators,
with $\mathbf{S}_i=\rho(g_i)$.

This last point is the source of the unifying reach of ASP. Holding the
algebra $\ccalA$ fixed while allowing the representation $(\ccalH,\rho)$ to
vary---equivalently, selecting the shift operator $\mathbf{S}$---recovers each
of the frameworks enumerated above as a special case of one and the same
abstract theory: $\mathbf{S}$ is realized as the time delay in discrete-time
signal processing, as a graph adjacency or Laplacian in graph signal
processing, as an integral operator in graphon signal processing, and so
forth. Since filtering, the Fourier transform, and the spectral
representation are formulated at the level of $\ccalA$ and its irreducible
representations~\cite{algSP1,algSP2}, they propagate to every signal model
sharing that algebra, regardless of the concrete nature of the data carried by
$\ccalH$. It is this algebraic layer that we leverage in what follows.

% =====================================================================
%  Instantiations of the ASM triplet (A, H, rho) -- bulleted paragraph
% =====================================================================

To illustrate better the preceding discussion, we collect below several
instantiations of the triplet $(\ccalA,\ccalH,\rho)$, each of which recovers an
established signal model as a special case of the same abstract construction.
\begin{itemize}[leftmargin=*,itemsep=0pt]
	
	\item \textbf{Discrete-time signal processing.} The filters form the
	algebra $\ccalA=\mbC[t]/\langle t^{N}-1\rangle$ of polynomials in the cyclic
	shift, signals are finite sequences in $\ccalH=\mbC^{N}$, and $\rho$ sends
	the generator $t$ to the cyclic shift $\mathbf{S}=\mathbf{C}\in\mbC^{N\times
		N}$. A filter is then $\rho(a)=\sum_{k}h_{k}\mathbf{C}^{k}$, i.e., a circular
	convolution, and the associated Fourier transform is the DFT~\cite{algSP1}.
	
	\item \textbf{Graph signal processing.} Here $\ccalA=\mbC[t]$ and
	$\ccalH=\mbC^{N}$ gathers the signals supported on the $N$ nodes of a graph.
	The homomorphism maps the generator to the graph shift operator,
	$\rho(t)=\mathbf{S}$, commonly an adjacency matrix $\mathbf{A}$ or a
	Laplacian $\mathbf{L}$, so that filters take the form
	$\rho(a)=\sum_{k}h_{k}\mathbf{S}^{k}$.
	
	\item \textbf{Graphon signal processing.} The algebra is again
	$\ccalA=\mbC[t]$, while the signal space becomes infinite-dimensional,
	$\ccalH=L^{2}([0,1])$. The generator is realized as the integral operator
	$\rho(t)=\boldsymbol{T}_{W}$ induced by a graphon $W:[0,1]^{2}\to[0,1]$, namely
	$(\boldsymbol{T}_{W}f)(x)=\int_{0}^{1}W(x,y)f(y)\,dy$, and filters are polynomials in the
	graphon shift $\boldsymbol{T}_{W}$~\cite{gphon_pooling_j,gphon_samp}.
	
	\item \textbf{Lie group signal processing.} The algebra $\ccalA$ is the
	convolution algebra of a Lie group $G$ (equivalently, a quotient of its
	universal enveloping algebra $\mathcal{U}(\mathfrak{g})$), signals reside in
	$\ccalH=L^{2}(G)$, and $\rho$ is a unitary representation of $G$ under which
	filtering is realized as group convolution~\cite{lga_j,lga_icassp}.
	
\end{itemize}

%%%%%%%%%%%%%%%%%%%%%%%%%%%%%%%%%%%%%%%%%%%%%%
%%%%%%%%%    S U B - S E C T I O N     %%%%%%%
%%%%%%%%%%%%%%%%%%%%%%%%%%%%%%%%%%%%%%%%%%%%%%

\subsection{Homomorphisms Between Signal Models}

%%----------------------------------------
%%-----------    F I G U R E    ----------
%%---------------------------------------- 

\begin{figure}
    \centering
        \input{\pathfigs/hom_design_basic_diagram.tex}
    \caption{Homomorphism between algebraic signal models. The linear map
    $\theta:\ccalH_{1}\to\ccalH_{2}$ makes the diagram commute for every
    filter $a\in\ccalA$: filtering in $(\ccalA,\ccalH_{1},\rho_{1})$ and then
    transferring with $\theta$ produces the same signal as transferring first
    and then filtering in $(\ccalA,\ccalH_{2},\rho_{2})$.}
    \label{fig_hom_diagram}
\end{figure}
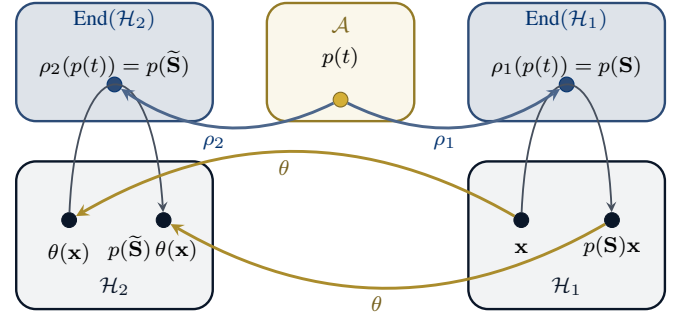

%%-------     End of Figure     ----------

The notion of homomorphism between signal models relies entirely on the
definition of homomorphism between the representations of an algebra. For our
discussion, we make this concept specific in the following definition, which
is depicted in Fig.~\ref{fig_hom_diagram}.

%%----------------------------------------
%%------   D E F I N I T I O N   ---------
%%----------------------------------------

\begin{definition}\label{def_hom_asm}
Let $(\mathcal{A},\mathcal{H}_{1},\rho_{1})$ and $(\mathcal{A},\mathcal{H}_{2},\rho_{2})$ be two algebraic signal models. Then, we say that the linear map $\theta:\mathcal{H}_{1} \to \mathcal{H}_{2}$ is a homomorphism if
\begin{equation}\label{eq_def_hom_asm_1}
 \theta\left( 
            \rho_{1}(a)
                    \mathbf{x}
       \right)
             =
              \rho_{2}(a)
                  \theta\left(
                            \mathbf{x}
                        \right)
                        ,
                        \quad
                        \forall~ a\in\mathcal{A}
                        ,~
                        \mathbf{x}\in\mathcal{H}_{1}
                        .
\end{equation}
\end{definition}

%%-------   End of Definition   ----------

It is important to emphasize that in the language of representation theory of algebras~\cite{repthysmbook,repthybigbook,barot2014introduction,folland2016course}, ~\eqref{eq_def_hom_asm_1} determines the
exact definition of homomorphic representations of the same algebra
$\mathcal{A}$. This is, $\theta$ in~\eqref{eq_def_hom_asm_1} is a homomorphism
between the representations $(\mathcal{H}_{1},\rho_{1})$ and
$(\mathcal{H}_{2},\rho_{2})$ of the algebra $\mathcal{A}$. Additionally,
since~\eqref{eq_def_hom_asm_1} holds for \emph{every}
$\mathbf{x}\in\ccalH_{1}$, it is equivalent to the operator identity
\begin{equation}\label{eq_def_hom_asm_2}
 \theta\,
     \rho_{1}(a)
     =
     \rho_{2}(a)\,
     \theta
     ,
     \qquad
     \forall~ a\in\ccalA
     ,
\end{equation}
and in what follows we use the two forms interchangeably, favoring the
operator form~\eqref{eq_def_hom_asm_2} whenever the signal plays no explicit
role in the argument.

While Definition~\ref{def_hom_asm} comes from representation theory~\cite{repthybigbook,repthysmbook}, it encapsulates a
notion of \textit{linear transferability} with practical implications. In particular, $\theta$ properly
transfers the information between the signal models
$(\mathcal{A},\mathcal{H}_{1},\rho_{1})$ and
$(\mathcal{A},\mathcal{H}_{2},\rho_{2})$ when $\theta$ is interchangeable with
the operation of filtering on each space. This is, given $\theta$, mapping
signals between $\mathcal{H}_{1}$ and $\mathcal{H}_{2}$, it is indifferent
where the filters in $\mathcal{A}$ are implemented,
$\text{End}(\mathcal{H}_{1})$ or $\text{End}(\mathcal{H}_{2})$, the result of
filtering a signal $\mathbf{x}\in \mathcal{H}_{1}$ and transferring it, or
transferring $\theta(\mathbf{x})\in\mathcal{H}_{2}$ and filtering it, is the
same. In Section~\ref{sec_finite_asm} we will formalize the notion of transferability.

Additionally, it is essential to emphasize that the condition
in~\eqref{eq_def_hom_asm_1} is identical to that required for
\textit{equivariance} when $\mathcal{A}$ is a group algebra and
$\mathcal{H}_{1}=\mathcal{H}_{2}$~\cite{kumar2024lie,lga_icassp}. We return to this connection in
Section~\ref{sec_eqvml}. Finally, notice that the zero mapping is a case of a
trivial homomorphism of representations of an algebra that, although it lacks
utility to map information between the signal models at hand, facilitates the
characterization of the space of homomorphisms as a vector space. We emphasize
this in the following corollary.

%%---------------------------------------
%%----------- COROLLARY -----------------
%%---------------------------------------

\begin{corollary}\label{cor_space_of_hom}
Let $(\mathcal{A},\mathcal{H}_{1},\rho_{1})$ and $(\mathcal{A},\mathcal{H}_{2},\rho_{2})$ be two algebraic signal models, let $\ccalA_{0}\subseteq\ccalA$ be a subset of filters, and let $\Theta_{\ccalA_{0}}$ be given by
\begin{multline}\label{eq_cor_space_of_hom_1}
\Theta_{\ccalA_{0}} = 
        \left\lbrace 
               \theta:\ccalH_{1}\to\ccalH_{2}\ \text{linear}
            \left\vert~
                 \theta\left( 
                     \rho_{1}(a)
                    \mathbf{x}
       \right)
             =
              \rho_{2}(a)
                  \theta\left(
                            \mathbf{x}
                        \right)
            \right.\right.
            \\
            \left.
            ~\forall~\mathbf{x}\in\mathcal{H}_{1},~ a\in\mathcal{A}_{0}
        \right\rbrace
        .
\end{multline}
Then, $\Theta_{\ccalA_{0}}$ is a vector space with the usual operations of
addition and multiplication by scalars.
\end{corollary}

\begin{proof}
    See Appendix~\ref{proof_cor_space_of_hom}
\end{proof}

%%-------- End of Corollary -------------

\noindent Notice that the dimension of $\Theta_{\ccalA_{0}}$ provides us with
a measure of the richness or degrees of freedom we have to choose $\theta$. The larger the dimension of
$\Theta_{\ccalA_{0}}$, the more diversity and flexibility we have to build a
$\theta$ that satisfies~\eqref{eq_def_hom_asm_1}. In particular, if the dimension of
$\Theta_{\ccalA_{0}}$ is $0$, the only homomorphism is the zero map itself.

In classical ASP, while the algebra $\mathcal{A}$ is arbitrary and can be
commutative or non-commutative, it is polynomial. This ensures that any
implementation of the elements in $\mathcal{A}$ is also a polynomial function
of the shift operators of the signal model. We highlight this fact in the
following corollary.

%%-----------------------------------
%%----------- COROLLARY -----------------
%%-----------------------------------

\begin{corollary}\label{corll_hom_asp_shiftoper}
Let $(\mathcal{A},\mathcal{H}_{1},\rho_{1})$ and $(\mathcal{A},\mathcal{H}_{2},\rho_{2})$ be two algebraic signal models, assume $\ccalA$ is generated by a single element $g$, and let $\mathbf{S}=\rho_{1}(g)$ and $\widetilde{\mathbf{S}}=\rho_{2}(g)$. Then, if $\theta: \mathcal{H}_{1}\to \mathcal{H}_{2}$ is a homomorphism as in Definition~\ref{def_hom_asm}, it follows that 
\begin{equation}\label{eq_corll_hom_asp_shiftoper_1}
     \theta\,
             p(\mathbf{S})
               =
               p\big(\widetilde{\mathbf{S}}\big)\,
               \theta
               ,
\end{equation}
where $p(g)\in\ccalA$ is any polynomial function of $g$.
\end{corollary}

\begin{proof}
     See Appendix~\ref{proof_corll_hom_asp_shiftoper}
\end{proof}

%%-------- End of Corollary -------------

This result emphasizes even more the fact that a homomorphism between signal
models is interchangeable with a polynomial operator whose independent
variables are interchanged between the shift operators of the two signal
models.

%% file: figures/hom_design_basic_diagram.tex
%%%%%%%%%%%%%%%%%%%%%%% Colors %%%%%%%%%%%%%%%%%%%%
% Colors come from asp_colors.sty (loaded in the main file)
% Style: navy = structure/signals, gold = abstract algebra & transfer,
%        light cividis fills = spaces (matches the paper's figure family)

\usetikzlibrary{positioning,decorations.pathreplacing,shapes}
\usetikzlibrary{arrows}

%%%%%%%%%%%%%%%%%% Settings of Blocks and picture %%%%%%%%%%%%%%%%%%%

\def\scale{1.3}
\def\unit{\scale cm}

% Block styles
\tikzstyle{set} = [rectangle,
rounded corners = 0.2*\unit,
inner sep=0pt,
draw,
anchor = center,
line width=0.9pt]

\tikzstyle{vectorspace} = [ set, color=ASPNavy,
fill=ASPVeryLightNavy,
minimum width  = 2*\unit,
minimum height = 1.5*\unit]

\tikzstyle{endomorphisms} = [ vectorspace, color=Cividis03,
fill=CividisLight03!35,
minimum height = 1.2*\unit]

\tikzstyle{algebra} = [endomorphisms, color=ASPGold!70!black,
fill=ASPLightGold,
minimum width = 1.5*\unit]

\tikzstyle{dot} = [ circle,
minimum width  = 0.15*\unit,
inner sep=0pt,
draw,
anchor = center ]

{\fontsize{8}{8}\selectfont

\begin{tikzpicture}[rounded corners]

	%%%%%%%%%%%%%%%%%%%%%%%%%%%%%%%%%%%%%%%%%%%%%%%%%%%%%%
	%%%%%%%%%%%%%%%%%%%%%%   ALGEBRA  %%%%%%%%%%%%%%%%%%%%%
	%%%%%%%%%%%%%%%%%%%%%%%%%%%%%%%%%%%%%%%%%%%%%%%%%%%%%%

	\path (0,0) node [algebra, anchor = north east] (LG) {};
	\path (LG.north) ++ (0,-0.1) node [below,color=ASPGold!60!black] {$\mathcal{A}$};
	\path (LG.center) ++ (0,0.3) node [below,color=black] {$p(t)$};

	% Element of the Algebra (the filter, in gold: abstract entity)
	\path (LG.south) ++ (0,0.3) node [dot, draw=ASPGold!60!black,
	                                  fill=ASPGold] (a) {};

%%%%%%%%%%%%%%%%%%%%%%%%%%%%%%%%%%%%%%%%%%%%%%%%%%%%%%%
%%%%%%%%%%%%%%%%%%% REPRESENTATION 1 %%%%%%%%%%%%%%%%%%%
%%%%%%%%%%%%%%%%%%%%%%%%%%%%%%%%%%%%%%%%%%%%%%%%%%%%%%%

	%%----------------------------------------------------------
	%%-------------  ENDOMORPHISMS OF H1  -----------------------
	%%----------------------------------------------------------

	\path (LG.east) ++ (2, 0)
	node [endomorphisms, anchor=center] (End1) {};
	\path (End1.north) ++ (0.0, 0) node [below, color=Cividis02] {$\text{End}(\mathcal{H}_1)$};

	% Adding an endomorphism element (the realized filter)
	  \path (End1.south) ++ (0.0, 0.5) node [dot, draw=Cividis02,
	                                          fill=Cividis03] (e1) {};
	\path (e1) node [above] {$\rho_{1}(p(t))=p(\mathbf{S})$};

	%%----------------------------------------------------------
	%%----------------  VECTOR SPACE H1  ------------------------
	%%----------------------------------------------------------

	\path (End1.south)++(0,-0.5) node [vectorspace, anchor=north] (M1) {};
	\path (M1.south) ++ (0, 0.1) node [above, color=ASPNavy] {$\mathcal{H}_1$};

	   % Element of the vector space
	   \path (M1) ++ (-0.6,0.2) node [dot, draw=ASPNavy, fill=ASPNavy] (x) {};
	   \path (x.south)++(0,-0.1) node [below, color=black] {$\mathbf{x}$};

	   % Element filtered in the vector space
	   \path (M1) ++ (0.6,0.2) node [dot, draw=ASPNavy, fill=ASPNavy] (ex) {};
	   \path (ex.south) node [below, color=black] {$p(\mathbf{S})\mathbf{x}$};

		% Arrow going from x to Ex (filtering inside H1)
		\path (e1)+(-0.5,0.55) coordinate (c1);
		\path (e1)+(0.5,0.55) coordinate (c2);
		\path [draw, -stealth, line width=0.7pt, color=ASPNavy!75]
		      (x) .. controls (c1) and (c2) .. (ex);

%%%%%%%%%%%%%%%%%%%%%%%%%%%%%%%%%%%%%%%%%%%%%%%%%%%%%%%
%%%%%%%%%%%%%%%%%%% REPRESENTATION 2 %%%%%%%%%%%%%%%%%%%
%%%%%%%%%%%%%%%%%%%%%%%%%%%%%%%%%%%%%%%%%%%%%%%%%%%%%%%

%%----------------------------------------------------------
%%-------------  ENDOMORPHISMS OF H2  -----------------------
%%----------------------------------------------------------

\path (LG.west) ++ (-2, 0)
node [endomorphisms, anchor=center] (End2) {};
\path (End2.north) ++ (0.0, 0) node [below, color=Cividis02] {$\text{End}(\mathcal{H}_2)$};

% Adding an endomorphism element (the realized filter)
 \path (End2.south) ++ (0.0, 0.5) node [dot, draw=Cividis02,
                                         fill=Cividis03] (e2) {};
\path (e2) node [above] {$\rho_{2}(p(t))=p(\widetilde{\mathbf{S}})$};

%%----------------------------------------------------------
%%----------------  VECTOR SPACE H2  ------------------------
%%----------------------------------------------------------

\path (End2.south)++(0,-0.5) node [vectorspace, anchor=north] (M2) {};
\path (M2.south) ++ (0, 0.1) node [above, color=ASPNavy] {$\mathcal{H}_2$};

   % Element of the vector space (the transferred signal)
   \path (M2) ++ (-0.6,0.2) node [dot, draw=ASPNavy, fill=ASPNavy] (y) {};
   \path (y.south)++(0,-0.1) node [below, color=black] {$\theta(\mathbf{x})$};

   % Element filtered in the vector space
   \path (M2) ++ (0.65,0.2) node [dot, draw=ASPNavy, fill=ASPNavy] (ey) {};
   \path (ey.south) node [below, xshift=-4pt, color=black] {$p(\widetilde{\mathbf{S}})\,\theta(\mathbf{x})$};

		% Arrow going from y to Ey (filtering inside H2)
		\path (e2)+(-0.5,0.55) coordinate (c1);
		\path (e2)+(0.5,0.55) coordinate (c2);
		\path [draw, -stealth, line width=0.7pt, color=ASPNavy!75]
		      (y) .. controls (c1) and (c2) .. (ey);

	%%%%%%%%%%%%%%%%%%%%%%%%%%%%%
	%%%%%%%%%%%%% ARROWS %%%%%%%%%
	%%%%%%%%%%%%%%%%%%%%%%%%%%%%%

	% Arrows from the Algebra to End(H1) and End(H2):
	% the homomorphisms rho_1, rho_2 realizing the abstract filter

	\path [draw, -stealth, line width = 1.1pt, Cividis03!85]
	(a) edge [bend right] node [below, color=Cividis02] {$\rho_{1}~~~$} (e1);

	\path [draw, -stealth, line width = 1.1pt, Cividis03!85]
	    (a) edge [bend left] node [below, color=Cividis02] {$\rho_{2}~~~$} (e2);

	% Homomorphism (transfer) arrows: gold, matching the figure family

	\path [draw, -stealth, line width = 1.1pt, color=ASPGold!80!black]
	   (x) edge [bend right] node [below, color=ASPGold!55!black] {$\theta~~~$} (y);

	\path [draw, -stealth, line width = 1.1pt, color=ASPGold!80!black]
	   (ex) edge [bend left] node [below, color=ASPGold!55!black] {$\theta~~~$} (ey);

\end{tikzpicture}

}

%% file: v09/sec_hom_trans_equiv_fsm.tex
%!TEX root =../compiler_paper.tex

%%%%%%%%%%%%%%%%%%%%%%%%%%%%%%%%%%%%%%%%%%%
%%%%%%%%%%%%% SECTION %%%%%%%%%%%%%%%%%%%%%
%%%%%%%%%%%%%%%%%%%%%%%%%%%%%%%%%%%%%%%%%%%

\section{Homomorphisms for Transferability in Finite ASMs}
\label{sec_finite_asm}

A cautious review of the condition in~\eqref{eq_def_hom_asm_1} emphasizes
that ensuring the existence of $\theta$ for all the elements $a\in\mathcal{A}$,
although possible, is quite restrictive, given the fact that when signals are
processed, one focuses on particular subclasses of filters. To this end, we
emphasize transferability and equivariance leveraging homomorphisms between
signal models restricted to subclasses of filters in $\mathcal{A}$. We make
this precise in the following definition.

%%----------------------------------------
%%------   D E F I N I T I O N   ---------
%%----------------------------------------

\begin{definition}[Transferability]
\label{def_transferability}
Let $(\mathcal{A},\mathcal{H}_{1},\rho_{1})$ and $(\mathcal{A},\mathcal{H}_{2},\rho_{2})$ be two algebraic signal models, and let $\mathcal{A}_{0}\subseteq\mathcal{A}$ be a subset of filters in $\mathcal{A}$. Then, we will say that the filter set $\mathcal{A}_{0}$ is transferable between $(\mathcal{A},\mathcal{H}_{1},\rho_{1})$ and $(\mathcal{A},\mathcal{H}_{2},\rho_{2})$ if there exists a nonzero linear map $\theta:\mathcal{H}_{1}\to\mathcal{H}_{2}$ such that 
\begin{equation}\label{eq_def_transferability_1}
 \theta\left( 
            \rho_{1}(a)
                    \mathbf{x}
       \right)
             =
              \rho_{2}(a)
                  \theta\left(
                            \mathbf{x}
                        \right)
                        ,
                        \quad
                        \forall~ a\in\mathcal{A}_{0},~\mathbf{x}\in\mathcal{H}_{1}
                        .
\end{equation}
\end{definition}

%%-------   End of Definition   ----------

\noindent This is, for our discussion, transferability exists when for a
subset of filters under consideration (being implemented in the signal
models), there is a nonzero linear map $\theta:\mathcal{H}_{1}\to \mathcal{H}_{2}$
that determines a homomorphism between signal models restricted to
$\ccalA_{0}$. The requirement that $\theta$ be nonzero is essential: the zero
map satisfies~\eqref{eq_def_transferability_1} for every $\ccalA_{0}$, so
without it the notion would be vacuous. Equivalently, in the notation of
Corollary~\ref{cor_space_of_hom}, the filter set $\ccalA_{0}$ is transferable
precisely when $\dim\left(\Theta_{\ccalA_{0}}\right)>0$.

It is worth pointing out that the notion of transferability in
Definition~\ref{def_transferability} has made no assumption on the nature of
$\mathcal{H}_{1}$ and $\mathcal{H}_{2}$. Then, while its physical
interpretation is self evident, we note that it contains
the notion of equivariance as a particular case when
$\mathcal{H}_{1}=\mathcal{H}_{2}$. We develop this connection in
Section~\ref{sec_eqvml}.

%%%%%%%%%%%%%%%%%%%%%%%%%%%%%%%%%%%%%%%%%%%%%%%%%%%
%%%%%%%%   S U B   -   S E C T I O N  %%%%%%%%%%%%%
%%%%%%%%%%%%%%%%%%%%%%%%%%%%%%%%%%%%%%%%%%%%%%%%%%%

\subsection{Conditions and Limits of Transferability: A Primer}

We now derive conditions for the existence of transferability between ASMs,
restricting our attention to signal models $(\mathcal{A},\mathcal{H},\rho)$
in which $\text{dim}(\mathcal{H})$ is finite and $\ccalA$ has a single
generator $g$. In this setting, each signal model has a single shift
operator. With this in place, we now state our first result.

%%---------------------------------------------------------------
%%---------------     T H E O R E M    --------------------------
%%---------------------------------------------------------------

\begin{theorem}\label{thm_trans_asp_finite}
Let $(\mathcal{A},\mathcal{H}_{1},\rho_{1})$ and $(\mathcal{A},\mathcal{H}_{2},\rho_{2})$ be algebraic signal models with $\dim(\mathcal{H}_1)=N_{1}<\infty$ and $\dim(\mathcal{H}_2)=N_{2}<\infty$, where $\mathcal{A}$ is generated by a single element $g$. Let $\lambda_{i}$ and $\widetilde{\lambda}_{j}$ denote, respectively, the $i$-th eigenvalue of $\rho_{1}(g)=\mathbf{S}$ and the $j$-th eigenvalue of $\rho_{2}(g)=\widetilde{\mathbf{S}}$. Suppose the linear map $\theta:\ccalH_{1}\to\ccalH_{2}$ satisfies
\begin{equation}\label{eq_thm_trans_asp_finite_1}
\theta\, p(\mathbf{S})\,\mathbf{x} 
      = 
       p\big(\widetilde{\mathbf{S}}\big)\,\theta\,\mathbf{x}, \qquad \forall~\, \mathbf{x}\in\mathcal{H}_{1}.    
\end{equation}
If $p(\lambda_{i})\neq p\big(\widetilde{\lambda}_{j}\big)$ for all $i=1,\ldots,N_{1}$ and $j=1,\ldots,N_{2}$, then $\theta\mathbf{x}=\mathbf{0}$ for all $\mathbf{x}\in\mathcal{H}_{1}$. Moreover, if we identify $\bbS$ and $\widetilde{\bbS}$ with matrices, $\dim\left(\Theta_{\ccalA_{0}}\right)$ in~\eqref{eq_cor_space_of_hom_1} with $\ccalA_{0}=\{p(g)\}$ equals the geometric multiplicity of the zero eigenvalue of the matrix
\begin{equation}\label{eq_thm_trans_asp_finite_2}
\mathbf{T}
     =
       \mathbf{I}_{N_{1}} \otimes p\big(\widetilde{\mathbf{S}}\big) \;-\; p(\mathbf{S})^{\mathsf{T}} \otimes \mathbf{I}_{N_{2}},   
\end{equation}
whose eigenvalues are precisely the differences $p\big(\widetilde{\lambda}_{j}\big)-p(\lambda_{i})$.
\end{theorem}

\begin{proof}
See Appendix~\ref{app_proof_thm_finite}.
\end{proof}

%%---------- End of Theorem -------------

\noindent  Theorem~\ref{thm_trans_asp_finite} shows that transferability is governed strongly by the number of terms for which $p(\lambda_{i})-p(\widetilde\lambda_{j})=0$. At one extreme, when $p(\lambda_{i})\neq p(\widetilde\lambda_{j})$ for \emph{every} pair $i=1,\ldots,N_1$, $j=1,\ldots,N_2$, transferability fails entirely as $\theta$ is forced to be the zero map. This can happen for two distinct reasons: a substantial change between domains, reflected in shift operators with very different spectra, or a filter $p(t)$ with high variability over a domain containing the spectra of both shift operators, which separates the eigenvalue images even when the spectra themselves are close. More generally, the richness available in the choice of $\theta$ in Definition~\ref{def_transferability} is directly controlled by \emph{how many} such terms vanish. For diagonalizable shift operators $\dim\left(\Theta_{\ccalA_{0}}\right)$ gives the precise count, while for defective ones the two quantities differ, since the theorem delivers the geometric rather than the algebraic multiplicity of the zero eigenvalue of $\mathbf{T}$; the correct refinement is given in Proposition~\ref{prop:dimjordan}. This yields a clear tradeoff: the richer we allow $\theta$ to be, the more coincidences $p(\lambda_{i})=p(\widetilde\lambda_{j})$ we must impose, which in turn requires either limited change in the eigenvalues of the shift operators across domains, or a filter $p(t)$ that is flat or constant over the portion of the spectral axis where the two spectra fail to align. In other words, richness in $\theta$ trades directly against how far apart the two domains, and the filter's response over their spectra, are allowed to be. This tradeoff also points to a design principle. If $p(t)$ is fixed in advance, or learned on a source domain, guaranteeing transferability to a target domain requires ensuring that at least some differences $p(\lambda_{i})-p(\widetilde\lambda_{j})$ vanish. One way to achieve this without sacrificing the filter's expressiveness is to restrict the richness of its amplitude response only where needed, while preserving high variability elsewhere. In the following section we provide a more rigorous characterization of the richness of $\theta$.

%%%%%%%%%%%%%%%%%%%%%%%%%%%%%%%%%%%%%%%%%%%%%%%%%%%%%%%%%
%%%%%%%%%    S E C T I O N     
%%%%%%%%%%%%%%%%%%%%%%%%%%%%%%%%%%%%%%%%%%%%%%%%%%%%%%%%%

\section{Degree of Transferability of $\theta$: Finite Dimensional Models}
\label{sec_deg_transf}

In the notion of transferability in Definition~\ref{def_transferability}, it
is clear that $\theta$ could allow trivial choices like the zero map. From our previous discussion, we also
know that the larger the dimension of $\Theta_{\ccalA_{0}}$, the more freedom
and the more enrichment we can provide to a specific map $\theta$. We now aim
to provide a more concrete characterization of such richness, understanding
the effect of $\theta$ on a given signal.

Throughout this section, and in order to measure the distortion introduced by
a homomorphism, we take $\ccalH_{1}$ and $\ccalH_{2}$ to be finite
dimensional complex Hilbert spaces, with inner products
$\langle\cdot,\cdot\rangle$ and induced norms $\|\cdot\|$. This additional
structure is what gives meaning to the notions of normality, adjoint,
orthogonal projection, and singular value used below, none of which are
available on a bare vector space.

%%%%%%%%%%%%%%%%%%%%%%%%%%%%%%%%%%%%%%%%%%%%%%%%%%%%%%%%%
%%%%%%%%%    S U B  -   S E C T I O N     

\subsection{When $\mathbf{S}$ and $\widetilde{\mathbf{S}}$ are Normal Operators}
\label{sec_normal_case}

Let $\mathbf{S}$ and
$\widetilde{\mathbf{S}}$ be \emph{normal} shift
operators (this standing assumption is relaxed in
Section~\ref{sec_nonnormal}) with spectral decompositions
\begin{equation}\label{eq_spectral_decomp_finite}
  \mathbf{S} = 
        \sum_{\lambda \in \sigma(\mathbf{S})} 
               \lambda\,
               \mathbf{P}_{\lambda},
  \qquad
  \widetilde{\mathbf{S}} =
        \sum_{\mu \in \sigma(\widetilde{\mathbf{S}})} 
               \mu\,
               \mathbf{Q}_{\mu},
\end{equation}
where $\mathbf{P}_\lambda$ and $\mathbf{Q}_\mu$ denote the orthogonal
spectral projections, with eigenspaces
$\mathcal{E}_\lambda = \ran\mathbf{P}_\lambda$ and
$\widetilde{\mathcal{E}}_\mu = \ran\mathbf{Q}_\mu$. Fix a polynomial
$p \in \mathbb{C}[t]$ (the \emph{filter}). We study the linear maps
$\theta : \ccalH_1 \to \ccalH_2$
satisfying~\eqref{eq_thm_trans_asp_finite_1}, and we write
$
\Theta_{p}
     :=
     \Theta_{\ccalA_{0}}
     \quad
     \text{with}
     \quad
     \ccalA_{0}
     =
     \{ p(g) \}
     ,
$
where $\Theta_{\ccalA_{0}}$ is the set in~\eqref{eq_cor_space_of_hom_1}. Then,
$\Theta_p$ is the space of homomorphisms of the fixed filter $p$. Now we introduce a definition that will allow us to characterize the effect of $\theta$ on the invariant subspaces associated to the signal models involved.

%%-----------------------------------
%%------------ FIGURE ---------------
%%-----------------------------------

\begin{figure*}
%\centering
       \begin{subfigure}{.49\linewidth}
          \resizebox{\textwidth}{!}{\input{\pathfigs/tikz_diagrams/fig_coincidence_classes.tex}}
          %\caption{}
       \end{subfigure}
       \begin{subfigure}{.49\linewidth}
          \resizebox{\textwidth}{!}{\input{\pathfigs/tikz_diagrams/fig_coincidence_nonflat.tex}}
          %\caption{}
       \end{subfigure}      
\caption{Coincidence classes arise from equality of the \emph{filtered}
eigenvalues, with or without flatness of the filter.
Left: A filter with a flat plateau: every eigenvalue
falling inside the plateau is mapped to the same value, so the coincidence
class is produced by the flatness itself.
Right: A non-flat, non-monotonic filter: aliasing
here arises purely from $p(t)$ having two local maxima, since the level
$c_{1}$ is crossed once ascending and once descending on the first hump, and
again while ascending into the second. The eigenvalues $\lambda_{1}$ and
$\lambda_{2}$ of $\mathbf{S}$ (navy dots), located near entirely different
extrema, still satisfy $p(\lambda_{1})=p(\lambda_{2})=c_{1}$, while a single
eigenvalue of $\widetilde{\mathbf{S}}$ (gold triangle) on the descending edge
of the first hump completes the class ($m_{c_{1}}=2$,
$\widetilde{m}_{c_{1}}=1$). The level $c_{2}$, reached only near the taller
second hump, gives an ordinary one-to-one coincidence
($m_{c_{2}}=\widetilde{m}_{c_{2}}=1$). An eigenvalue of $\mathbf{S}$ past the
second hump is annihilated, as its filtered image matches no eigenvalue of
$\widetilde{\mathbf{S}}$.}
\label{fig_coincidence}
\end{figure*}
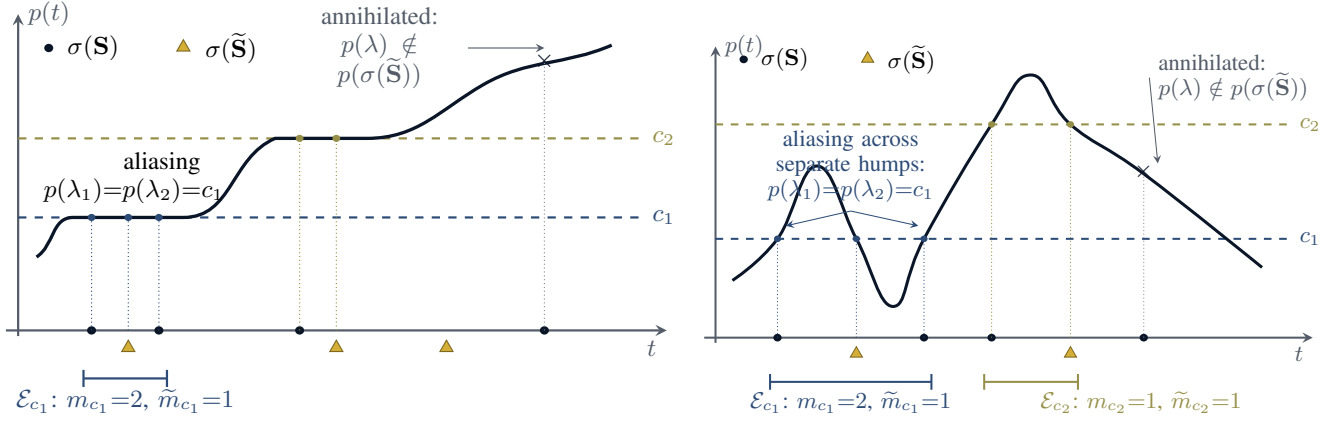

%%--------- End of Figure -----------

%%-------------------------------------------------
%%------      D E F I N I T I O N      ------------
%%-------------------------------------------------

\begin{definition}[Coincidence values and coincidence spaces]\label{def:coincidence}
The \emph{coincidence set} of the pair
$(\mathbf{S},\widetilde{\mathbf{S}})$ under $p$ is
\begin{equation}\label{eq_coincidence_set}
  \ccalC_{p} \;:=\; p\big(\sigma(\mathbf{S})\big) \,\cap\, p\big(\sigma(\widetilde{\mathbf{S}})\big)
  \;\subset\; \mathbb{C}.
\end{equation}
For each $c \in \ccalC_{p}$, we define the coincidence spaces
\begin{equation}\label{eq_coincidence_space_1}
  \ccalE_c :=
          \bigoplus_{\substack{\lambda \in \sigma(\mathbf{S}) \\ p(\lambda) = c}}
               \mathcal{E}_{\lambda}
          = \ker\big(p(\mathbf{S}) - c\, \mathbf{I}\big),
\end{equation}
\begin{equation}\label{eq_coincidence_space_2}
\widetilde{\ccalE}_{c} :=
       \bigoplus_{\substack{\mu \in \sigma(\widetilde{\mathbf{S}}) \\ p(\mu) = c}}
             \widetilde{\mathcal{E}}_{\mu}
     = \ker\big(p(\widetilde{\mathbf{S}}) - c\, \mathbf{I}\big),  
\end{equation}
with \emph{coincidence multiplicities}
$m_c := \dim \ccalE_c$ and $\widetilde{m}_c := \dim \widetilde{\ccalE}_{c}$.
\end{definition}

%%------      End of Definition      -------------- 

\noindent One particular aspect of Definition~\ref{def:coincidence} deserves emphasis. The filter $p$ may \emph{alias} distinct eigenvalues of $\mathbf{S}$ (or of
$\widetilde{\mathbf{S}}$) into a single value $c$. Then, the space $\mathcal{E}_c$ merges several eigenspaces of $\mathbf{S}$. In this context, notice that the existence of such coincidence spaces may be the result of the flatness of the filter on a portion of the spectral axes, but such flatness is not a requirement for the non-emptiness of $\ccalE_{c}$ -- see Fig.~\ref{fig_coincidence}. As will become evident in the following result, encapsulating the invariant subspaces of $p(\bbS)$ in $\ccalE_{c}$ will allow us to have a concrete object of study to understand the effect of $\theta$, just in the same way that invariances allow us to understand the behavior of an algebraic filter on the Fourier representation. We emphasize this idea in the following result.

%%-------------------------------------------------
%%------      L E M M A       ---------------------
%%-------------------------------------------------

\begin{lemma}\label{lem:support}
Let $\theta \in \Theta_{p}$. Then
\begin{equation}\label{eq:support}
\theta
    =
     \sum_{c \,\in\, \ccalC_{p}} 
        \widetilde{\mathbf{\Pi}}_{c}\,
        \theta\,
        \mathbf{\Pi}_c 
        ,
\end{equation}
where $\mathbf{\Pi}_c$ and $\widetilde{\mathbf{\Pi}}_c$ denote the orthogonal projections onto $\ccalE_c$ and $\widetilde{\ccalE}_{c}$, respectively. In particular, $\theta(\ccalE_{c}) \subseteq \widetilde{\ccalE}_{c}$ for every
$c \in \ccalC_{p}$, and $\theta$ annihilates every eigenspace
$\mathcal{E}_\lambda$ of $\mathbf{S}$ with
$p(\lambda) \notin p(\sigma(\widetilde{\mathbf{S}}))$. Conversely, for any linear
$\mathbf{A}:\ccalH_{1}\to\ccalH_{2}$, the map
$\sum_{c\in\ccalC_{p}}\widetilde{\mathbf{\Pi}}_{c}\,\mathbf{A}\,\mathbf{\Pi}_{c}$
belongs to $\Theta_p$, so~\eqref{eq:support} characterizes $\Theta_{p}$.
\end{lemma}

\begin{proof}
    See Appendix~\ref{proof_lem_support}.
\end{proof}

%%---------         End of Lemma       ------------

\noindent Lemma~\ref{lem:support} provides an extension of the spectral theorem for the homomorphism $\theta$, interrelating the invariances between the two algebraic signal models associated with repeated eigenvalues of $p(\bbS)$ and $p(\widetilde{\bbS})$, and its action is well defined on any signal $\bbx\in\ccalH_1$ in the source ASM.

With the description of the action of $\theta$ as in~\eqref{eq:support} we
introduce some terminology that will facilitate the study of the properties
of homomorphisms in general. For $\theta \in \Theta_{p}$ and
$c \in \ccalC_{p}$, the \emph{block map at $c$} is the restriction
$\theta_c := \theta\big|_{\ccalE_{c}}: \ccalE_c \longrightarrow
\widetilde{\ccalE}_{c}$, which is well defined by Lemma~\ref{lem:support}
and is represented, in orthonormal bases of $\ccalE_c$ and
$\widetilde{\ccalE}_{c}$, by an $\widetilde{m}_c \times m_c$ complex matrix.
By~\eqref{eq:support}, $\theta$ is completely determined by, and decomposes
orthogonally into, its blocks $\{\theta_c\}_{c \in \ccalC_{p}}$. Notice that
the summand $\widetilde{\mathbf{\Pi}}_{c}\,\theta\,\mathbf{\Pi}_c$
in~\eqref{eq:support} is an operator on the ambient spaces, while
$\theta_{c}$ is the intrinsic map between the subspaces $\ccalE_c$ and
$\widetilde{\ccalE}_c$ themselves. The two determine each other without loss
of information, precisely because Lemma~\ref{lem:support} guarantees
$\theta(\ccalE_c)\subseteq\widetilde{\ccalE}_c$. We make this correspondence
explicit in Appendix~\ref{app_block_maps}, and in what follows we use the two
descriptions interchangeably.

We now introduce a definition to properly establish a measure for the degree of transferability achievable with a homomorphism.

%%-------------------------------------------------
%%------      D E F I N I T I O N      ------------
%%-------------------------------------------------

\begin{definition}[Spectral transfer efficiency]\label{def:eta}
Assume $\ccalC_{p}\neq\emptyset$, which by Lemma~\ref{lem:support} is
necessary for $\Theta_{p}$ to contain a nonzero element. For
$\theta\in\Theta_{p}$ and $c \in \ccalC_{p}$, let us define
\begin{equation}\label{eq_extreme_gains}
  \sigma_{\max}(\theta_c) := \max_{\substack{\mathbf{x} \in \ccalE_c \\ \|\mathbf{x}\| = 1}}
    \big\|\theta_c\, \mathbf{x}\big\|,
  \quad
  \sigma_{\min}(\theta_c) := \min_{\substack{\mathbf{x} \in \ccalE_c \\ \|\mathbf{x}\| = 1}}
    \big\|\theta_c\, \mathbf{x}\big\|.
\end{equation}
Then, the \emph{transfer efficiency at $c$} is
\begin{equation}\label{eq:etac}
  \eta_c(\theta) \;:=\;
  \begin{dcases}
    \frac{\sigma_{\min}^2(\theta_c)}{\sigma_{\max}^2(\theta_c)}
      & \text{if } \theta_c \neq 0, \\[1mm]
    \;0 & \text{if } \theta_c = 0,
  \end{dcases}
  \qquad \eta_c(\theta) \in [0,1],
\end{equation}
and the \emph{(global) richness} of $\theta$ is
\begin{equation}\label{eq:eta}
  \eta(\theta) \;:=\; \min_{c \,\in\, \ccalC_{p}} \eta_c(\theta) \;\in\; [0,1].
\end{equation}
\end{definition}

%%------      End of Definition      -------------- 

\noindent As a note of clarification in Definition~\ref{def:eta}, notice that $\sigma_{\max}(\theta_c)$ and $\sigma_{\min}(\theta_c)$ are, respectively, the square roots of the
largest and smallest eigenvalues of $\theta_c^{*}\theta_c$, the minimum being taken over all of $\ccalE_c$, so that
$\sigma_{\min}(\theta_c)=0$ whenever $\ker\theta_c \neq \{0\}$.

%%----------------------------------------
%%-----------    F I G U R E    ----------
%%---------------------------------------- 

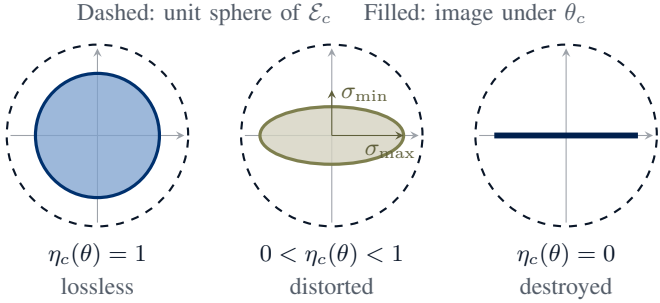
\begin{figure}
    \centering
        \input{\pathfigs/tikz_diagrams/fig_transfer_efficiency.tex}
    \caption{Geometric picture of the spectral transfer efficiency
$\eta_{c}(\theta)$: the unit sphere of $\mathcal{E}_{c}$ (dashed) mapped
by $\theta_{c}$ into its image (filled) in
$\widetilde{\mathcal{E}}_{c}$. Left ($\eta_{c}(\theta)=1$): a scaled
isometry maps the sphere onto a uniformly shrunk circle -- every
direction has the same gain
$\sigma_{\min}(\theta_{c})=\sigma_{\max}(\theta_{c})$, and transfer is
lossless. Middle ($0<\eta_{c}(\theta)<1$): the image is an ellipse with
distinct semi-axes $\sigma_{\max}(\theta_{c}),\sigma_{\min}(\theta_{c})$,
so gains differ and the signal is distorted. Right
($\eta_{c}(\theta)=0$): the image collapses to a segment,
$\sigma_{\min}(\theta_{c})=0$, and that direction's information is
destroyed before reaching $\widetilde{\mathcal{E}}_{c}$.}
\label{fig_richness_measure}
\end{figure}

%%-------     End of Figure     ----------

Definition~\ref{def:eta} requires no positivity assumption on the singular
values of $\theta$: whenever $\theta_c \neq 0$ we have
$\sigma_{\max}(\theta_c) > 0$, so the quotient in \eqref{eq:etac} is
defined, and a vanishing \emph{smallest} singular value appears in the
numerator, correctly reporting $\eta_c(\theta) = 0$. Rank deficiency of a
block is thus recorded as zero efficiency rather than as an undefined
expression. In particular, the condition $\eta(\theta) > 0$ is equivalent
to the injectivity of every block $\theta_c$.

Regarding the interpretability of the terms introduced in Definition~\ref{def:eta}, notice that $\eta_c(\theta) = 1$ if and only if $\theta_c$ is a nonzero scaled
isometry on $\ccalE_c$: every direction of the coincidence space at $c$ is
transferred with the same gain, i.e., transfer at the filtered frequency
$c$ is undistorted. At the other extreme, $\eta_c(\theta) = 0$ means that
some direction in $\ccalE_c$ is annihilated. Intermediate values quantify the
spectral distortion of the transfer within $\ccalE_c$. The measure is
\emph{intrinsic to the homomorphism structure}: it is evaluated exactly on
the $\{\ccalE_c\}_{c\in\ccalC_{p}}$ on which, by
Lemma~\ref{lem:support}, nontrivial action of $\theta$ is algebraically
possible. Notice that the zero map $\theta = 0$ satisfies \eqref{eq_thm_trans_asp_finite_1} and has
$\eta(\theta) = 0$ by convention. A rank-one map, e.g. a matrix with all
entries equal to a nonzero constant (when it happens to be a homomorphism), has
$\sigma_{\min}(\theta_c) = 0$ on every block of dimension $m_c \geq 2$ and
hence $\eta(\theta) = 0$. Both informationally degenerate cases are thus
detected by $\eta$. Figure~\ref{fig_richness_measure} provides an illustration of the interpretation of $\eta_{c}$.

Now, we turn our attention to the dimension of the space of homomorphisms and we show how it relates the dimensions of the block maps defined on the coincidence spaces.

%%---------------------------------------------------------------
%%---------------     P R O P O S I T I O N    ------------------
%%---------------------------------------------------------------

\begin{proposition}\label{prop:dimension}
For any polynomial $p(t)\in\mbC[t]$ we have that
\begin{equation}\label{eq_dimension_formula}
\Theta_p 
    \cong \bigoplus_{c \in \ccalC_{p}}
     \operatorname{Hom}\big(\ccalE_{c},\widetilde{\ccalE}_{c}\big),
  \quad\text{so that}\quad
  \dim \Theta_p =\sum_{c \in \ccalC_{p}} m_c\, \widetilde{m}_c .
\end{equation}
Consequently, the number of degrees of freedom available to construct
$\theta$ is maximized by maximizing the coincidences
$p(\lambda_i) = p(\widetilde{\lambda}_j)$ between eigenvalues
$\lambda_i \in \sigma(\mathbf{S})$ and
$\widetilde{\lambda}_j \in \sigma(\widetilde{\mathbf{S}})$, and the maximal
achievable rank of $\theta\in\Theta_{p}$ is
\begin{equation}\label{eq_max_rank}
\max_{\theta \in \Theta_p} \text{rank}(\theta)
      =
        \sum_{c\in\ccalC_{p}}
              \min\left(
                     m_{c},\widetilde{m}_{c}
                  \right).
\end{equation}
\end{proposition}

\begin{proof}
    See Appendix~\ref{proof_prop_dimension}.
\end{proof}

%%---------- End of Proposition -------------

Proposition~\ref{prop:dimension} refines the count of
Theorem~\ref{thm_trans_asp_finite}: for normal (more generally,
diagonalizable) shifts, the geometric multiplicity of the zero eigenvalue of $\mathbf{T}$ in~\eqref{eq_thm_trans_asp_finite_2} is exactly
$\sum_{c\in\ccalC_{p}} m_c\widetilde{m}_c$, since the zero eigenvalues of
$\mathbf{T}$ are indexed by the pairs $(i,j)$ with
$p(\lambda_i)=p(\widetilde{\lambda}_j)$, and there are
$m_c\,\widetilde{m}_c$ such pairs for each coincidence value $c$. When the
shifts are defective the two counts differ, and the correct refinement will be
given in Proposition~\ref{prop:dimjordan}.

To end this subsection, we present a result where we put emphasis on the conditions under which the maximum dimension of $\Theta_p$ is achieved in relationship with the transfer efficiency and the global richness of a homomorphisms.

%%---------------------------------------------------------------
%%---------------     P R O P O S I T I O N    ------------------
%%---------------------------------------------------------------

\begin{proposition}[Achievability of maximal richness]\label{prop:achievability}
Suppose $\ccalC_{p} \neq \emptyset$. Then:
\begin{enumerate}
  \item[\textup{(i)}] If
    $m_c > \widetilde{m}_c$ for some $c \in \ccalC_{p}$, then
    $\ker \theta_c \neq \{0\}$ for \emph{every} $\theta \in \Theta_p$, so
    $\eta_c(\theta) = 0$ and hence $\eta(\theta) = 0$ identically on
    $\Theta_p$, i.e. a dimension mismatch at any shared filtered frequency forces
    information loss, and no choice of homomorphism can repair it.
  \item[\textup{(ii)}]
    $\displaystyle\max_{\theta \in \Theta_p} \eta(\theta) = 1$ if and only if
    $m_c \leq \widetilde{m}_c$ for all $c \in \ccalC_{p}$. In that case it
    suffices to choose each block $\theta_c$ to be a scaled isometry
    $\ccalE_c \hookrightarrow \widetilde{\ccalE}_{c}$.
  \item[\textup{(iii)}] There exists an
    \emph{injective} $\theta \in \Theta_p$ with $\eta(\theta) = 1$ if and only
    if
    \begin{equation}\label{eq:injectivity}
      p\big(\sigma(\mathbf{S})\big) \subseteq p\big(\sigma(\widetilde{\mathbf{S}})\big)
      \quad\text{and}\quad
      m_c \leq \widetilde{m}_c \quad \forall\, c \in \ccalC_{p} .
    \end{equation}
    In that case $\theta$ may be chosen with
    $\theta^{*}\theta = \mathbf{I}$, i.e., a global isometry of $\ccalH_1$
    into $\ccalH_2$.
\end{enumerate}
\end{proposition}

\begin{proof}
    See Appendix~\ref{proof_prop_achievability}.
\end{proof}

%%---------- End of Proposition -------------

%%%%%%%%%%%%%%%%%%%%%%%%%%%%%%%%%%%%%%%%%%%%%%%%%%%%%%%%%
%%%%%%%%%    S U B  -   S E C T I O N     
%%%%%%%%%%%%%%%%%%%%%%%%%%%%%%%%%%%%%%%%%%%%%%%%%%%%%%%%%

\subsection{Beyond Normality of $\mathbf{S}$ and $\widetilde{\mathbf{S}}$}
\label{sec_nonnormal}

The normality of $\mathbf{S}$ and $\widetilde{\mathbf{S}}$ was used at
exactly two points of the preceding analysis: it makes the spectral
projections $\mathbf{\Pi}_c, \widetilde{\mathbf{\Pi}}_c$ \emph{orthogonal}, and $p(\mathbf{S})$ acts on $\ccalE_c$ as
the exact scalar $c$. Neither is needed for the algebraic skeleton, and we
now extend the analysis to shifts that are not normal, treating two
scenarios separately.

%%%%%%%%%%%%%%%%%%%%%%%%%%%%%%%%%%%%%%%%%%%%%%%%%%%%%%%%%
%%%%%%%%%    S U B  -  S U B  -   S E C T I O N     
%%%%%%%%%%%%%%%%%%%%%%%%%%%%%%%%%%%%%%%%%%%%%%%%%%%%%%%%%

\subsubsection{\textbf{Diagonalizable, non-normal shifts}}

Suppose now that $\mathbf{S}$ and $\widetilde{\mathbf{S}}$ are
diagonalizable but not necessarily normal (e.g., directed graph shifts).
The spectral projections onto the coincidence spaces still exist, but they
are now the \emph{oblique} (Riesz) projections: $\mathbf{\Pi}_{c}$ projects
onto $\ccalE_{c}$ along $\bigoplus_{c'\neq c}\ccalE_{c'}$, which is no
longer the orthogonal complement of $\ccalE_{c}$. Equivalently, and without
reference to any basis,
\begin{equation}\label{eq_lagrange_projection}
\mathbf{\Pi}_{c}
     =
     \ell_{c}\big(p(\mathbf{S})\big)
     ,
     \qquad
     \widetilde{\mathbf{\Pi}}_{c}
     =
     \ell_{c}\big(p(\widetilde{\mathbf{S}})\big)
     ,
\end{equation}
where $\ell_{c}$ is the Lagrange polynomial taking the value $1$ at $c$ and
$0$ at the remaining points of
$p\big(\sigma(\mathbf{S})\big)\cup p\big(\sigma(\widetilde{\mathbf{S}})\big)$.
Each projection is therefore a polynomial in the corresponding filtered
shift, so the identities $p(\mathbf{S})\mathbf{\Pi}_c = c\,\mathbf{\Pi}_c$ and
$\widetilde{\mathbf{\Pi}}_{c}\, p(\widetilde{\mathbf{S}}) = c\,\widetilde{\mathbf{\Pi}}_{c}$
hold for the same reason as in the normal case, and they are all the proof
of Lemma~\ref{lem:support} used. Since $\ell_{c}$ is one and the same
polynomial on both sides, the intertwining
relation~\eqref{eq_thm_trans_asp_finite_1} passes directly to the
projections, $\theta\,\mathbf{\Pi}_{c} = \widetilde{\mathbf{\Pi}}_{c}\,\theta$,
which gives the support decomposition~\eqref{eq:support} on summing over
$c\in\ccalC_{p}$. Consequently, the following results still hold.

%%---------------------------------------------------------------
%%---------------     P R O P O S I T I O N    ------------------
%%---------------------------------------------------------------

\begin{proposition}\label{prop:diag}
Let $\mathbf{S}$ and $\widetilde{\mathbf{S}}$ be diagonalizable, and let
$\mathbf{\Pi}_{c},\widetilde{\mathbf{\Pi}}_{c}$ be the oblique spectral
projections of \eqref{eq_lagrange_projection}. Then
Lemma~\ref{lem:support} and Proposition~\ref{prop:dimension} hold verbatim,
as their statements involve no metric notion. Proposition~\ref{prop:achievability}
also holds verbatim, provided the extreme gains
$\sigma_{\min}(\theta_{c})$ and $\sigma_{\max}(\theta_{c})$ of
Definition~\ref{def:eta} are measured in the \emph{adapted} inner products
\begin{equation}\label{eq_adapted_inner_product}
\langle\mathbf{x},\mathbf{y}\rangle_{\oplus}
     :=
     \sum_{c\,\in\,\ccalC_{p}}
     \big\langle
        \mathbf{\Pi}_{c}\mathbf{x},\,\mathbf{\Pi}_{c}\mathbf{y}
     \big\rangle
     ,
\end{equation}
and its counterpart on $\ccalH_{2}$ built from
$\widetilde{\mathbf{\Pi}}_{c}$, which render the coincidence spaces
mutually orthogonal and coincide with the ambient inner products precisely
when $\mathbf{S}$ and $\widetilde{\mathbf{S}}$ are normal.
\end{proposition}
\begin{proof}
    See Appendix~\ref{proof_prop_diag}.
\end{proof}

%%---------------      End of Proposition   ---------------------

Proposition~\ref{prop:diag} preserves the algebraic content but degrades the
metric one. The decomposition
$\theta\mathbf{x} = \sum_{c\in\ccalC_p} \theta_c\,\mathbf{\Pi}_c\mathbf{x}$
survives because the proof of Lemma~\ref{lem:support} used only the
commutation identities above, never the orthogonality of the projections.
What that proof does \emph{not} give, once the projections cease to be
orthogonal, is the Pythagorean identity
$\|\theta\mathbf{x}\|^{2}=\sum_{c}\|\theta_{c}\mathbf{\Pi}_{c}\mathbf{x}\|^{2}$.
Expanding the norm of the sum instead produces
\begin{equation}\label{eq_cross_terms}
\|\theta\mathbf{x}\|^{2}
=
\sum_{c\,\in\,\ccalC_{p}}
     \big\|\theta_{c}\mathbf{\Pi}_{c}\mathbf{x}\big\|^{2}
\;+\;
\sum_{c\neq c'}
     \big\langle
          \theta_{c}\mathbf{\Pi}_{c}\mathbf{x}
          ,\,
          \theta_{c'}\mathbf{\Pi}_{c'}\mathbf{x}
     \big\rangle
     ,
\end{equation}
whose cross terms vanish only because
$\theta_{c}\mathbf{\Pi}_{c}\mathbf{x}\in\widetilde{\ccalE}_{c}$ and
$\theta_{c'}\mathbf{\Pi}_{c'}\mathbf{x}\in\widetilde{\ccalE}_{c'}$ are
orthogonal, which holds when $\widetilde{\mathbf{S}}$ is normal but fails in
general once the $\{\widetilde{\ccalE}_{c}\}_{c\in\ccalC_{p}}$ form only an
\emph{oblique} (complementary but non-orthogonal) direct sum of
$\ccalH_{2}$. The same failure occurs on the source side, where
$\|\mathbf{x}\|^{2}=\sum_{c}\|\mathbf{\Pi}_{c}\mathbf{x}\|^{2}$ likewise
requires $\ccalH_{1}=\bigoplus_{c}\ccalE_{c}$ to be an orthogonal, not
merely direct, sum. Both failures are governed by the discrepancy between
the ambient inner product and the adapted
one~\eqref{eq_adapted_inner_product}, encoded by the positive definite
\emph{metric operators}
\begin{equation}\label{eq_metric_operator}
\mathbf{G}
     :=
     \sum_{c\,\in\,\ccalC_{p}}
        \mathbf{\Pi}_{c}^{*}\mathbf{\Pi}_{c}
     ,
     \qquad
\widetilde{\mathbf{G}}
     :=
     \sum_{c\,\in\,\ccalC_{p}}
        \widetilde{\mathbf{\Pi}}_{c}^{*}\widetilde{\mathbf{\Pi}}_{c}
     ,
\end{equation}
which satisfy $\langle\mathbf{G}\mathbf{x},\mathbf{x}\rangle=\|\mathbf{x}\|_{\oplus}^{2}$
and reduce to the identity precisely when the corresponding coincidence
spaces are mutually orthogonal, as is the case whenever the shift is normal.

%%%%%%%%%%%%%%%%%%%%%%%%%%%%%%%%%%%%%%%%%%%%%%%%%%%%%%%%%
%%%%%%%%%    S U B  -  S U B  -   S E C T I O N     
%%%%%%%%%%%%%%%%%%%%%%%%%%%%%%%%%%%%%%%%%%%%%%%%%%%%%%%%%

\subsubsection{\textbf{Defective shifts and the role of $p'$}}

When $\mathbf{S}$ or $\widetilde{\mathbf{S}}$ has nontrivial Jordan
structure, the support lemma survives at the level of \emph{generalized}
eigenspaces. Throughout this subsection we therefore redefine
\begin{equation}\label{eq_generalized_coincidence}
\ccalE_{c}:=\ker\!\big((p(\mathbf{S})-c\mathbf{I})^{N_{1}}\big),
\qquad
\widetilde{\ccalE}_{c}:=\ker\!\big((p(\widetilde{\mathbf{S}})-c\mathbf{I})^{N_{2}}\big),
\end{equation}
with $m_{c},\widetilde m_{c}$ their dimensions, this is, the algebraic rather
than the geometric multiplicities of $c$. The two coincide precisely in the
semisimple case, so this extends Definition~\ref{def:coincidence} rather than
conflicting with it. Taking $\mathbf{\Pi}_{c},\widetilde{\mathbf{\Pi}}_{c}$ to be the Riesz
projections onto these spaces --- necessarily oblique, since a defective
operator admits no orthogonal spectral projections --- the argument used
above applies unchanged, with each projection again a polynomial in the
corresponding filtered shift, now given by Hermite rather than Lagrange
interpolation: $\mathbf{\Pi}_{c}=q_{c}(p(\mathbf{S}))$ and
$\widetilde{\mathbf{\Pi}}_{c}=q_{c}(p(\widetilde{\mathbf{S}}))$, for the
polynomial $q_{c}$ equal to $1$ at $c$ and to $0$ at the remaining points of
$p(\sigma(\mathbf{S}))\cup p(\sigma(\widetilde{\mathbf{S}}))$, in each case
to the order of the largest Jordan block occurring at that point in either
$p(\mathbf{S})$ or $p(\widetilde{\mathbf{S}})$. Since $q_{c}$ is one and the
same polynomial on both sides, $\theta\,\mathbf{\Pi}_{c}=\widetilde{\mathbf{\Pi}}_{c}\,\theta$,
and summing over $c\in\ccalC_{p}$ gives the decomposition~\eqref{eq:support}
with blocks $\theta_{c}:\ccalE_{c}\to\widetilde{\ccalE}_{c}$, exactly as in
Lemma~\ref{lem:support}. The blocks, however, are \emph{no
longer free}. We can see this by writing
\begin{equation}\label{eq_nilpotent_parts}
  p(\mathbf{S})\big|_{\ccalE_c} = c\,\mathbf{I} + \mathbf{N}_c,
  \qquad
  p(\widetilde{\mathbf{S}})\big|_{\widetilde{\ccalE}_{c}} = c\,\mathbf{I} + \widetilde{\mathbf{N}}_c,
\end{equation}
with $\mathbf{N}_c, \widetilde{\mathbf{N}}_c$ nilpotent. Then, each block must
additionally intertwine the nilpotent parts,
\begin{equation}\label{eq:nilpotent}
  \theta_c\, \mathbf{N}_c \;=\; \widetilde{\mathbf{N}}_c\, \theta_c .
\end{equation}
The effect of this on the dimension of the space of homomorphisms is presented in the following result.

%%---------------------------------------------------------------
%%---------------     P R O P O S I T I O N    ------------------
%%---------------------------------------------------------------

\begin{proposition}[Dimension formula, defective case]\label{prop:dimjordan}
Let the Jordan block sizes of $p(\mathbf{S})$ at $c$ be
$p_1 \geq p_2 \geq \cdots$ and those of $p(\widetilde{\mathbf{S}})$ at $c$ be
$q_1 \geq q_2 \geq \cdots$. Then
\begin{equation}\label{eq:dimjordan}
  \dim \Theta_p
  \;=\;
  \sum_{c \,\in\, \ccalC_{p}} \;\sum_{i,\,j} \min\big(p_i,\, q_j\big),
\end{equation}
which reduces to $\sum_c m_c \widetilde m_c$ exactly in the
case where $p_i \equiv q_j \equiv 1$. Moreover, a map satisfying
\eqref{eq:nilpotent} with $\mathbf{N}_c \neq 0$ or
$\widetilde{\mathbf{N}}_c \neq 0$ is a scaled isometry only under exact
matching of the Jordan structures, so $\eta(\theta) = 1$ is in general
\emph{unachievable}, i.e. defectiveness is an intrinsic obstruction to lossless
transfer.
\end{proposition}

\begin{proof}
   See Appendix~\ref{proof_prop_dimjordan}. 
\end{proof}

%%------       End of   P R O P O S I T I O N         -----------

Unlike the case where $p_i \equiv q_j \equiv 1$, in which $m_c\leq\widetilde m_c$
already suffices for $\eta_c(\theta)=1$
(Proposition~\ref{prop:achievability}(ii)), a dimension surplus
$\widetilde m_c>m_c$ does not compensate for mismatched Jordan structure.
For instance, a single Jordan block of size $p_1=5$ cannot be intertwined
injectively --- let alone isometrically --- into any direct sum of blocks of
size $1$, however many are available: every such map must annihilate the
range of $\mathbf{N}_{c}$, which is four-dimensional, so it reads off a
single coordinate of the source and has rank at most $1$. The loss of
freedom quantified by Proposition~\ref{prop:dimjordan} can therefore be
considerably more severe than a comparison of $m_c$ and $\widetilde m_c$
alone would suggest. The count~\eqref{eq:dimjordan} is the classical
dimension of the solution space of the intertwining equation for Jordan
matrices~\cite{horn1991topics}, and its consequence for design is that
eigenvalue coincidence $p(\lambda_i) = p(\widetilde{\lambda}_j)$ no longer
suffices: \emph{Jordan structure compatibility} enters, and mismatched
nilpotent orders silently kill entries of the blocks. There is, however, an
unexpected design lever, since the Jordan structure of $p(\mathbf{S})$ at
$c = p(\lambda)$ is not that of $\mathbf{S}$ at $\lambda$, but is governed by
the derivatives of the filter.

%%-------------------------------------------------
%%------      L E M M A       ---------------------
%%-------------------------------------------------

\begin{lemma}[Filtered Jordan structure]\label{lem:pprime}
Let $\mathbf{J}_k(\lambda)$ be a Jordan block of $\mathbf{S}$ of size $k$ at
$\lambda$, and let $r \geq 1$ be the order of the first nonvanishing
derivative of $p$ at $\lambda$, i.e.,
$p'(\lambda) = \cdots = p^{(r-1)}(\lambda) = 0$ and
$p^{(r)}(\lambda) \neq 0$. Then $p\big(\mathbf{J}_k(\lambda)\big)$ is similar
to a direct sum of $r$ Jordan blocks at $p(\lambda)$: $(k \bmod r)$
blocks of size $\lceil k/r \rceil$ and $r - (k \bmod r)$ blocks of size
$\lfloor k/r \rfloor$. In particular:
\begin{enumerate}
\item[\textup{(i)}] if $p'(\lambda) \neq 0$ \textup{(}$r = 1$\textup{)},
    the block passes to $p(\mathbf{S})$ intact;
\item[\textup{(ii)}] if $r \geq k$, the block splits completely into
    $k$ blocks of size $1$, i.e., $p(\mathbf{S})$ is \emph{semisimple} on the
    generalized eigenspace of $\lambda$.
\end{enumerate}
\end{lemma}

\begin{proof}
    See Appendix~\ref{app_proof_pprime}
\end{proof}

%%---------         End of Lemma       ------------

\noindent This result shows that the filter itself can \emph{repair}
defectiveness, in a sense worth making precise.

A filter $p$ for which $p-p(\lambda)$ has a zero of order $r$ at a defective
eigenvalue $\lambda$ does not merely shrink the Jordan block there, it
\emph{fragments} it into $r$ smaller ones. This matters for
Proposition~\ref{prop:dimjordan} because that count sums $\min(p_i,q_j)$ over
\emph{pairs} of blocks, so splitting increases the number of contributing
pairs. A single source block of size $2$ paired with a single target block of
size $2$ contributes only $\min(2,2)=2$ to~\eqref{eq:dimjordan}, half of the
semisimple count $m_c\widetilde{m}_c=4$, the shortfall being exactly the
price of the nilpotent constraint~\eqref{eq:nilpotent}; if the filter instead
splits both blocks completely ($r\geq2$ on both sides), the same pair of
coincidence spaces contributes $\sum_{i,j}\min(1,1)=4=m_c\widetilde{m}_c$ and
the full count is recovered.

Splitting does more than enlarge a sum. Once every block at a coincidence
class $c$ has size $1$, the nilpotent parts vanish and the
constraint~\eqref{eq:nilpotent} collapses to $0=0$, leaving $\theta_c$ free
to be \emph{any} element of
$\operatorname{Hom}(\ccalE_c,\widetilde{\ccalE}_c)$, exactly as in
Section~\ref{sec_normal_case}. Consequently, once $r$ exceeds the largest
Jordan block size at a given eigenvalue on \emph{both} shifts, the filtered
operators are semisimple at that value and every result of this section
applies to that coincidence class as if the shifts had been semisimple to
begin with. The repair is purely \emph{algebraic}, however. Derivative
flattening acts within the generalized eigenspace at a single eigenvalue,
splitting its Jordan blocks without merging it with any other eigenvalue, so
the partition into coincidence classes, and with it the obliqueness measured
by $\mathbf{G}$ and $\widetilde{\mathbf{G}}$ in~\eqref{eq_metric_operator},
is unchanged, and the penalty of Proposition~\ref{prop:ricdiag} still applies
after the Jordan structure has been repaired. Aliasing, by contrast, does
reduce it: merging eigenspaces into a common class removes the boundary
across which their obliqueness was measured, and a filter constant on the
whole spectrum gives $\mathbf{G}=\mathbf{I}$.

The two mechanisms therefore act on entirely different data. Aliasing
enlarges $\Theta_p$ by \emph{merging distinct eigenvalues} into a single
class, coarsening the partition of
$\sigma(\mathbf{S})\cup\sigma(\widetilde{\mathbf{S}})$ that determines
$\ccalC_p$; derivative flattening enlarges it by \emph{splitting a single
eigenvalue's Jordan block} from within, refining the internal structure of a
class already fixed. Yet both are properties of the filter $p$ alone, and
both enlarge the space of admissible transfers, reinforcing the view
developed throughout this section of the filter as the principal design
variable of transferability.

%% file: figures/tikz_diagrams/fig_coincidence_classes.tex
% Fig: coincidence classes of the filtered spectra; aliasing and annihilation
\begin{tikzpicture}[>=stealth, xscale=0.78, yscale=0.72, font=\small]

% axes
\draw[->, thick, ASPNavy!70] (-0.2,0) -- (10.6,0) node[below left]{$t$};
\draw[->, thick, ASPNavy!70] (0,-0.1) -- (0,5.6) node[right]{$p(t)$};

% coincidence levels
\draw[dashed, Cividis03, thick] (0,2) -- (10.2,2) node[right, text=Cividis03]{$c_{1}$};
\draw[dashed, Cividis08!80!black, thick] (0,3.4) -- (10.2,3.4) node[right, text=Cividis08!80!black]{$c_{2}$};

% filter curve
\draw[very thick, ASPNavy]
  (0.3,1.3) to[out=35,in=180] (0.9,2) -- (2.7,2)
  to[out=0,in=205] (4.2,3.4) -- (5.7,3.4)
  to[out=0,in=212] (7.6,4.35) to[out=32,in=205] (9.7,5.05);
%\node[text=ASPNavy] at (9.0,5.35) {$p(t)$};

% eigenvalues of S (navy dots) and tilde S (gold triangles) on the axis
\foreach \x in {1.2, 2.3, 4.6, 8.6}{
  \fill[ASPNavy] (\x,0) circle (2.1pt);}
\foreach \x in {1.8, 5.2, 7.0}{
  \node[fill=ASPGold, draw=ASPGold!60!black, isosceles triangle,
        isosceles triangle apex angle=60, rotate=90, inner sep=1.1pt] at (\x,-0.32) {};}
        
% images on the curve
\foreach \x in {1.2, 2.3}{ \draw[densely dotted, Cividis03] (\x,0) -- (\x,2);
  \fill[Cividis03] (\x,2) circle (1.8pt);}
\draw[densely dotted, Cividis03] (1.8,0) -- (1.8,2);
\fill[Cividis03] (1.8,2) circle (1.8pt);
\draw[densely dotted, Cividis08!80!black] (4.6,0) -- (4.6,3.4);
\fill[Cividis08!80!black] (4.6,3.4) circle (1.8pt);
\draw[densely dotted, Cividis08!80!black] (5.2,0) -- (5.2,3.4);
\fill[Cividis08!80!black] (5.2,3.4) circle (1.8pt);

% annihilated / unreachable
\draw[densely dotted, ASPNavy!50] (8.6,0) -- (8.6,4.77);
\node[ASPNavy, font=\small] at (8.6,4.77) {$\times$};
\draw[->, ASPNavy!70] (7.35,5) node[left, align=center, text width=2cm]
  {annihilated:\\ $p(\lambda)\notin p(\sigma(\widetilde{\mathbf{S}}))$} -- (8.6,5);
  
% class annotations
\draw[Cividis03, thick, |-|] (1.05,-0.85) -- (2.45,-0.85)
  node[midway, below, text=Cividis03]{$\mathcal{E}_{c_{1}}$: $m_{c_1}{=}2$, $\widetilde{m}_{c_1}{=}1$};

\node[align=right,text width=2.0cm] at (1.7,2.7) {aliasing\\ 
$p(\lambda_1){=}p(\lambda_2){=}c_1$};

%  \node[right, align=right, text width=2.0cm]{aliasing:\\ $p(\lambda_1){=}p(\lambda_2){=}c_1$} (0.7,5);
  
% legend
\fill[ASPNavy] (0.5,5) circle (2.1pt); \node[anchor=west] at (0.65,5) {$\sigma(\mathbf{S})$};
\node[fill=ASPGold, draw=ASPGold!60!black, isosceles triangle,
      isosceles triangle apex angle=60, rotate=90, inner sep=1.1pt] at (2.7,5) {};
\node[anchor=west] at (2.9,5) {$\sigma(\widetilde{\mathbf{S}})$};
\end{tikzpicture}

%% file: figures/tikz_diagrams/fig_coincidence_nonflat.tex
% Fig: coincidence classes from a NON-FLAT filter -- aliasing via
% non-monotonicity (a hump crossed going up and coming down) rather
% than via a flat plateau.
\begin{tikzpicture}[>=stealth, xscale=0.78, yscale=0.72, font=\small]
% axes
\draw[->, thick, ASPNavy!70] (-0.2,0) -- (10.6,0) node[below left]{$t$};
\draw[->, thick, ASPNavy!70] (0,-0.1) -- (0,5.6) node[right]{$p(t)$};

% coincidence levels
\draw[dashed, Cividis03, thick] (0,1.9) -- (10.2,1.9) node[right, text=Cividis03]{$c_{1}$};
\draw[dashed, Cividis08!80!black, thick] (0,4.1) -- (10.2,4.1) node[right, text=Cividis08!80!black]{$c_{2}$};

% filter curve: two humps, guaranteed to pass through every marker
% below since the spline interpolates exactly through each listed point
\draw[very thick, ASPNavy] plot [smooth, tension=0.75] coordinates {
  (0.3,1.1) (1.1,1.9) (1.8,3.3) (2.5,1.9) (3.15,0.6)
  (3.7,1.9) (4.9,4.1) (5.6,5.05) (6.3,4.1) (7.6,3.15) (9.7,1.35)};
%\node[text=ASPNavy] at (9.05,1.75) {$p(t)$};

% ------------------------------------------------------------------
% class c1: hump A (ascending + descending) and hump B (ascending)
% ------------------------------------------------------------------
% axis markers: S eigenvalues (navy dots)
\fill[ASPNavy] (1.1,0) circle (2.1pt);
\fill[ASPNavy] (3.7,0) circle (2.1pt);
% axis marker: tilde-S eigenvalue (gold triangle)
\node[fill=ASPGold, draw=ASPGold!60!black, isosceles triangle,
      isosceles triangle apex angle=60, rotate=90, inner sep=1.1pt] at (2.5,-0.32) {};
% dotted connectors + images at c1
\foreach \x in {1.1, 3.7}{
  \draw[densely dotted, Cividis03] (\x,0) -- (\x,1.9);
  \fill[Cividis03] (\x,1.9) circle (1.8pt);}
\draw[densely dotted, Cividis03] (2.5,0) -- (2.5,1.9);
\fill[Cividis03] (2.5,1.9) circle (1.8pt);

% ------------------------------------------------------------------
% class c2: hump B only (ascending + descending, closer to its peak)
% ------------------------------------------------------------------
% axis markers
\fill[ASPNavy] (4.9,0) circle (2.1pt);
\node[fill=ASPGold, draw=ASPGold!60!black, isosceles triangle,
      isosceles triangle apex angle=60, rotate=90, inner sep=1.1pt] at (6.3,-0.32) {};
% dotted connectors + images at c2
\foreach \x in {4.9, 6.3}{
  \draw[densely dotted, Cividis08!80!black] (\x,0) -- (\x,4.1);
  \fill[Cividis08!80!black] (\x,4.1) circle (1.8pt);}

% ------------------------------------------------------------------
% annihilated eigenvalue of S: image matches neither c1 nor c2
% ------------------------------------------------------------------
\fill[ASPNavy] (7.6,0) circle (2.1pt);
\draw[densely dotted, ASPNavy!50] (7.6,0) -- (7.6,3.15);
\node[ASPNavy, font=\small] at (7.6,3.2) {$\times$};
\draw[->, ASPNavy!70] (7.7,5) node[right, align=left, text width=3cm]
  {annihilated:\\ $p(\lambda)\notin p(\sigma(\widetilde{\mathbf{S}}))$} -- (7.85,3.35);

% ------------------------------------------------------------------
% aliasing callout: two S-eigenvalues from DIFFERENT humps land on c1
% ------------------------------------------------------------------
\node[align=center, text width=3.5cm, text=Cividis03] (aliastxt)
  at (2.4,3.3) {aliasing across separate humps:\\ $p(\lambda_1){=}p(\lambda_2){=}c_1$};
\draw[->, Cividis03] (aliastxt.south) to[] (1.25,2.15);
\draw[->, Cividis03] (aliastxt.south) to[] (3.6,2.15);

% ------------------------------------------------------------------
% coincidence-space brackets (below axis)
% ------------------------------------------------------------------
\draw[Cividis03, thick, |-|] (0.95,-0.85) -- (3.85,-0.85)
  node[midway, below, text=Cividis03]
  {$\mathcal{E}_{c_{1}}$: $m_{c_1}{=}2$, $\widetilde{m}_{c_1}{=}1$};
\draw[Cividis08!80!black, thick, |-|] (4.75,-0.85) -- (6.45,-0.85)
  node[midway, below right, text=Cividis08!80!black]
  {$\mathcal{E}_{c_{2}}$: $m_{c_2}{=}1$, $\widetilde{m}_{c_2}{=}1$};

% legend
\fill[ASPNavy] (0.5,5.35) circle (2.1pt);
\node[anchor=west] at (0.65,5.35) {$\sigma(\mathbf{S})$};
\node[fill=ASPGold, draw=ASPGold!60!black, isosceles triangle,
      isosceles triangle apex angle=60, rotate=90, inner sep=1.1pt] at (2.7,5.35) {};
\node[anchor=west] at (2.9,5.35) {$\sigma(\widetilde{\mathbf{S}})$};
\end{tikzpicture}

%% file: figures/tikz_diagrams/fig_transfer_efficiency.tex
% Fig: spectral transfer efficiency eta_c -- image of the unit sphere of E_c
\begin{tikzpicture}[>=stealth, font=\small]
\def\panel#1#2{
  \begin{scope}[shift={(#1,0)}]
  \draw[->, ASPNavy!40] (-1.15,0) -- (1.15,0);
  \draw[->, ASPNavy!40] (0,-1.15) -- (0,1.15);
  \draw[dashed, thick, ASPNavy] (0,0) circle (1.2);
  #2
  \end{scope}}
% panel 1: eta = 1
\panel{0}{
  \fill[CividisLight02, opacity=0.75] (0,0) circle (0.82);
  \draw[very thick, Cividis02] (0,0) circle (0.82);
  \node[text=ASPNavy, below] at (0,-1.3) {$\eta_{c}(\theta)=1$};
  \node[text=ASPNavy!70, below] at (0,-1.75) {lossless};}
% panel 2: 0 < eta < 1
\panel{3.1}{
  \fill[CividisLight08, opacity=0.8] (0,0) ellipse (0.95 and 0.38);
  \draw[very thick, Cividis07!80!black] (0,0) ellipse (0.95 and 0.38);
  \draw[->, Cividis07!60!black] (0,0) -- (0.95,0) node[pos=0.33, below right]{$\sigma_{\max}$};
  \draw[->, Cividis07!60!black] (0,0) -- (0,0.6) node[pos=0.9, right]{$\sigma_{\min}$};
  \node[text=ASPNavy, below] at (0,-1.3) {$0<\eta_{c}(\theta)<1$};
  \node[text=ASPNavy!70, below] at (0,-1.75) {distorted};}
% panel 3: eta = 0
\panel{6.2}{
  \draw[line width=2.2pt, Cividis01] (-0.95,0) -- (0.95,0);
  \node[text=ASPNavy, below] at (0,-1.3) {$\eta_{c}(\theta)=0$};
  \node[text=ASPNavy!70, below] at (0,-1.75) {destroyed};}
% shared legend
\node[align=left, text=ASPNavy!70] at (3.1,1.6)
  {Dashed: unit sphere of $\mathcal{E}_{c}$ \quad Filled: image under $\theta_{c}$};
\end{tikzpicture}

%% file: v09/sec_implications_in_samp.tex
%!TEX root =../compiler_paper.tex

%%%%%%%%%%%%%%%%%%%%%%%%%%%%%%%%%%%%%%%%%%%%%%
%%%%%%%%%%%%%%%%%% SECTION %%%%%%%%%%%%%%%%%%%
%%%%%%%%%%%%%%%%%%%%%%%%%%%%%%%%%%%%%%%%%%%%%%

\subsection{Sampling, Pooling, and Dimensionality Reduction}
\label{sec_sampling}
When considering scenarios where
$\dim(\mathcal{H}_{1})>\dim(\mathcal{H}_{2})$ with
$\theta: \mathcal{H}_{1}\to\mathcal{H}_{2}$, the map $\theta$ has a natural
interpretation as a sampling and/or dimensionality reduction
operator. Then, every result presented in the previous section has a clean and direct interpretation. In this context, there are a few extra insights and observations worth discussing when $\theta$ arises from a concrete sampling scheme.

First, notice that if $N_{2}=\dim(\ccalH_{2})<\dim(\ccalH_{1})=N_{1}$, then every
$\theta\in\Theta_{p}$ satisfies
$\dim\left(\ker\theta\right)\geq N_{1}-N_{2}$. Therefore a subspace of dimension at
least $N_{1}-N_{2}$ is annihilated regardless of how $\theta$ is chosen.
Consequently, lossless transfer in the sense of
Proposition~\ref{prop:achievability}(iii) is impossible globally, and the
correct design objective is to steer the annihilated subspace \emph{away from
the signal class of interest}. In particular, for signals bandlimited to a
subset of coincidence classes $\ccalK\subseteq\ccalC_{p}$ ---this is, signals
in $\bigoplus_{c\in\ccalK}\ccalE_{c}$--- the conditions of
Proposition~\ref{prop:achievability}(iii) restricted to $\ccalK$, namely
$m_{c}\leq\widetilde{m}_{c}$ for all $c\in\ccalK$, guarantee the existence of
$\theta\in\Theta_{p}$ that is a scaled isometry on the bandlimited class. Sampling is lossless \emph{on the band}, with the out-of-band content
absorbing the mandatory annihilation.

%%----------------------------------------
%%-----------    F I G U R E    ----------
%%---------------------------------------- 

\begin{figure}
    \centering
        \input{\pathfigs/tikz_diagrams/fig_interlacing_design.tex}
\caption{Cauchy interlacing as a design constraint under node
subsampling. Navy dots are eigenvalues of $\mathbf{S}$ ($N_{1}=6$ nodes) and
gold triangles are two eigenvalues of $\widetilde{\mathbf{S}}$ after
removing $N_{1}-N_{2}=2$ nodes. Interlacing guarantees
$\widetilde{\lambda}_{i}\in[\lambda_{i+N_{1}-N_{2}},\lambda_{i}]$
regardless of which nodes are removed, shown shaded for $i=1,4$. Making
$p(t)$ constant on each bracket guarantees
$p(\widetilde{\lambda}_{i})=p(\lambda_{i})$ independently of the sampling
set. Elsewhere, $p$ retains full
spectral resolution.}
    \label{fig_interlacing_design}
\end{figure}
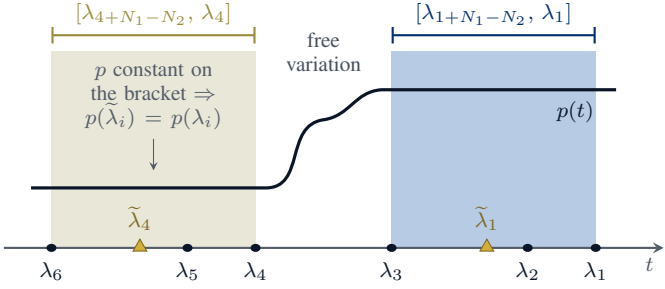

%%-------     End of Figure     ----------

The scenario in which $\widetilde{\mathbf{S}}$ is obtained by node
subsampling admits a sharper statement, since in that case the spectra of the
two shift operators are not independent, they are coupled by eigenvalue
interlacing. Making this precise requires coordinates, as the notion of node
subsampling refers to a distinguished basis of the signal space. To this end,
suppose $\mathbf{S}$ is Hermitian ---a strengthening of the standing normality
assumption, needed here so that both spectra are real and can be ordered---
and that, identifying $\ccalH_{1}$ with $\mbC^{N_{1}}$ through the node basis,
$\widetilde{\mathbf{S}}\in\mbC^{N_{2}\times N_{2}}$ is the principal submatrix
of $\mathbf{S}$ obtained by retaining $N_{2}$ nodes of the original domain.
Ordering the eigenvalues decreasingly, the Cauchy
interlacing theorem~\cite{chung1997spectral,brouwer2011spectra,horn2012matrix} guarantees
\begin{equation}\label{eq_interlacing}
\lambda_{i+N_{1}-N_{2}}
     \leq
     \widetilde{\lambda}_{i}
     \leq
     \lambda_{i}
     ,
     \qquad
     i=1,\ldots,N_{2}
     .
\end{equation}
Two consequences follow for the design problem of a filter given a specific homomorphism. First, the exact coincidences
$\widetilde{\lambda}_{j}=\lambda_{i}$ required when $p$ is injective on
$\sigma(\mathbf{S})\cup\sigma(\widetilde{\mathbf{S}})$ are \emph{not generic}.
Subsampling perturbs each eigenvalue within its
interlacing bracket
$[\lambda_{i+N_{1}-N_{2}},\lambda_{i}]$, so that for such a filter the only
homomorphism is, generically, $\theta=0$. Second,
interlacing \emph{localizes} where the filter must be flat. If $p$ is
constant on each interlacing bracket associated with the frequencies to be
retained, then $p(\widetilde{\lambda}_{i})=p(\lambda_{i})$ is guaranteed for
those frequencies \emph{regardless of which nodes are sampled}. This is, the
brackets in~\eqref{eq_interlacing} are precisely the regions of the spectral
axis over which discriminability must be surrendered to obtain
transferability that is robust to the choice of the sampling set, while $p$
remains free to vary arbitrarily between brackets. The interlacing theorem
thereby converts the abstract coincidence requirement of
Theorem~\ref{thm_trans_asp_finite} into an explicit, computable design
constraint on the amplitude response of $p$. Figure~\ref{fig_interlacing_design} illustrates these ideas.

%% file: figures/tikz_diagrams/fig_interlacing_design.tex
% Fig: Cauchy interlacing localizes where the filter must be flat
\begin{tikzpicture}[>=stealth, xscale=0.9, yscale=0.72, font=\footnotesize]
% shaded interlacing brackets (retained frequencies)
\fill[CividisLight09, opacity=0.65] (1,0) rectangle (4,3.6);
\fill[CividisLight02, opacity=0.55] (6,0) rectangle (9,3.6);
% axis
\draw[->, thick, ASPNavy!70] (0.3,0) -- (10,0) node[below left]{$t$};
% filter: flat on brackets, free in between
\draw[very thick, ASPNavy]
  (0.7,1.1) -- (4.15,1.1)
  to[out=0,in=190] (5.0,2.35) to[out=10,in=185] (5.85,2.9)
  -- (9.3,2.9);
\node[text=ASPNavy] at (8.7,2.5) {$p(t)$};
% eigenvalues of S
\foreach \x in {1.0, 3.0, 4.0, 6.0, 8.0, 9.0}{ \fill[ASPNavy] (\x,0) circle (2.1pt);}
\node[below=2pt, ASPNavy] at (9.0,0) {$\lambda_{1}$};
\node[below=2pt, ASPNavy] at (8.0,0) {$\lambda_{2}$};
\node[below=2pt, ASPNavy] at (6.0,0) {$\lambda_{3}$};
\node[below=2pt, ASPNavy] at (4.0,0) {$\lambda_{4}$};
\node[below=2pt, ASPNavy] at (3.0,0) {$\lambda_{5}$};
\node[below=2pt, ASPNavy] at (1.0,0) {$\lambda_{6}$};
% subsampled eigenvalues inside brackets
\node[fill=ASPGold, draw=ASPGold!60!black, isosceles triangle,
      isosceles triangle apex angle=60, rotate=90, inner sep=1.2pt] at (7.4,0) {};
\node[above=1pt, ASPGold!60!black] at (7.4,0.1) {$\widetilde{\lambda}_{1}$};
\node[fill=ASPGold, draw=ASPGold!60!black, isosceles triangle,
      isosceles triangle apex angle=60, rotate=90, inner sep=1.2pt] at (2.3,0) {};
\node[above=1pt, ASPGold!60!black] at (2.3,0.1) {$\widetilde{\lambda}_{4}$};
% bracket extents
\draw[Cividis02, thick, |-|] (6,3.9) -- (9,3.9)
  node[midway, above, text=Cividis02]{$[\lambda_{1+N_1-N_2},\,\lambda_{1}]$};
\draw[Cividis09!70!black, thick, |-|] (1,3.9) -- (4,3.9)
  node[midway, above, text=Cividis09!70!black]{$[\lambda_{4+N_1-N_2},\,\lambda_{4}]$};
% annotations
\node[text=ASPNavy!75, align=center] at (5.0,3.6) {free\\ variation};
\draw[->, ASPNavy!70] (2.5,2.0) node[above, text width=2.3cm, align=center]
  {$p$ constant on the bracket $\Rightarrow$ $p(\widetilde{\lambda}_i)=p(\lambda_i)$} -- (2.5,1.42);
\end{tikzpicture}

%% file: v09/sec_implications_in_cs.tex
%!TEX root =../compiler_paper.tex

%%%%%%%%%%%%%%%%%%%%%%%%%%%%%%%%%%%%%%%%%%%%%%
%%%%%%%%%%%%%%%%%% SECTION %%%%%%%%%%%%%%%%%%%
%%%%%%%%%%%%%%%%%%%%%%%%%%%%%%%%%%%%%%%%%%%%%%

\subsection{Compressed Sensing: Recovery Guarantees for Measurements with Homomorphisms}
\label{sec_cs}

When $\dim(\ccalH_{2})\leq\dim(\ccalH_{1})$, a map $\theta\in\Theta_{p}$ can
be regarded as a \emph{measurement operator} in the sense of compressed
sensing. Given a signal $\mathbf{x}\in\ccalH_{1}$, one observes
$\mathbf{y}=\theta\,\mathbf{x}\in\ccalH_{2}$ and aims to recover
$\mathbf{x}$ under a sparsity assumption~\cite{eldar2015sampling}. Now we will show that the block structure of
Lemma~\ref{lem:support} determines the two standard figures of merit of
$\theta$ as a measurement operator -- its restricted isometry constant and
its coherence -- directly from the spectral data of the signal models, with
the richness $\eta(\theta)$ playing the leading role. We assume
$\mathbf{S}$ and $\widetilde{\mathbf{S}}$ normal until
Section~\ref{sec_cs_nonnormal}, which relaxes this hypothesis.

Let us recall that the restricted isometry constant (RIC) of order $s$ of a
linear map $\mathbf{A}$, denoted $\delta_s(\mathbf{A})$, is the smallest $\delta \geq 0$
with
$(1-\delta)\|\mathbf{x}\|^2 \leq \|\mathbf{A}\mathbf{x}\|^2 \leq (1+\delta)\|\mathbf{x}\|^2$
for all $s$-sparse $\mathbf{x}$~\cite{eldar2015sampling}. The smaller
$\delta_{s}$, the easier it is to guarantee uniqueness and quality of the
reconstruction of $\bbx$ from $\bby=\theta\bbx$. Notice that $s$-sparsity
presupposes a basis of $\ccalH_{1}$. The bound we establish below is insensitive to that
choice, since it follows from a two-sided frame bound valid for
\emph{every} $\mathbf{x}$ in the relevant subspace, and a specific basis is
fixed only in Section~\ref{sec:coherence}, where coherence requires one.

Because the coincidence structure is indexed by classes, it is natural to
state the guarantee on a \emph{band}, that is, on the signals supported on a
prescribed subset of coincidence classes. Given a nonempty
$\ccalK\subseteq\ccalC_{p}$, let
\begin{equation}\label{eq_band_subspace}
\ccalH_{\ccalK}
     :=
     \bigoplus_{c\,\in\,\ccalK}
        \ccalE_{c}
     \;\subseteq\;
     \ccalH_{1}
     ,
     \qquad
\eta_{\ccalK}(\theta)
     :=
     \min_{c\,\in\,\ccalK}
        \eta_{c}(\theta)
     ,
\end{equation}
and write $\delta_{s}^{\ccalK}(\mathbf{A})$ for the restricted isometry
constant of order $s$ computed over the $s$-sparse signals lying in
$\ccalH_{\ccalK}$. Taking $\ccalK=\ccalC_{p}$ recovers the unrestricted
notions.

%%---------------------------------------------------------------
%%---------------     P R O P O S I T I O N    ------------------
%%---------------------------------------------------------------

\begin{proposition}[Richness controls the RIC on a band]\label{prop:ric}
Let $\ccalK\subseteq\ccalC_{p}$ be nonempty with
$m_{c}\leq\widetilde m_{c}$ for every $c\in\ccalK$, and let
$\theta\in\Theta_{p}$ have $\theta_{c}$ injective for every $c\in\ccalK$,
with the block gains normalized so that $\sigma_{\max}(\theta_{c})=1$ for
all $c\in\ccalK$. Then the rescaled map
$\theta':=\tfrac{\sqrt{2}}{\sqrt{1+\eta_{\ccalK}(\theta)}}\,\theta$
satisfies
\begin{equation}\label{eq:ricbound}
  \delta_{s}^{\ccalK}(\theta') \;\leq\;
  \frac{1 - \eta_{\ccalK}(\theta)}{\,1 + \eta_{\ccalK}(\theta)\,}
  \quad\forall\, s \leq \dim\ccalH_{\ccalK} .
\end{equation}
\end{proposition}

\begin{proof}
    See Appendix~\ref{proof_prop_ric}.
\end{proof}

%%---------- End of Proposition -------------

Two remarks are in order. First, without the normalization
$\sigma_{\max}(\theta_c)\equiv1$ the frame bounds become
$\alpha=\min_c\sigma_{\min}^2(\theta_c)$ and
$\beta=\max_c\sigma_{\max}^2(\theta_c)$, and one only has
$\alpha/\beta\leq\eta(\theta)$: a homomorphism can be perfectly rich on each
block ($\eta=1$) and still be badly conditioned globally if the blocks carry
very different gains. Since block scalings are free parameters of $\Theta_p$
(Proposition~\ref{prop:dimension}), gain balancing is always available and
should be regarded as part of the design of $\theta$. Second, the hypotheses
constrain only $\ccalK$, so that
$\dim\ccalH_{\ccalK}=\sum_{c\in\ccalK}m_{c}\leq\sum_{c\in\ccalK}\widetilde m_{c}\leq N_{2}$
while $N_{1}$ remains unconstrained: the map compresses $\ccalH_{1}$ into a
strictly smaller $\ccalH_{2}$ while acting as a scaled isometry on the band,
which is the regime of interest in compressed sensing. Taking
$\ccalK=\ccalC_{p}$ together with~\eqref{eq:injectivity} instead forces
$N_{1}\leq N_{2}$, so the unrestricted bound describes the lossless case.
This is the recovery counterpart of the design principle of
Section~\ref{sec_sampling}, where the out-of-band content absorbs the
annihilation that $N_{2}<N_{1}$ makes unavoidable.

%%%%%%%%%%%%%%%%%%%%%%%%%%%%%%%%%%%%%%%%%%%%%%%%%%%%%%%%%
%%%%%%%%%    S U B  -   S E C T I O N     
%%%%%%%%%%%%%%%%%%%%%%%%%%%%%%%%%%%%%%%%%%%%%%%%%%%%%%%%%

\subsubsection{Coherence of Intertwining Measurements}\label{sec:coherence}
In some applications it is common to work with the \emph{coherence} of the
measurement matrix rather than its restricted isometry
constant~\cite{eldar2015sampling,apm_coherence,apm_sr}, and the block
structure of Lemma~\ref{lem:support} makes it equally transparent. We fix an
orthonormal eigenbasis $\{\mathbf{v}_j\}_{j=1}^{N_1}$ of $\mathbf{S}$,
$\mathbf{S} \mathbf{v}_j = \lambda_j \mathbf{v}_j$, as the \emph{sparsity
basis}, so that signals are $s$-sparse when they have at most $s$ active
frequencies; this is the natural choice for bandlimited-type models in ASP,
and it matters, as coherence is basis dependent. We assume $\theta \in
\Theta_p$ has no vanishing columns, $\theta \mathbf{v}_j \neq 0$ for all $j$
(equivalently, by Lemma~\ref{lem:support}, that
$p(\sigma(\mathbf{S}))\subseteq p(\sigma(\widetilde{\mathbf{S}}))$; on a band
$\ccalK$ it suffices that this hold for the indices $j$ with
$p(\lambda_{j})\in\ccalK$), so that the coherence
\begin{equation}\label{eq_coherence_def}
  \mu(\theta) \;:=\; \max_{j \neq k}
  \frac{\big|\langle \theta \mathbf{v}_j,\, \theta \mathbf{v}_k\rangle\big|}
       {\|\theta \mathbf{v}_j\|\,\|\theta \mathbf{v}_k\|}
\end{equation}
is well defined.

It is worth being precise about what~\eqref{eq_coherence_def} means, since
$\theta$ is a linear map (Definition~\ref{def_hom_asm}) and not intrinsically
a matrix: the \emph{columns} of a linear map are the images of a chosen basis
of the domain, so the vectors $\theta\mathbf{v}_j$ are exactly the columns of
$\theta$ with respect to the eigenbasis just fixed. They are the same object
a reader accustomed to coherence for a sensing matrix $\mathbf{A}$ would
write as $\mathbf{A}\mathbf{e}_j$, with the canonical basis replaced by the
one in which sparsity is defined here. As Lemma~\ref{lem:gram} shows next,
this is also the only basis in which the coherence inherits the
coincidence-class structure already established for $\eta(\theta)$.

%%-------------------------------------------------
%%------      L E M M A       ---------------------
%%-------------------------------------------------

\begin{lemma}\label{lem:gram}
Let $\theta \in \Theta_p$. Then
$\langle \theta \mathbf{v}_j, \theta \mathbf{v}_k \rangle = 0$ whenever
$p(\lambda_j) \neq p(\lambda_k)$, i.e. columns associated with different
coincidence classes are automatically orthogonal. Consequently the Gram
matrix $\theta^{*}\theta$ is block-diagonal over $\ccalC_{p}$ in the sparsity
basis, and
\begin{equation}\label{eq:mumax}
\mu(\theta) 
     =
       \max_{c \,\in\, \ccalC_{p}}
          \mu(\theta_c),
\end{equation}
where $\mu(\theta_c)$ denotes the coherence of the block $\theta_c$
viewed as a matrix with columns
$\{\theta \mathbf{v}_j : p(\lambda_j) = c\}$, with the convention
$\mu(\theta_c):=0$ when $m_c=1$.
\end{lemma}

\begin{proof}
    See Appendix~\ref{proof_lem_gram}.
\end{proof}

%%---------         End of Lemma       ------------

This is a direct consequence of Lemma~\ref{lem:support}, traced through one
further fact. That lemma confines the image of any source direction at
coincidence value $c$ to the single matching target subspace
$\widetilde{\ccalE}_c$, and since $p(\widetilde{\mathbf{S}})$ is normal
its spectral decomposition~\eqref{eq_spectral_decomp_finite} makes the
target coincidence spaces mutually orthogonal. Routing therefore becomes
alignment: two outputs $\theta\mathbf{v}_j\in\widetilde{\ccalE}_c$ and
$\theta\mathbf{v}_k\in\widetilde{\ccalE}_{c'}$, $c\neq c'$, lie in
orthogonal subspaces and are orthogonal to each other exactly, not
approximately, for every $\theta\in\Theta_p$. Coherence is thus a
\emph{purely intra-class} quantity and cross-class coherence vanishes for
free: no element of $\Theta_p$ needs to be designed, or even checked, to
guarantee it. It is worth noting the asymmetry with $\eta(\theta)$, which is
per-class \emph{by definition} (Definition~\ref{def:eta}), whereas $\mu$ in
\eqref{eq_coherence_def} is a maximum over \emph{every} pair $j\neq k$ in
the whole domain and collapses to the per-class structure only as a
theorem. That a quantity with no awareness of the coincidence classes turns
out to respect them exactly is a consequence of the algebra alone.

%%---------------------------------------------------------------
%%---------------     P R O P O S I T I O N    ------------------
%%---------------------------------------------------------------

\begin{proposition}[Achievable coherence]\label{prop:mudesign}
\leavevmode
\begin{enumerate}
  \item[\textup{(i)}] If $m_c \leq \widetilde m_c$ for all
    $c \in \ccalC_{p}$, there exists $\theta \in \Theta_p$ with
    $\mu(\theta) = 0$, obtained by choosing each block a scaled isometry, so
    that its columns are orthogonal. This is the same condition under which
    $\eta(\theta) = 1$ is achievable
    \textup{(Proposition~\ref{prop:achievability})}.
  \item[\textup{(ii)}] If $m_c > \widetilde m_c$ for some $c$, then every
    $\theta \in \Theta_p$ \textup{(}without vanishing columns\textup{)}
    satisfies
    \begin{equation}\label{eq:welch}
      \mu(\theta) \;\geq\; \mu(\theta_c) \;\geq\;
      \sqrt{\frac{m_c - \widetilde m_c}{\widetilde m_c\,(m_c - 1)}}
      \;>\; 0.
    \end{equation}
\end{enumerate}
\end{proposition}

\begin{proof}
    See Appendix~\ref{proof_prop_mudesign}
\end{proof}

%%---------- End of Proposition -------------

\noindent Both parts stem from a single fact. Fix a class $c$ with
$m_c>\widetilde m_c$; since $\theta_c:\ccalE_c\to\widetilde{\ccalE}_c$ maps
an $m_c$-dimensional space into a strictly smaller one, rank-nullity forces
$\theta_c$ to be non-injective for \emph{every} $\theta\in\Theta_p$. In the
language of singular values this reads $\sigma_{\min}(\theta_c)=0$, forcing
$\eta(\theta)=0$ identically on $\Theta_p$
(Proposition~\ref{prop:achievability}(i)); in the language of inner
products, the same rank deficiency bounds the rank of the Gram matrix of the
normalized columns and drives the Welch bound, so $\mu(\theta)>0$
identically on $\Theta_p$. Richness and coherence are therefore not
independently obstructed; they are two readouts of one algebraic fact.
Notice also that part~(i) sits in the same non-compressive regime as the
unrestricted form of Proposition~\ref{prop:ric}: requiring
$m_{c}\leq\widetilde m_{c}$ for \emph{every} $c\in\ccalC_{p}$ forces
$N_{1}\leq N_{2}$, so $\mu(\theta)=0$ is attainable only when no
dimensionality reduction takes place. Equivalently, and more directly,
$\mu(\theta)=0$ means the columns $\{\theta\mathbf{v}_{j}\}$ are nonzero and
pairwise orthogonal, hence linearly independent, so they cannot number more
than $\dim\ccalH_{2}$. The band version follows by the same
argument restricted to $\ccalK$, giving
$\max_{c\in\ccalK}\mu(\theta_{c})=0$ together with
$\eta_{\ccalK}(\theta)=1$, while part~(ii) shows that outside this regime
a non zero lower bound is unavoidable.

%%---------------------------------------------------------------
%%---------------     P R O P O S I T I O N    ------------------
%%---------------------------------------------------------------

\begin{proposition}[Coherence vs.\ richness]\label{prop:mueta}
Fix $c \in \ccalC_{p}$ and suppose the columns of $\theta_c$ have equal norms.
\begin{enumerate}
  \item[\textup{(i)}] If $\mu(\theta_c) < \dfrac{1}{m_c - 1}$, then
    \begin{equation}\label{eq:gershgorin}
      \eta_c(\theta) \;\geq\;
      \frac{1 - (m_c - 1)\,\mu(\theta_c)}{1 + (m_c - 1)\,\mu(\theta_c)} .
    \end{equation}
  \item[\textup{(ii)}] Conversely, $\eta_c(\theta) = 1$ implies
    $\mu(\theta_c) = 0$.
\end{enumerate}
\end{proposition}

\begin{proof}
    See Appendix~\ref{proof_prop_mueta}
\end{proof}

%%---------- End of Proposition -------------

The equal-norm hypothesis is necessary because coherence normalizes the
columns and is therefore blind to unequal column gains, whereas $\eta$
penalizes them: $\mu(\theta_c)=0$ alone implies orthogonal columns but
yields $\eta_c(\theta)=1$ only after the column norms are equalized.

To close, the block structure yields a statement stronger than any single
global bound. Since
$\widetilde{\mathbf{\Pi}}_c\, \theta\mathbf{x} = \theta_c\, \mathbf{\Pi}_c \mathbf{x}$,
the measurement $\mathbf{y} = \theta\mathbf{x}$ decouples,
\begin{equation}\label{eq_decoupling}
  \mathbf{y}_c \;=\; \theta_c\, \mathbf{x}_c,
  \qquad
  \mathbf{y}_c := \widetilde{\mathbf{\Pi}}_c\,\mathbf{y},\quad
  \mathbf{x}_c := \mathbf{\Pi}_c\,\mathbf{x},
  \qquad c \in \ccalC_{p} ,
\end{equation}
so sparse recovery through a homomorphism is \emph{separable across
coincidence classes}. If $\mathbf{x}$ has $s_c$ active frequencies in class
$c$, basis pursuit and orthogonal matching pursuit recover $\mathbf{x}_c$
exactly whenever~\cite{foucart2013mathematical}
\begin{equation}\label{eq:perclass}
  \mu(\theta_c) \;<\; \frac{1}{2 s_c - 1},
  \qquad \forall\, c \in \ccalC_{p} ,
\end{equation}
a strictly weaker requirement than the global condition
$\mu(\theta) < 1/(2s - 1)$ with $s = \sum_c s_c$. The algebra does not
merely bound the coherence, it \emph{localizes} the recovery problem to
small blocks, on which the coherence is cheap to compute and the sparsity
budget is only the local one.

%%%%%%%%%%%%%%%%%%%%%%%%%%%%%%%%%%%%%%%%%%%%%%%%%%%%%%%%%
%%%%%%%%%    S U B  -   S E C T I O N     
%%%%%%%%%%%%%%%%%%%%%%%%%%%%%%%%%%%%%%%%%%%%%%%%%%%%%%%%%

\subsubsection{Recovery Guarantees Without Normality}
\label{sec_cs_nonnormal}

The guarantees above rely on the normality of the shift operators through the
orthogonality of the coincidence decomposition. As established in
Section~\ref{sec_nonnormal}, when $\mathbf{S}$ and $\widetilde{\mathbf{S}}$
are merely diagonalizable the \emph{algebraic} theory holds verbatim with
oblique projections, and what degrades is the metric statement: the
decomposition $\ccalH_1 = \bigoplus_c \ccalE_c$ is no longer orthogonal, so
cross terms appear in $\|\theta\mathbf{x}\|^2$ and the obliqueness of the
coincidence decomposition, encoded by the metric operators
of~\eqref{eq_metric_operator}, enters the frame bounds. The following result
quantifies this degradation for the full domain, that is, for
$\ccalK=\ccalC_{p}$ in the notation of~\eqref{eq_band_subspace}.

%%---------------------------------------------------------------
%%---------------     P R O P O S I T I O N    ------------------
%%---------------------------------------------------------------
\begin{proposition}\label{prop:ricdiag}
Let $\mathbf{S},\widetilde{\mathbf{S}}$ be diagonalizable, and let
$\mathbf{G}$ and $\widetilde{\mathbf{G}}$ be the positive definite
\emph{metric operators} of~\eqref{eq_metric_operator}, which satisfy
$\langle\mathbf{G}\mathbf{x},\mathbf{x}\rangle
=\|\mathbf{x}\|_{\oplus}^{2}$ and reduce to the identity exactly when the
corresponding shift is normal. Let $\theta\in\Theta_{p}$
satisfy~\eqref{eq:injectivity}, with its blocks normalized so that
$\sigma_{\max}(\theta_{c})=1$ for all $c\in\ccalC_{p}$ in the adapted inner
products of Proposition~\ref{prop:diag}, and let $\eta(\theta)$ be computed
there. Then
\begin{equation}\label{eq:ricdiag}
  \frac{\lambda_{\min}(\mathbf{G})}{\lambda_{\max}(\widetilde{\mathbf{G}})}\;
  \eta(\theta)\,\|\mathbf{x}\|^{2}
  \;\leq\;
  \|\theta\,\mathbf{x}\|^{2}
  \;\leq\;
  \frac{\lambda_{\max}(\mathbf{G})}{\lambda_{\min}(\widetilde{\mathbf{G}})}\,
  \|\mathbf{x}\|^{2},
\end{equation}
$\forall~\mathbf{x}\in\ccalH_{1},$ so that, after centering the frame bounds by a global rescaling,
\begin{equation}\label{eq:ricdiag2}
  \delta_{s}(\theta)
  \;\leq\;
  \frac{1-\dfrac{\eta(\theta)}
        {\kappa(\mathbf{G})\,\kappa(\widetilde{\mathbf{G}})}}
       {1+\dfrac{\eta(\theta)}
        {\kappa(\mathbf{G})\,\kappa(\widetilde{\mathbf{G}})}}\,,
  \qquad
  \kappa(\mathbf{G}):=
  \frac{\lambda_{\max}(\mathbf{G})}{\lambda_{\min}(\mathbf{G})} .
\end{equation}
\end{proposition}
\begin{proof}
    See Appendix~\ref{proof_prop_ricdiag}.
\end{proof}
%%---------- End of Proposition -------------

\noindent The factor $\kappa(\mathbf{G})\,\kappa(\widetilde{\mathbf{G}})\geq1$,
with equality exactly when both shifts are normal, is the
\emph{departure-from-normality penalty}: two domains may admit an
algebraically rich homomorphism ($\eta(\theta)=1$) whose transfer quality
in the ambient metric is nonetheless poor, because the coincidence spaces
are far from orthogonal. Whether $\eta$ should be measured in the adapted
inner products (making it independent of the ambient metric, as in
Proposition~\ref{prop:diag}) or in the ambient ones (making it metrically
honest, at the price of losing the clean block theory) is a genuine
modeling choice, and \eqref{eq:ricdiag2} quantifies the gap between the
two. As with Proposition~\ref{prop:ric}, the band-restricted version follows
by the same argument applied on $\ccalH_{\ccalK}$, with
$\eta_{\ccalK}(\theta)$ in place of $\eta(\theta)$ and the metric operators
of~\eqref{eq_metric_operator} summed over $\ccalK$.

%% file: v09/sec_implications_in_eqvml.tex
%!TEX root =../compiler_paper.tex

%%%%%%%%%%%%%%%%%%%%%%%%%%%%%%%%%%%%%%%%%%%%%%
%%%%%%%%%%%%%%%%%% SECTION %%%%%%%%%%%%%%%%%%%
%%%%%%%%%%%%%%%%%%%%%%%%%%%%%%%%%%%%%%%%%%%%%%

\subsection{Equivariance and Transferable Machine Learning}
\label{sec_eqvml}

% %%----------------------------------------
% %%-----------    F I G U R E    ----------
% %%---------------------------------------- 

% \begin{figure}
%     \centering
%         \input{\pathfigs/tikz_diagrams/fig_equivariance_hierarchy.tex}
%     \caption{CS decoupling}
%     %\label{fig_hom_diagram}
% \end{figure}

% %%-------     End of Figure     ----------

The homomorphisms studied in this paper are equivariant maps, and making
this connection explicit clarifies both the scope of our results and their
consequences for machine learning architectures built on ASMs. We begin by
distinguishing two notions that the literature refers to with the same word.

In its classical, single-domain form, equivariance concerns maps
$\varphi:\ccalH\to\ccalH$ commuting with the action of a symmetry on
\emph{one} space, $\varphi\,\rho(a)=\rho(a)\,\varphi$. The equivariant
linear maps are then the commutant of the representation, and for the
polynomial algebra they include the filters themselves. In its
representation-theoretic form, the \emph{same abstract} symmetry acts on
\emph{two different} spaces through two representations, and the
equivariant maps are those satisfying
$\theta\,\rho_{1}(a)=\rho_{2}(a)\,\theta$ -- this is, exactly the
homomorphisms of Definition~\ref{def_hom_asm}. The key observation is that
$\mathbf{S}$ and $\widetilde{\mathbf{S}}$ need not be equal for the second
notion to apply: neither shift is the symmetry itself, instead both are concrete
realizations of the same abstract generator $g\in\ccalA$. The two notions
coincide when $\ccalH_{1}=\ccalH_{2}$ and $\rho_{1}=\rho_{2}$, so that one
and the same formalism contains \emph{filters} (equivariance with the
domain fixed) and \emph{transferability maps} (equivariance across
domains) as the diagonal and off-diagonal cases of one definition.

With the terminology fixed, the constructions of
Section~\ref{sec_deg_transf} organize into a hierarchy of symmetry
constraints. Define the space of \emph{shift-equivariant} maps as 
\begin{equation}\label{eq_shift_equivariant}
\Theta_{t}
     :=
     \left\lbrace
          \theta
          \left\vert~
          \theta\,\mathbf{S}
          =
          \widetilde{\mathbf{S}}\,\theta
          \right.
     \right\rbrace
     ,
\end{equation}
this is, $\Theta_{p_{0}}$ for the identity filter $p_{0}(t)=t$. Since
$\theta\mathbf{S}=\widetilde{\mathbf{S}}\theta$ implies
$\theta\,q(\mathbf{S})=q(\widetilde{\mathbf{S}})\,\theta$ for every
polynomial $q$ (Corollary~\ref{corll_hom_asp_shiftoper}),
\begin{equation}\label{eq_hierarchy}
\Theta_{t}
     =
     \bigcap_{q\,\in\,\mbC[t]}
          \Theta_{q}
     \;\subseteq\;
     \Theta_{p}
     \qquad
     \text{for every fixed filter } p
     .
\end{equation}
In this language, membership in $\Theta_{p}$ is \emph{equivariance modulo
$p$}: $\theta$ is required to commute with the domain actions only after
the filter has been applied, i.e., only up to the aliasing that $p$
induces on the spectra. Full shift-equivariance is the most rigid symmetry
constraint; $p$-equivariance is its controlled relaxation. Because
$p_{0}(t)=t$ never aliases, the entire theory of
Section~\ref{sec_deg_transf} specializes to $\Theta_{t}$ with the
coincidence data replaced by the \emph{raw} spectral data: the coincidence
set becomes
$\sigma_{r}:=\sigma(\mathbf{S})\cap\sigma(\widetilde{\mathbf{S}})$, the
coincidence spaces become the eigenspaces
$\ccalE_{\lambda},\widetilde{\ccalE}_{\lambda}$ with multiplicities
$m_{\lambda},\widetilde{m}_{\lambda}$, and in particular
\begin{equation}\label{eq_dim_gap}
\dim\Theta_{t}
     =
     \sum_{\lambda\,\in\,\sigma_{r}}
          m_{\lambda}\,\widetilde{m}_{\lambda}
     \;\;\leq\;\;
     \sum_{c\,\in\,\ccalC_{p}}
          m_{c}\,\widetilde{m}_{c}
     =
     \dim\Theta_{p}
     ,
\end{equation}
with equality if and only if $p$ is injective on
$\sigma(\mathbf{S})\cup\sigma(\widetilde{\mathbf{S}})$, this is, if and
only if $p$ does not alias. The gap between the two dimensions is
therefore an exact count of the degrees of freedom \emph{purchased by
relaxing equivariance} and every identification
$p(\lambda)=p(\mu)$ with $\lambda\neq\mu$ unlocks entries of $\theta$
that full shift-equivariance forbids.

The interaction with the richness $\eta(\theta)$ is where the tension
between symmetry and transfer becomes visible. For
$\theta\in\Theta_{t}$, each block $\theta_{c}$ is itself block-diagonal
with respect to the refinement of $\ccalE_{c}$ into raw eigenspaces,
$\theta(\ccalE_{\lambda})\subseteq\widetilde{\ccalE}_{\lambda}$. In
particular, an equivariant map must \emph{annihilate}
$\ccalE_{\lambda}$ whenever
$\lambda\notin\sigma(\widetilde{\mathbf{S}})$ -- even if
$p(\lambda)\in\ccalC_{p}$, so that a mere $p$-intertwiner could have
transferred that content into the eigenspace of some $\mu\neq\lambda$
with $p(\mu)=p(\lambda)$.

%%---------------------------------------------------------------
%%---------------     P R O P O S I T I O N    ------------------
%%---------------------------------------------------------------

\begin{proposition}[Richness under full equivariance]
\label{prop_equivariance}
Assume $\ccalC_{p}\neq\emptyset$ and that
$\mathbf{S},\widetilde{\mathbf{S}}$ are normal.
\begin{enumerate}
\item[\textup{(i)}] If some $\lambda\in\sigma(\mathbf{S})$ with $p(\lambda)\in\ccalC_{p}$, or
    equivalently sharing its filtered value with some (possibly
    different) $\mu\in\sigma(\widetilde{\mathbf{S}})$, satisfies
    either $\lambda\notin\sigma(\widetilde{\mathbf{S}})$ or
    $m_{\lambda}>\widetilde{m}_{\lambda}$, then $\eta(\theta)=0$ for
    \emph{every} $\theta\in\Theta_{t}$, even when $\Theta_{p}$ contains
    maps with $\eta(\theta)=1$.
\item[\textup{(ii)}] $\displaystyle\max_{\theta\in\Theta_{t}}\eta(\theta)=1$ if and only if
    every $\lambda\in\sigma(\mathbf{S})$ with $p(\lambda)\in\ccalC_{p}$
    itself satisfies $\lambda\in\sigma(\widetilde{\mathbf{S}})$ and
    $m_{\lambda}\leq\widetilde{m}_{\lambda}$, i.e. the raw eigenvalue, not
    merely a filtered match for it. In that case it suffices to choose
    $\theta\vert_{\ccalE_{\lambda}}$ a scaled isometry into
    $\widetilde{\ccalE}_{\lambda}$ with a \emph{common} scale within each
    coincidence class.
\end{enumerate}
\end{proposition}
\begin{proof}
    See Appendix~\ref{proof_prop_equivariance}
\end{proof}
%%---------- End of Proposition -------------

\noindent Propositions~\ref{prop:achievability} and \ref{prop_equivariance} compare
directly: maximal richness over $\Theta_{p}$ requires only the
\emph{class-wise} condition $m_{c}\leq\widetilde{m}_{c}$, while maximal
richness over $\Theta_{t}$ requires the \emph{eigenvalue-wise} condition
$\lambda\in\sigma(\widetilde{\mathbf{S}})$,
$m_{\lambda}\leq\widetilde{m}_{\lambda}$. The minimal example is already
decisive: if $\sigma(\mathbf{S})=\{\lambda\}$ and
$\sigma(\widetilde{\mathbf{S}})=\{\mu\}$ with $\lambda\neq\mu$ but
$p(\lambda)=p(\mu)$, then $\Theta_{t}=\{0\}$ while $\Theta_{p}$ contains
isometries with $\eta(\theta)=1$. Relaxing equivariance to
$p$-equivariance is thus not a technical weakening but the mechanism by
which transfer between spectrally mismatched domains becomes possible at
all: aliasing is \emph{controlled symmetry breaking}, and the richness
gained is exactly the content of~\eqref{eq_dim_gap}.

Proposition~\ref{prop_equivariance} has direct implications on general neural networks on arbitrary domains. In particular, under the lens of algebraic neural networks~\cite{parada_algnn,algnn_nc_j}, the linear part of every
layer is a filter $q_{\ell}(\mathbf{S})$, and transferring a trained
network from $(\ccalA,\ccalH_{1},\rho_{1})$ to
$(\ccalA,\ccalH_{2},\rho_{2})$ amounts to finding a map $\theta$ that
commutes with the filters the architecture actually uses. The
hierarchy~\eqref{eq_hierarchy} makes the design space explicit, i.e. demanding
equivariance with respect to the full algebra yields the rigid space
$\Theta_{t}$ and the strong spectral matching of
Proposition~\ref{prop_equivariance}, whereas demanding it only for the
deployed filter bank $\{p_{1},\ldots,p_{L}\}$ yields
$\bigcap_{\ell}\Theta_{p_{\ell}}$, whose coincidence classes are the joint
level sets of $(p_{1},\ldots,p_{L})$ on the spectra -- intermediate
between the two extremes, and strictly larger than $\Theta_{t}$ whenever
the deployed filters share level sets. The richness $\eta(\theta)$ of the
chosen $\theta$ then quantifies the per-layer information transfer at
the linear stage.

%% file: v09/sec_discussion.tex
%!TEX root =../compiler_paper.tex

%%%%%%%%%%%%%%%%%%%%%%%%%%%%%%%%%%%%%%%%%%%%%%
%%%%%%%%%%%%%%%%%% SECTION %%%%%%%%%%%%%%%%%%%
%%%%%%%%%%%%%%%%%%%%%%%%%%%%%%%%%%%%%%%%%%%%%%

\section{Discussion and Conclusions}
\label{sec_discussion}

This paper set out to answer a question that is easy to state but, as
Theorem~\ref{thm_trans_asp_finite} and its consequences show, has a
genuinely rich answer: given two algebraic signal models built on the
same algebra, when can information processed on one be transferred to
the other, and how much of that information survives the transfer? We
showed that the existence of a homomorphism $\theta$ between two signal
models is governed entirely by the coincidences
$p(\lambda_i)=p(\widetilde\lambda_j)$ between the filtered spectra of
their shift operators (Theorem~\ref{thm_trans_asp_finite}). Additionally, we showed that existence alone is insufficient for the characterization of transferability as the space of homomorphisms
$\Theta_p$ always contains the zero map and other informationally
degenerate elements, so a theory of transferability needs a second,
independent axis of measurement. The spectral transfer efficiency
$\eta(\theta)$ supplies that axis. Once introduced, it forces the whole
theory into a clean direct sum decomposition where every homomorphism decomposes,
without loss, into a direct sum of linear independent maps over the coincidence classes of the
filtered spectra (Lemma~\ref{lem:support}). Additionally, we show that this
structure is what makes $\eta(\theta)$, the dimension of $\Theta_p$, the
restricted isometry constant, the coherence, and the achievability of
equivariance all computable, and all governed by the identical
combinatorial data: which eigenvalues of $\mathbf{S}$ and
$\widetilde{\mathbf{S}}$ the filter $p$ identifies with one another.

We showed that while sampling and pooling, compressed sensing, and equivariant machine
learning look, on the surface, like three separate application areas, in
this framework they are three readings of the same coincidence-class
structure. Eigenvalue interlacing
tells a designer exactly where a filter must be flat for transferability
to survive subsampling regardless of which nodes are removed. The
restricted isometry constant and the coherence of a homomorphism
(Section~\ref{sec_cs}) are controlled, block by block in a direct sum decomposition, by the same
$\eta(\theta)$ that governs lossless transfer, and sparse recovery
inherits the same decoupling that the spectral support lemma imposes on
$\theta$ itself. And the hierarchy of equivariance constraints
(Section~\ref{sec_eqvml}) shows that spectral aliasing, which in
Section~\ref{sec_deg_transf} first appears as a loss of resolution, is exactly the controlled symmetry breaking that makes transfer between
structurally mismatched domains possible at all. In each case, the
filter $p$ is not a passive component of the processing pipeline, it is
the principal design variable through which transferability, recovery,
and equivariance are simultaneously negotiated.

It is important to point out that the theory developed quantifies exact and worst-case scenarios, and this carries costs worth stating plainly. Exactness means the results
characterize \emph{lossless} transfer and its precise degradation, but
say nothing about the common practical situation in which $\mathbf{S}$
and $\widetilde{\mathbf{S}}$ are close but not exactly related through
shared spectrum, which is the perturbative regime addressed for finite graphs
converging to a common limit by the graphon-transferability literature
mentioned in the Introduction. Worst-case aggregation means
$\eta(\theta)=\min_{c}\eta_c(\theta)$ is the correct quantity for the
uniform recovery guarantees of Section~\ref{sec_cs}, but it is, by
construction, indifferent to how small or how consequential the worst
coincidence class happens to be: a homomorphism that is excellent on
every class but one is reported as having no richness at all. Finally,
the sharpest form of every result in Section~\ref{sec_deg_transf}
assumes normal shift operators.

% Section~\ref{sec_nonnormal} shows the
% algebraic skeleton survives diagonalizability and even defectiveness,
% but the metric guarantees degrade by a conditioning penalty in the
% former case and require joint Jordan-structure compatibility, not merely
% eigenvalue coincidence, in the latter.

These limitations point directly to the directions we consider most promising. The theory developed here is restricted to finite-dimensional signal models. When $\ccalH_1$ or $\ccalH_2$ is infinite-dimensional, additional questions arise in settings such as graphon and Lie-group models, where the spectral \emph{type} of an operator, rather than merely its spectral values, becomes a fundamental obstruction. These questions are substantial enough to warrant a separate treatment and are therefore deferred to a companion paper.

Within the finite-dimensional setting, three further extensions stand
out. First, the results of Section~\ref{sec_deg_transf} are stated for
algebras with a single generator, matching the classical ASP filters of
a single shift. Extending the coincidence-class and dimension theory to
several noncommuting generators would connect this framework directly to
multigraph and Lie-group signal processing, where filters are
polynomials in more than one operator. Second, an approximate version of
Definition~\ref{def_transferability}, replacing exact homomorphisms with
$\|\theta\,p(\mathbf{S})-p(\widetilde{\mathbf{S}})\,\theta\|\leq\epsilon$,
would connect the exact theory developed here to the perturbative
graphon-transferability results already in the literature, and would let
$\eta(\theta)$ be tracked as a function of $\epsilon$ rather than only
at $\epsilon=0$. 

Third, the aggregation $\eta(\theta)=\min_{c}\eta_{c}(\theta)$ is a
definitional choice, not a derived one and the per-class structure is built
into Definition~\ref{def:eta} from the outset, whereas for coherence the
same structure is a theorem (Lemma~\ref{lem:gram}) about a quantity
defined globally. The minimum is the correct choice for the uniform
guarantees of Proposition~\ref{prop:ric}, and cannot be weakened there
without making that bound false. It is, however, open to supplement:
complementary aggregations, dimension-weighted, or quantile-trimmed so
as to disregard a small and known fraction of the domain, would report
typical rather than worst-case transfer quality, and deserve a fuller
treatment for applications that can tolerate degradation confined to a
few coincidence classes.

The picture that emerges is that transferability, equivariance,
sampling, and compressed sensing, four notions ordinarily developed
with separate tools, are different consequences of one algebraic fact:
two representations of the same algebra share exactly as much structure
as their filtered spectra coincide, and the filter is the one object
common to all four notions through which that sharing can be designed,
measured, and, when necessary, repaired.

%% file: v09/sec_appendix.tex
%!TEX root =../compiler_paper.tex

%%%%%%%%%%%%%%%%%%%%%%%%%%%%%%%%%%%%%%%%%%%%%%
%%%%%%%%%%%%%%%%%% APPENDIX %%%%%%%%%%%%%%%%%%
%%%%%%%%%%%%%%%%%%%%%%%%%%%%%%%%%%%%%%%%%%%%%%
\section{Proofs}

%%%%%%%%%%%%%%%%%%%%%%%%%%%%%%%%%%%%%%%%%%%%%%
%%%%%%%%%%%%%%%%%%%%%%%%%%%%%%%%%%%%%%%%%%%%%%
\subsection{Proof of Corollary~\ref{cor_space_of_hom}}
\label{proof_cor_space_of_hom}

\begin{proof}
The zero map belongs to $\Theta_{\ccalA_{0}}$, so the set is nonempty. Let
$\theta_{1},\theta_{2}\in\Theta_{\ccalA_{0}}$ and $\alpha,\beta\in\mbC$. Then,
for every $a\in\ccalA_{0}$,
$
(\alpha\theta_{1}+\beta\theta_{2})\rho_{1}(a)
=\alpha\,\theta_{1}\rho_{1}(a)+\beta\,\theta_{2}\rho_{1}(a)
=\alpha\,\rho_{2}(a)\theta_{1}+\beta\,\rho_{2}(a)\theta_{2}
=\rho_{2}(a)(\alpha\theta_{1}+\beta\theta_{2})
$,
so $\alpha\theta_{1}+\beta\theta_{2}\in\Theta_{\ccalA_{0}}$. Hence
$\Theta_{\ccalA_{0}}$ is a linear subspace of the space of linear maps from
$\ccalH_{1}$ to $\ccalH_{2}$.
\end{proof}

%%%%%%%%%%%%%%%%%%%%%%%%%%%%%%%%%%%%%%%%%%%%%%
%%%%%%%%%%%%%%%%%%%%%%%%%%%%%%%%%%%%%%%%%%%%%%
\subsection{Proof of Corollary~\ref{corll_hom_asp_shiftoper}}
\label{proof_corll_hom_asp_shiftoper}

\begin{proof}
Since $\rho_{1}$ and $\rho_{2}$ are unital homomorphisms, iterated
application of~\eqref{eq_hom_def} to the monomial $a=g^{k}$ gives
$\rho_{1}(g^{k})=\rho_{1}(g)^{k}=\mathbf{S}^{k}$ and
$\rho_{2}(g^{k})=\rho_{2}(g)^{k}=\widetilde{\mathbf{S}}^{k}$ for every
$k\geq0$, the case $k=0$ being the unitality of $\rho_{1}$ and $\rho_{2}$.
Writing $p(g)=\sum_{k}h_{k}g^{k}$ and using the linearity of $\rho_{1}$ and
$\rho_{2}$, it follows that
$\rho_{1}\left(p(g)\right)=\sum_{k}h_{k}\mathbf{S}^{k}=p(\mathbf{S})$ and
$\rho_{2}\left(p(g)\right)=\sum_{k}h_{k}\widetilde{\mathbf{S}}^{k}
=p(\widetilde{\mathbf{S}})$. The claim is then the operator
form~\eqref{eq_def_hom_asm_2} of Definition~\ref{def_hom_asm} applied to the
filter $a=p(g)\in\ccalA$.
\end{proof}

%%%%%%%%%%%%%%%%%%%%%%%%%%%%%%%%%%%%%%%%%%%%%%
%%%%%%%%%%%%%%%%%%%%%%%%%%%%%%%%%%%%%%%%%%%%%%
\subsection{Proof of Theorem~\ref{thm_trans_asp_finite}}
\label{app_proof_thm_finite}

\begin{proof}
Throughout the proof we identify $\theta$, $\mathbf{S}$ and
$\widetilde{\mathbf{S}}$ with matrices, as in the statement. The argument
relies on the vectorization identity
$\mathrm{vec}\left(\mathbf{A}\mathbf{X}\mathbf{B}\right)
=\left(\mathbf{B}^{\mathsf{T}}\otimes\mathbf{A}\right)\mathrm{vec}(\mathbf{X})$,
valid for any matrices of compatible dimensions~\cite{horn2012matrix}, where
$\mathrm{vec}(\cdot)$ stacks the columns of its argument into a single
vector.

Since~\eqref{eq_thm_trans_asp_finite_1} holds for every
$\mathbf{x}\in\ccalH_{1}$, it is equivalent to the matrix equation
$\theta\, p(\mathbf{S})-p(\widetilde{\mathbf{S}})\,\theta=\mathbf{0}$.
Applying the vectorization identity to each term, with
$\theta\,p(\mathbf{S})=\mathbf{I}_{N_{2}}\,\theta\,p(\mathbf{S})$ and
$p(\widetilde{\mathbf{S}})\,\theta
=p(\widetilde{\mathbf{S}})\,\theta\,\mathbf{I}_{N_{1}}$, we obtain
\begin{equation}\label{eq_app_vec}
\mathrm{vec}\left(
       \theta\, p(\mathbf{S})-p(\widetilde{\mathbf{S}})\,\theta
       \right)
       =
       -\,\mathbf{T}\,
       \mathrm{vec}(\theta)
       ,
\end{equation}
with $\mathbf{T}$ as in~\eqref{eq_thm_trans_asp_finite_2}. Hence,
$\theta\in\Theta_{\ccalA_{0}}$ with $\ccalA_{0}=\{p(g)\}$ if and only if
$\mathrm{vec}(\theta)\in\ker(\mathbf{T})$, and since $\mathrm{vec}$ is a
linear bijection,
$\dim\left(\Theta_{\ccalA_{0}}\right)=\dim\ker(\mathbf{T})$, which is the
geometric multiplicity of the zero eigenvalue of $\mathbf{T}$.

To compute the eigenvalues of $\mathbf{T}$, let
$\mathbf{U}_{1}^{*}\,p(\mathbf{S})^{\mathsf{T}}\,\mathbf{U}_{1}=\mathbf{R}_{1}$
and
$\mathbf{U}_{2}^{*}\,p(\widetilde{\mathbf{S}})\,\mathbf{U}_{2}=\mathbf{R}_{2}$
be Schur triangularizations, with $\mathbf{U}_{1},\mathbf{U}_{2}$ unitary
and $\mathbf{R}_{1},\mathbf{R}_{2}$ upper triangular carrying the
eigenvalues $\{p(\lambda_{i})\}_{i=1}^{N_{1}}$ and
$\{p(\widetilde{\lambda}_{j})\}_{j=1}^{N_{2}}$ on their diagonals -- note
that $p(\mathbf{S})^{\mathsf{T}}$ and $p(\mathbf{S})$ share the same
eigenvalues. Then
\begin{equation}
\left(\mathbf{U}_{1}\otimes\mathbf{U}_{2}\right)^{*}
\mathbf{T}
\left(\mathbf{U}_{1}\otimes\mathbf{U}_{2}\right)
=
\mathbf{I}_{N_{1}}\otimes\mathbf{R}_{2}
-
\mathbf{R}_{1}\otimes\mathbf{I}_{N_{2}}
,
\end{equation}
which is upper triangular with diagonal entries
$p(\widetilde{\lambda}_{j})-p(\lambda_{i})$, for $i=1,\ldots,N_{1}$ and
$j=1,\ldots,N_{2}$. These are therefore the eigenvalues of $\mathbf{T}$.

Finally, if $p(\lambda_{i})\neq p(\widetilde{\lambda}_{j})$ for all pairs
$(i,j)$, then all eigenvalues of $\mathbf{T}$ are nonzero, so $\mathbf{T}$
is invertible, $\ker(\mathbf{T})=\{0\}$, and
$\mathrm{vec}(\theta)=\mathbf{0}$, this is, $\theta=\mathbf{0}$.
\end{proof}

%%%%%%%%%%%%%%%%%%%%%%%%%%%%%%%%%%%%%%%%%%%%%%
%%%%%%%%%%%%%%%%%%%%%%%%%%%%%%%%%%%%%%%%%%%%%%
\subsection{Proof of Lemma~\ref{lem:support}}
\label{proof_lem_support}

\begin{proof}
Let $c \in p(\sigma(\mathbf{S}))$ and
$c' \in p(\sigma(\widetilde{\mathbf{S}}))$. Multiplying
\eqref{eq_thm_trans_asp_finite_1} by $\widetilde{\mathbf{\Pi}}_{c'}$ on the
left and $\mathbf{\Pi}_{c}$ on the right, and using
$p(\mathbf{S})\mathbf{\Pi}_c = c\,\mathbf{\Pi}_c$ and
$\widetilde{\mathbf{\Pi}}_{c'}\, p(\widetilde{\mathbf{S}}) = c'\,\widetilde{\mathbf{\Pi}}_{c'}$, we obtain
\begin{equation}
  c \;\widetilde{\mathbf{\Pi}}_{c'}\, \theta\, \mathbf{\Pi}_{c}
  \;=\;
  c' \;\widetilde{\mathbf{\Pi}}_{c'}\, \theta\, \mathbf{\Pi}_{c},
  \qquad\text{i.e.}\qquad
  (c - c')\, \widetilde{\mathbf{\Pi}}_{c'}\, \theta\, \mathbf{\Pi}_{c} = 0 .
\end{equation}
Hence $\widetilde{\mathbf{\Pi}}_{c'} \theta \mathbf{\Pi}_c = 0$ whenever
$c \neq c'$, and summing
$\theta = \sum_{c,c'} \widetilde{\mathbf{\Pi}}_{c'}\theta \mathbf{\Pi}_c$ over
all pairs leaves only the diagonal terms $c = c' \in \ccalC_{p}$. The
converse is a direct computation on each block.
\end{proof}

%%%%%%%%%%%%%%%%%%%%%%%%%%%%%%%%%%%%%%%%%%%%%%
%%%%%%%%%%%%%%%%%%%%%%%%%%%%%%%%%%%%%%%%%%%%%%
\subsection{Block Maps and Their Ambient Extensions}
\label{app_block_maps}

Here we make explicit the correspondence between the block map $\theta_{c}$
of Section~\ref{sec_deg_transf} and the summand
$\widetilde{\mathbf{\Pi}}_{c}\,\theta\,\mathbf{\Pi}_c$
of~\eqref{eq:support}.

For $\theta \in \Theta_{p}$ and $c \in \ccalC_{p}$, the \emph{block map at
$c$} is the restriction
$\theta_c := \theta\big|_{\ccalE_{c}}: \ccalE_c \longrightarrow \widetilde{\ccalE}_{c}$,
which is well defined by Lemma~\ref{lem:support} and is represented, in
orthonormal bases of $\ccalE_c$ and $\widetilde{\ccalE}_{c}$, by an
$\widetilde{m}_c \times m_c$ complex matrix. By \eqref{eq:support},
$\theta$ is completely determined by, and decomposes orthogonally into,
its blocks $\{\theta_c\}_{c \in \ccalC_{p}}$ with a careful distinction. The summand $\widetilde{\mathbf{\Pi}}_{c}\,\theta\,\mathbf{\Pi}_c$ in~\eqref{eq:support} determines the block map $\theta_c$, however the former is an operator on the ambient
spaces, while the latter is the intrinsic map between the subspaces
$\ccalE_c$ and $\widetilde{\ccalE}_c$ themselves. The two are related as
follows. Let $\iota_c : \ccalE_c \hookrightarrow \mathcal{H}_1$ and $\widetilde{\iota}_c : \widetilde{\ccalE}_c \hookrightarrow \mathcal{H}_{2}$ denote the canonical isometric inclusions of $\ccalE_c$ and
$\widetilde{\ccalE}_c$ into their respective ambient spaces, so that the
spectral projectors factor as
\begin{equation}\label{eq:proj_factor}
    \mathbf{\Pi}_c = \iota_c\, \iota_c^{*},
    \qquad
    \widetilde{\mathbf{\Pi}}_c = \widetilde{\iota}_c\, \widetilde{\iota}_c^{*}
    ,
\end{equation}
where $\iota_{c}^{*}$ is the adjoint of $\iota_{c}$. Then
\begin{equation}\label{eq:block_vs_ambient}
    \widetilde{\mathbf{\Pi}}_{c}\, \theta\, \mathbf{\Pi}_c
    \;=\;
    \widetilde{\iota}_c\, \theta_c\, \iota_c^{*},
    \qquad\text{equivalently}\qquad
    \theta_c
    \;=\;
    \widetilde{\iota}_c^{*}\, \theta\, \iota_c .
\end{equation}
That is, $\widetilde{\mathbf{\Pi}}_{c}\theta\mathbf{\Pi}_c$ is the
zero-extension of $\theta_c$ to the ambient spaces, while $\theta_c$ is
the compression of $\widetilde{\mathbf{\Pi}}_{c}\theta\mathbf{\Pi}_c$ to
$\ccalE_c$ and $\widetilde{\ccalE}_c$. This compression is lossless
precisely because of Lemma~\ref{lem:support}: since
$\theta(\ccalE_c) \subseteq \widetilde{\ccalE}_c$, the projector
$\widetilde{\mathbf{\Pi}}_c$ acts as the identity on $\theta(\ccalE_c)$,
so no information is discarded in passing from the ambient operator
$\widetilde{\mathbf{\Pi}}_{c}\theta\mathbf{\Pi}_c$ to the intrinsic block
$\theta_c$, nor is any spurious information introduced in the reverse
direction. Fixing orthonormal bases of $\ccalE_c$ and
$\widetilde{\ccalE}_c$ identifies $\iota_c$ and $\widetilde{\iota}_c$
with the corresponding basis-embedding matrices, and $\theta_c$ with the
$\widetilde{m}_c \times m_c$ matrix representative used throughout.

%%%%%%%%%%%%%%%%%%%%%%%%%%%%%%%%%%%%%%%%%%%%%%
%%%%%%%%%%%%%%%%%%%%%%%%%%%%%%%%%%%%%%%%%%%%%%
\subsection{Proof of Proposition~\ref{prop:dimension}}
\label{proof_prop_dimension}

\begin{proof}
By Lemma~\ref{lem:support} the map
$\theta \mapsto (\theta_c)_{c\in\ccalC_{p}}$ is a linear bijection from
$\Theta_p$ onto
$\bigoplus_c \operatorname{Hom}(\ccalE_c,\widetilde{\ccalE}_{c})$, and
$\dim \operatorname{Hom}(\ccalE_c,\widetilde{\ccalE}_{c}) = m_c \widetilde{m}_c$.
Ranks add across the blocks, since
$\ccalH_{2}=\bigoplus_{c}\widetilde{\ccalE}_{c}$ is a direct sum, and
$\text{rank}(\theta_c)\leq\min(m_c,\widetilde{m}_c)$, with equality attained
by choosing each $\theta_{c}$ of full rank. Summing over $c$
gives~\eqref{eq_max_rank}.
\end{proof}

%%%%%%%%%%%%%%%%%%%%%%%%%%%%%%%%%%%%%%%%%%%%%%
%%%%%%%%%%%%%%%%%%%%%%%%%%%%%%%%%%%%%%%%%%%%%%
\subsection{Proof of Proposition~\ref{prop:achievability}}
\label{proof_prop_achievability}

\begin{proof}
(i) If $m_c > \widetilde{m}_c$, then $\ker \theta_c \neq \{0\}$ by
dimension counting, so $\sigma_{\min}(\theta_c) = 0$ and
$\eta_c(\theta) = 0$ (this covers $\theta_c = 0$ as well).

(ii) Necessity follows from (i). For sufficiency, if
$m_c \leq \widetilde{m}_c$ choose $\theta_c$ a nonzero scaled isometry of
$\ccalE_c$ into $\widetilde{\ccalE}_{c}$; then
$\sigma_{\min}(\theta_c) = \sigma_{\max}(\theta_c) > 0$ and
$\eta_c(\theta) = 1$ for every $c$.

(iii) If \eqref{eq:injectivity} holds, choose each $\theta_c$ an isometry,
that is, with unit scale rather than the arbitrary scale permitted in (ii),
since $\theta^{*}\theta=\sum_{c}\sigma_{c}^{2}\mathbf{\Pi}_{c}$ equals
$\mathbf{I}$ only when every $\sigma_{c}=1$. The first condition
in \eqref{eq:injectivity} guarantees that
no eigenspace of $\mathbf{S}$ lies outside
$\bigoplus_{c\in\ccalC_{p}} \ccalE_c$, hence none is annihilated by
Lemma~\ref{lem:support}, and $\theta^{*}\theta = \mathbf{I}$
follows from the orthogonality of the blocks. Conversely, if
$p(\lambda) \notin p(\sigma(\widetilde{\mathbf{S}}))$ for some
$\lambda \in \sigma(\mathbf{S})$, then
$\mathcal{E}_\lambda \subseteq \ker\theta$ for every
$\theta \in \Theta_p$; and if $m_c > \widetilde{m}_c$ for some $c$, apply (i).
Either obstruction rules out injectivity.
\end{proof}

%%%%%%%%%%%%%%%%%%%%%%%%%%%%%%%%%%%%%%%%%%%%%%
%%%%%%%%%%%%%%%%%%%%%%%%%%%%%%%%%%%%%%%%%%%%%%
\subsection{Proof of Proposition~\ref{prop:diag}}
\label{proof_prop_diag}

\begin{proof}
The proof of Lemma~\ref{lem:support} uses only the identities
$p(\mathbf{S})\mathbf{\Pi}_c=c\,\mathbf{\Pi}_c$,
$\widetilde{\mathbf{\Pi}}_{c'}\,p(\widetilde{\mathbf{S}})=c'\,\widetilde{\mathbf{\Pi}}_{c'}$,
and the resolutions of the identity
$\sum_{c}\mathbf{\Pi}_c=\mathbf{I}$,
$\sum_{c}\widetilde{\mathbf{\Pi}}_c=\mathbf{I}$ -- the latter holding
for \emph{any} diagonalizable operator, since its spectral projections are
built from a full eigenbasis regardless of normality. All four identities
were verified in Section~\ref{sec_nonnormal} for the oblique Riesz
projections of~\eqref{eq_lagrange_projection}, so the conclusion of
Lemma~\ref{lem:support}, and with it the well-definedness of $\theta_c$,
transfer unchanged. The proof of Proposition~\ref{prop:dimension} is then
purely combinatorial, it identifies $\Theta_p$ with
$\bigoplus_c\operatorname{Hom}(\ccalE_c,\widetilde{\ccalE}_c)$ through
Lemma~\ref{lem:support} and counts dimensions, so it is likewise
unaffected. Neither statement refers to a metric, which is why both hold
verbatim.

For Proposition~\ref{prop:achievability} the situation is different, as its
conclusions are stated in terms of the extreme gains of
Definition~\ref{def:eta}, which do depend on the inner products. Equip
$\ccalH_{1}$ with the adapted inner product
$\langle\cdot,\cdot\rangle_{\oplus}$ of~\eqref{eq_adapted_inner_product}
and $\ccalH_{2}$ with its counterpart built from
$\widetilde{\mathbf{\Pi}}_{c}$. By construction
$\|\mathbf{x}\|_{\oplus}^{2}=\sum_{c}\|\mathbf{\Pi}_{c}\mathbf{x}\|^{2}$,
so distinct coincidence spaces are mutually orthogonal in
$\langle\cdot,\cdot\rangle_{\oplus}$: for
$\mathbf{u}\in\ccalE_{c}$ and $\mathbf{w}\in\ccalE_{c'}$ with $c\neq c'$
one has $\mathbf{\Pi}_{c''}\mathbf{u}=\delta_{c,c''}\mathbf{u}$ and
$\mathbf{\Pi}_{c''}\mathbf{w}=\delta_{c',c''}\mathbf{w}$, whence
$\langle\mathbf{u},\mathbf{w}\rangle_{\oplus}=0$; the same computation
applies on $\ccalH_{2}$. The decomposition
$\ccalH_{1}=\bigoplus_{c}\ccalE_{c}$ is therefore orthogonal in the adapted
inner product, exactly as it was in the ambient one under normality. Since
the dimension-counting and isometric-embedding arguments in the proof of
Proposition~\ref{prop:achievability} are stated entirely in terms of the
blocks and of orthogonality of the coincidence spaces, they apply verbatim
once the gains of Definition~\ref{def:eta} are computed with respect to
$\langle\cdot,\cdot\rangle_{\oplus}$. The isometry
$\theta^{*}\theta=\mathbf{I}$ produced by
Proposition~\ref{prop:achievability}(iii) is understood in this same sense,
with the adjoint taken with respect to the adapted inner products; its
translation into a bound in the ambient inner products is given separately
in Proposition~\ref{prop:ricdiag}. Finally, when $\mathbf{S}$ and
$\widetilde{\mathbf{S}}$ are normal the projections are orthogonal, so
$\mathbf{\Pi}_{c}^{*}\mathbf{\Pi}_{c}=\mathbf{\Pi}_{c}$ and
$\|\mathbf{x}\|_{\oplus}^{2}=\sum_{c}\langle\mathbf{\Pi}_{c}\mathbf{x},\mathbf{x}\rangle
=\|\mathbf{x}\|^{2}$, so the adapted and ambient inner products coincide,
as claimed.
\end{proof}

%%%%%%%%%%%%%%%%%%%%%%%%%%%%%%%%%%%%%%%%%%%%%%
%%%%%%%%%%%%%%%%%%%%%%%%%%%%%%%%%%%%%%%%%%%%%%
\subsection{Proof of Proposition~\ref{prop:dimjordan}}
\label{proof_prop_dimjordan}

\begin{proof}
Fix $c\in\ccalC_p$ and write $\mathbf N:=\mathbf N_c$,
$\widetilde{\mathbf N}:=\widetilde{\mathbf N}_c$. By the Jordan
decomposition of a nilpotent operator, $\ccalE_c=\bigoplus_i V_i$ and
$\widetilde{\ccalE}_c=\bigoplus_j\widetilde V_j$, with
$\mathbf N|_{V_i}$ a single nilpotent Jordan block of size $p_i$ on $V_i$
and $\widetilde{\mathbf N}|_{\widetilde V_j}$ one of size $q_j$ on
$\widetilde V_j$. Writing $\theta_c$ in block form with respect to these
decompositions, $\theta_c=(\theta_c^{(i,j)})_{i,j}$ with
$\theta_c^{(i,j)}:V_i\to\widetilde V_j$, relation~\eqref{eq:nilpotent}
decouples into independent relations on each pair, since $\mathbf N$ and
$\widetilde{\mathbf N}$ act block-diagonally with respect to
$\{V_i\}$ and $\{\widetilde V_j\}$ respectively:
\begin{equation}\label{eq_decoupled_jordan}
  \theta_c^{(i,j)}\, \mathbf N_{p_i} \;=\; \mathbf N_{q_j}\, \theta_c^{(i,j)},
  \qquad \forall\, i,j,
\end{equation}
where $\mathbf N_{p_i},\mathbf N_{q_j}$ are the standard nilpotent Jordan
blocks of sizes $p_i,q_j$. It remains to compute
$\dim\{X:V_i\to\widetilde V_j \mid X\mathbf N_{p_i}=\mathbf N_{q_j}X\}$.
Identify $(V_i,\mathbf N_{p_i})$ with the $\mathbb C[t]$-module
$\mathbb C[t]/(t^{p_i})$, $t$ acting as $\mathbf N_{p_i}$, and similarly
$(\widetilde V_j,\mathbf N_{q_j})$ with $\mathbb C[t]/(t^{q_j})$. A
$\mathbb C[t]$-module homomorphism out of the cyclic module
$\mathbb C[t]/(t^{p_i})$ is determined freely by the image of its
generator, subject only to annihilation by $t^{p_i}$. The annihilator of
$t^{p_i}$ in $\mathbb C[t]/(t^{q_j})$ is the submodule
$(t^{\max(q_j-p_i,0)})/(t^{q_j})$, of dimension
$q_j-\max(q_j-p_i,0)=\min(p_i,q_j)$. Hence
$\dim\{X:V_i\to\widetilde V_j\mid X\mathbf N_{p_i}=\mathbf N_{q_j}X\}=\min(p_i,q_j)$,
recovering the classical solution count for the Sylvester equation
restricted to a single pair of Jordan blocks~\cite{horn1991topics}. Summing
\eqref{eq_decoupled_jordan} over $(i,j)$, and the resulting count over
$c\in\ccalC_p$ via Lemma~\ref{lem:support} in its generalized-eigenspace
form, gives~\eqref{eq:dimjordan}. In the semisimple case
$p_i\equiv q_j\equiv1$, $\min(p_i,q_j)=1$ for every pair, and the double
sum collapses to $(\#\{i\})(\#\{j\})=m_c\widetilde m_c$, recovering
Proposition~\ref{prop:dimension}.

For the second claim, suppose first $m_c=\widetilde m_c$ and the
partitions match termwise, $p_i=q_i$ for all $i$. Sending each Jordan
basis vector of $V_i$ to the corresponding basis vector of the
identically-sized $\widetilde V_i$ defines a unitary $\mathbf U$ with
$\mathbf U\mathbf N=\widetilde{\mathbf N}\mathbf U$ by direct check on
each generator, so $\theta_c=\mathbf U$ is a nonzero scaled isometry satisfying~\eqref{eq:nilpotent}, and $\eta_c(\theta)=1$.
Conversely, suppose $m_c=\widetilde m_c$ and $\theta_c$ is a nonzero
scaled isometry, $\theta_c^{*}\theta_c=\sigma^2\mathbf I$ for some
$\sigma>0$. Then $\mathbf U:=\theta_c/\sigma$ satisfies
$\mathbf U^{*}\mathbf U=\mathbf I$. Since
$\dim\ccalE_c=\dim\widetilde{\ccalE}_c$, $\mathbf U$ is a bijective
isometry, hence unitary, and $\mathbf U\mathbf N\mathbf U^{-1}=\widetilde{\mathbf N}$.
Conjugation by an invertible operator sends $\ker(\mathbf N^k)$ onto
$\ker(\widetilde{\mathbf N}^k)$ exactly, so
$\dim\ker(\mathbf N^k)=\dim\ker(\widetilde{\mathbf N}^k)$ for every
$k\geq1$. Since $\dim\ker(\mathbf N^k)=\sum_i\min(p_i,k)$, the counts
$\#\{i:p_i\geq k\}=\dim\ker(\mathbf N^k)-\dim\ker(\mathbf N^{k-1})$
recover the partition $\{p_i\}$ exactly from these dimensions, and
likewise for $\{q_i\}$. The equality of $\dim\ker(\mathbf N^k)$ and
$\dim\ker(\widetilde{\mathbf N}^k)$ for every $k$ therefore forces
$\{p_i\}=\{q_i\}$ termwise. Hence, whenever $m_c=\widetilde m_c$ but the
Jordan partitions do not coincide -- which occurs, in particular, whenever
exactly one of $\mathbf N_c,\widetilde{\mathbf N}_c$ is nonzero, or both
are nonzero with different block sizes -- no scaled isometry exists,
so $\eta_c(\theta)<1$ for every $\theta\in\Theta_p$, and therefore
$\eta(\theta)<1$: $\eta(\theta)=1$ is unachievable.
\end{proof}

%%%%%%%%%%%%%%%%%%%%%%%%%%%%%%%%%%%%%%%%%%%%%%
%%%%%%%%%%%%%%%%%%%%%%%%%%%%%%%%%%%%%%%%%%%%%%
\subsection{Proof of Lemma~\ref{lem:pprime}}
\label{app_proof_pprime}

\begin{proof}
Write $\mathbf{J}=\mathbf{J}_k(\lambda)=\lambda\mathbf{I}+\mathbf{N}$, where
$\mathbf{N}$ is the $k\times k$ nilpotent shift ($\mathbf{N}^{k}=\mathbf 0$,
$\mathbf{N}^{k-1}\neq\mathbf 0$). Since $p$ is a polynomial, its Taylor
expansion about $\lambda$ is exact and truncates at $\mathbf{N}^{k-1}$:
\begin{equation}\label{eq_app_taylor}
p(\mathbf{J})
=
\sum_{i=0}^{k-1}
     \frac{p^{(i)}(\lambda)}{i!}\,
     \mathbf{N}^{i}
     .
\end{equation}
By hypothesis $p^{(i)}(\lambda)=0$ for $1\leq i\leq r-1$ and
$p^{(r)}(\lambda)\neq0$, so~\eqref{eq_app_taylor} reduces to
\begin{equation}
p(\mathbf{J})
-
p(\lambda)\mathbf{I}
=
\mathbf{M}
:=
\sum_{i=r}^{k-1}
     c_{i}\,
     \mathbf{N}^{i}
     ,
\qquad
c_{i}:=\frac{p^{(i)}(\lambda)}{i!}
     ,
     \quad
     c_{r}\neq0
     .
\end{equation}
Adding the scalar multiple of the identity $p(\lambda)\mathbf{I}$ does not
alter Jordan block sizes, so it suffices to determine the Jordan type of
the nilpotent matrix $\mathbf{M}$; this Jordan type is then that of
$p(\mathbf{J})$ at the eigenvalue $p(\lambda)$.

\smallskip
\noindent\textbf{Step 1: the rank sequence of $\mathbf{M}$ depends only on
$k$ and $r$.} Factor
$\mathbf{M}=c_{r}\mathbf{N}^{r}\,\mathbf{U}$, where
\begin{equation}
\mathbf{U}
:=
\mathbf{I}
+
\sum_{i=r+1}^{k-1}
     \frac{c_{i}}{c_{r}}\,
     \mathbf{N}^{i-r}
     .
\end{equation}
As a polynomial in $\mathbf{N}$ with constant term $\mathbf{I}$,
$\mathbf{U}$ is upper triangular with every diagonal entry equal to $1$
(in the basis in which $\mathbf{N}$ is the standard shift), hence
$\det\mathbf{U}=1$ and $\mathbf{U}$ is invertible; moreover $\mathbf{U}$
commutes with $\mathbf{N}$, being a polynomial in it. Consequently, for
every $j\geq0$,
\begin{equation}
\mathbf{M}^{j}
=
c_{r}^{\,j}\,
\mathbf{N}^{rj}\,
\mathbf{U}^{j}
     ,
\end{equation}
and since $\mathbf{U}^{j}$ is invertible, right multiplication by it does
not change rank:
\begin{equation}\label{eq_app_rankM}
\operatorname{rank}\big(\mathbf{M}^{j}\big)
=
\operatorname{rank}\big(\mathbf{N}^{rj}\big)
=
\max(k-rj,\,0)
     ,
\end{equation}
using the standard fact that a single nilpotent Jordan block of size $k$
satisfies $\operatorname{rank}(\mathbf{N}^{i})=\max(k-i,0)$. In particular,
\eqref{eq_app_rankM} does not depend on the coefficients
$c_{r+1},\ldots,c_{k-1}$, only on $k$ and $r$.

\smallskip
\noindent\textbf{Step 2: the rank sequence determines the Jordan type.}
For a nilpotent matrix with Jordan block sizes $b_{1}\geq b_{2}\geq\cdots\geq b_{m}$
(a partition of $k$), ranks add across blocks, so
\begin{equation}
\operatorname{rank}\big(\mathbf{M}^{j}\big)
=
\sum_{\ell=1}^{m}
     \max(b_{\ell}-j,\,0)
     ,
     \qquad
     j=0,1,2,\ldots
     ,
\end{equation}
and this correspondence is injective: the differences
\begin{equation}
\#\{\ell:b_{\ell}\geq j\}
=
\operatorname{rank}\big(\mathbf{M}^{j-1}\big)
-
\operatorname{rank}\big(\mathbf{M}^{j}\big)
     ,
     \qquad
     j\geq1,
\end{equation}
recover, for every $j$, the number of blocks of size at
least $j$, which determines the partition $\{b_{\ell}\}$ uniquely (the
same recovery principle used in the proof of
Proposition~\ref{prop:dimjordan}). It therefore suffices to exhibit a
partition of $k$ into $r$ parts reproducing~\eqref{eq_app_rankM}, since
uniqueness then forces it to be the Jordan type of $\mathbf{M}$.

\smallskip
\noindent\textbf{Step 3: verifying the candidate partition.} Write
$k=qr+s$ with $q=\lfloor k/r\rfloor$ and $s=k\bmod r$, $0\leq s<r$, and
take the partition consisting of $s$ parts equal to $q+1=\lceil k/r\rceil$
and $r-s$ parts equal to $q=\lfloor k/r\rfloor$ (parts equal to $0$, which
occur only if $q=0$, are simply absent). Its rank sequence is
\begin{equation}
R(j)
:=
s\max(q+1-j,0)
+
(r-s)\max(q-j,0)
     .
\end{equation}
For $0\leq j\leq q$, every part is at least $q\geq j$, so both maxima are
attained without truncation and
$R(j)=s(q+1-j)+(r-s)(q-j)=rq+s-rj=k-rj$, which equals $\max(k-rj,0)$
since $k-rj\geq k-rq=s\geq0$ in this range. For $j\geq q+1$, every part
satisfies $b_{\ell}\leq q+1\leq j$, so $R(j)=0$; and
$\max(k-rj,0)=\max(k-r(q+1)-r(j-q-1),0)=\max(s-r-r(j-q-1),0)=0$ as well,
since $s<r$ makes $s-r<0$ and the remaining term only decreases it
further. Hence $R(j)=\max(k-rj,0)=\operatorname{rank}(\mathbf{M}^{j})$
for every $j\geq0$, and by the uniqueness of Step~2 this candidate
partition is exactly the Jordan type of $\mathbf{M}$, hence of
$p(\mathbf{J})-p(\lambda)\mathbf{I}$, hence of $p(\mathbf{J})$ at the
eigenvalue $p(\lambda)$: $s=k\bmod r$ blocks of size $\lceil k/r\rceil$
and $r-s$ blocks of size $\lfloor k/r\rfloor$, as claimed.

\smallskip
\noindent\textbf{The two special cases.} Setting $r=1$ gives $q=k$,
$s=0$: the formula returns a single block of size $k$, i.e., the block is
unchanged -- statement (i). Setting $r\geq k$ gives $q=0$,
$s=k\bmod r=k$ (since $k<r$, or $k=r$ forcing $s=0,q=1$, a degenerate
instance of the same count): in either case the partition collapses to
$k$ blocks of size $1$ and (if $r>k$) $r-k$ vacuous blocks of size $0$,
i.e., $p(\mathbf{J})$ is semisimple on the generalized eigenspace of
$\lambda$ -- statement (ii). This is consistent with
$\mathbf{M}=\mathbf{0}$ directly whenever $r\geq k$, since every term
$\mathbf{N}^{i}$ with $i\geq r\geq k$ vanishes in~\eqref{eq_app_taylor}.
\end{proof}

%%%%%%%%%%%%%%%%%%%%%%%%%%%%%%%%%%%%%%%%%%%%%%
%%%%%%%%%%%%%%%%%%%%%%%%%%%%%%%%%%%%%%%%%%%%%%
\subsection{Proof of Proposition~\ref{prop:ric}}
\label{proof_prop_ric}

\begin{proof}
Fix $\mathbf{x}\in\ccalH_{\ccalK}$. By the definition
of~\eqref{eq_band_subspace} we have $\mathbf{\Pi}_{c}\mathbf{x}=0$ for every
$c\notin\ccalK$, so
$\mathbf{x}=\sum_{c\in\ccalK}\mathbf{\Pi}_c\mathbf{x}$. Since $p(\mathbf{S})$ is
normal, the spectral decomposition~\eqref{eq_spectral_decomp_finite} makes
the coincidence spaces $\{\ccalE_{c}\}_{c\in\ccalC_{p}}$ mutually
orthogonal and the $\mathbf{\Pi}_c$ orthogonal projections, so this
decomposition is orthogonal and
$\|\mathbf{x}\|^2=\sum_{c\in\ccalK}\|\mathbf{\Pi}_c\mathbf{x}\|^2$.
By Lemma~\ref{lem:support}, $\theta\mathbf{x}=\sum_{c\in\ccalK}\theta_c\mathbf{\Pi}_c\mathbf{x}$
with each term $\theta_c\mathbf{\Pi}_c\mathbf{x}\in\widetilde{\ccalE}_c$. Since
$p(\widetilde{\mathbf{S}})$ is normal, the target coincidence spaces
$\widetilde{\ccalE}_c$ are likewise mutually orthogonal, so this sum is also
orthogonal, giving
\begin{equation}
\left\Vert 
      \theta\mathbf{x}
\right\Vert^2 
        = 
         \sum_{c\,\in\,\ccalK} 
              \left\Vert 
                   \theta_c\, \mathbf{\Pi}_c \mathbf{x}
              \right\Vert^2 .
\end{equation}
Taking into account the definition of $\sigma_{\min}(\theta_c)$, we have
$\sigma_{\min}(\theta_c)\|\mathbf{y}\|\leq\|\theta_c\mathbf{y}\|\leq\sigma_{\max}(\theta_c)\|\mathbf{y}\|$
for every $\mathbf{y}\in\ccalE_c$ (the bound holds by definition for unit
vectors, hence for all vectors by homogeneity of the linear map
$\theta_c$). Applying this with $\mathbf{y}=\mathbf{\Pi}_c\mathbf{x}$ and
squaring,
\begin{equation}
  \sigma_{\min}^2(\theta_c)\,\|\mathbf{\Pi}_c\mathbf{x}\|^2
  \;\leq\;
  \|\theta_c\,\mathbf{\Pi}_c\mathbf{x}\|^2
  \;\leq\;
  \sigma_{\max}^2(\theta_c)\,\|\mathbf{\Pi}_c\mathbf{x}\|^2 .
\end{equation}
Under the stated gain normalization $\sigma_{\max}(\theta_c)=1$ for
$c\in\ccalK$,
Definition~\ref{def:eta} gives $\sigma_{\min}^2(\theta_c)=\eta_c(\theta)$
by~\eqref{eq:etac}, and $\eta_c(\theta)\geq\eta_{\ccalK}(\theta)$
by~\eqref{eq_band_subspace}. Then, by substitution it follows that
\begin{equation}
  \eta_{\ccalK}(\theta)\,\|\mathbf{\Pi}_c\mathbf{x}\|^2
  \;\leq\;
  \|\theta_c\,\mathbf{\Pi}_c\mathbf{x}\|^2
  \;\leq\;
  \|\mathbf{\Pi}_c\mathbf{x}\|^2 .
\end{equation}
Summing over $c\in\ccalK$ and using the two orthogonal decompositions
established above,
\begin{equation}\label{eq_frame_bound_proof}
  \eta_{\ccalK}(\theta)\, \|\mathbf{x}\|^2 \;\leq\; \|\theta\, \mathbf{x}\|^2 \;\leq\; \|\mathbf{x}\|^2,
  \qquad \forall\, \mathbf{x} \in \ccalH_{\ccalK} .
\end{equation}

This is a two-sided frame bound with constants
$(\alpha,\beta)=(\eta_{\ccalK}(\theta),1)$, valid for \emph{every}
$\mathbf{x}\in\ccalH_{\ccalK}$ and hence, in particular, uniformly over every
sparsity level $s\leq\dim\ccalH_{\ccalK}$ and under any sparsity model on the
band. Rescaling
$\theta\mapsto\gamma\,\theta$ with
$\gamma=\sqrt{2}/\sqrt{\alpha+\beta}=\sqrt{2}/\sqrt{1+\eta_{\ccalK}(\theta)}$, which
is the map $\theta'$ of the statement,
multiplies~\eqref{eq_frame_bound_proof} by $\gamma^{2}$, giving
\begin{equation}
  \frac{2\alpha}{\alpha+\beta}\,\|\mathbf{x}\|^2
  \;\leq\;
  \|\theta'\,\mathbf{x}\|^2
  \;\leq\;
  \frac{2\beta}{\alpha+\beta}\,\|\mathbf{x}\|^2
  ,
\end{equation}
that is, $(1-\delta)\|\mathbf{x}\|^2\leq\|\theta'\mathbf{x}\|^2\leq(1+\delta)\|\mathbf{x}\|^2$
for every $\mathbf{x}\in\ccalH_{\ccalK}$, with
$\delta=(\beta-\alpha)/(\beta+\alpha)=(1-\eta_{\ccalK}(\theta))/(1+\eta_{\ccalK}(\theta))$,
since $1\mp\delta=2\alpha/(\alpha+\beta)$ or $2\beta/(\alpha+\beta)$
respectively. This is exactly the restricted isometry condition on the band
with this $\delta$, holding for every $s\leq\dim\ccalH_{\ccalK}$, which
gives~\eqref{eq:ricbound}.

Finally, the hypothesis $m_{c}\leq\widetilde m_{c}$ for $c\in\ccalK$ is what
makes the statement non-vacuous: it is exactly the condition under which
each $\theta_{c}$, $c\in\ccalK$, can be chosen injective
(Proposition~\ref{prop:achievability}), hence with
$\sigma_{\min}(\theta_{c})>0$ and $\eta_{\ccalK}(\theta)>0$.
\end{proof}

%%%%%%%%%%%%%%%%%%%%%%%%%%%%%%%%%%%%%%%%%%%%%%
%%%%%%%%%%%%%%%%%%%%%%%%%%%%%%%%%%%%%%%%%%%%%%
\subsection{Proof of Lemma~\ref{lem:gram}}
\label{proof_lem_gram}

\begin{proof}
By Lemma~\ref{lem:support},
$\theta \mathbf{v}_j \in \widetilde{\ccalE}_{c}$ with $c = p(\lambda_j)$,
and the coincidence spaces
$\{\widetilde{\ccalE}_{c}\}_{c \in \ccalC_{p}}$ are mutually orthogonal
since $p(\widetilde{\mathbf{S}})$ is normal. Inner products between columns
in different classes therefore vanish, and the maximum defining
$\mu(\theta)$ is attained within a single class.
\end{proof}

%%%%%%%%%%%%%%%%%%%%%%%%%%%%%%%%%%%%%%%%%%%%%%
%%%%%%%%%%%%%%%%%%%%%%%%%%%%%%%%%%%%%%%%%%%%%%
\subsection{Proof of Proposition~\ref{prop:mudesign}}
\label{proof_prop_mudesign}

\begin{proof}
(i) Suppose $m_c\leq\widetilde m_c$ for every $c\in\ccalC_p$. By
Proposition~\ref{prop:achievability}(ii), there exists $\theta\in\Theta_p$
with $\eta(\theta)=1$, obtained by choosing each block $\theta_c$ a
nonzero scaled isometry, $\theta_c^{*}\theta_c=\sigma_c^{2}\mathbf{I}$ for
some $\sigma_c>0$. In particular $\theta_c$ is injective, so
$\theta\mathbf{v}_j=\theta_c\mathbf{v}_j\neq0$ for every eigenvector
$\mathbf{v}_j\in\ccalE_c$, and $\theta$ has no vanishing columns. Since
$\theta_c$ is a scaled isometry, it preserves orthogonality on $\ccalE_c$:
for $j\neq k$ with $p(\lambda_j)=p(\lambda_k)=c$, the vectors
$\mathbf{v}_j,\mathbf{v}_k$ are orthonormal, so
\begin{equation}
\langle\theta_c\mathbf{v}_j,\theta_c\mathbf{v}_k\rangle
=
\sigma_c^{2}\langle\mathbf{v}_j,\mathbf{v}_k\rangle
=
0
     ,
\end{equation}
hence $\mu(\theta_c)=0$ for every $c\in\ccalC_p$. By
Lemma~\ref{lem:gram}, $\mu(\theta)=\max_c\mu(\theta_c)=0$.

(ii) Follows from~\eqref{eq:mumax} in Lemma~\ref{lem:gram} and the derivation of the Welch bound when considering linear maps with normalized columns~\cite{foucart2013mathematical,eldar2015sampling}.
\end{proof}

%%%%%%%%%%%%%%%%%%%%%%%%%%%%%%%%%%%%%%%%%%%%%%
%%%%%%%%%%%%%%%%%%%%%%%%%%%%%%%%%%%%%%%%%%%%%%
\subsection{Proof of Proposition~\ref{prop:mueta}}
\label{proof_prop_mueta}

\begin{proof}
(i) Normalize the columns to unit norm, which by the equal-norm hypothesis
is a single scalar rescaling of $\theta_{c}$ and therefore leaves
$\eta_{c}(\theta)$ unchanged, and let
$\mathbf{\Gamma}_c := \theta_c^{*}\theta_c$ denote the Gram operator of the
block, which then has unit diagonal and off-diagonal
entries of modulus at most $\mu(\theta_c)$. By Gershgorin's theorem its
eigenvalues lie in
$[\,1 - (m_c-1)\mu(\theta_c),\; 1 + (m_c-1)\mu(\theta_c)\,]$, and
$\eta_c(\theta)$ is the ratio of the extreme eigenvalues of
$\mathbf{\Gamma}_c$.
(ii) If $\eta_c(\theta) = 1$, then $\theta_c$ is a nonzero scaled
isometry, so $\mathbf{\Gamma}_c$ is a positive multiple of the identity and all
off-diagonal inner products vanish.
\end{proof}

%%%%%%%%%%%%%%%%%%%%%%%%%%%%%%%%%%%%%%%%%%%%%%
%%%%%%%%%%%%%%%%%%%%%%%%%%%%%%%%%%%%%%%%%%%%%%
\subsection{Proof of Proposition~\ref{prop:ricdiag}}
\label{proof_prop_ricdiag}

\begin{proof}
We argue for $\ccalK=\ccalC_{p}$, the case stated in
Proposition~\ref{prop:ricdiag}; the band-restricted version is obtained by
replacing every sum over $\ccalC_{p}$ by a sum over $\ccalK$, every
$\eta(\theta)$ by $\eta_{\ccalK}(\theta)$, and $\ccalH_{1}$ by
$\ccalH_{\ccalK}$ throughout.

Both metric operators in~\eqref{eq_metric_operator} are positive definite.
Indeed, $\langle\mathbf{G}\mathbf{x},\mathbf{x}\rangle
=\sum_{c}\|\mathbf{\Pi}_{c}\mathbf{x}\|^{2}=\|\mathbf{x}\|_{\oplus}^{2}\geq0$,
and if it vanishes then every $\mathbf{\Pi}_{c}\mathbf{x}=0$, so
$\mathbf{x}=\sum_{c}\mathbf{\Pi}_{c}\mathbf{x}=0$; the same argument applies
to $\widetilde{\mathbf{G}}$. Consequently, the Rayleigh quotient bounds for
the self-adjoint operator $\mathbf{G}$ give
\begin{equation}\label{eq_app_rayleigh}
\lambda_{\min}(\mathbf{G})\,\|\mathbf{x}\|^{2}
     \;\leq\;
     \|\mathbf{x}\|_{\oplus}^{2}
     \;\leq\;
     \lambda_{\max}(\mathbf{G})\,\|\mathbf{x}\|^{2}
     ,
     \qquad \forall\,\mathbf{x}\in\ccalH_{1},
\end{equation}
and likewise
$\lambda_{\min}(\widetilde{\mathbf{G}})\|\mathbf{z}\|^{2}\leq
\|\mathbf{z}\|_{\oplus}^{2}\leq
\lambda_{\max}(\widetilde{\mathbf{G}})\|\mathbf{z}\|^{2}$ for every
$\mathbf{z}\in\ccalH_{2}$.

By Proposition~\ref{prop:diag}, the coincidence decomposition is orthogonal
in the adapted inner products, so the proof of
Proposition~\ref{prop:ric} applies verbatim there, with $\ccalK=\ccalC_{p}$
and $\ccalH_{\ccalK}=\ccalH_{1}$ by~\eqref{eq:injectivity}, and yields the
frame bound
\begin{equation}\label{eq_app_frame_adapted}
\eta(\theta)\,\|\mathbf{x}\|_{\oplus}^{2}
      \;\leq\;
      \|\theta\,\mathbf{x}\|_{\oplus}^{2}
      \;\leq\;
      \|\mathbf{x}\|_{\oplus}^{2}
      ,
      \qquad \forall\,\mathbf{x}\in\ccalH_{1},
\end{equation}
under the stated normalization $\sigma_{\max}(\theta_{c})=1$, the norms on
the left and right being those of $\ccalH_{1}$ and the middle one that of
$\ccalH_{2}$. Combining~\eqref{eq_app_frame_adapted}
with~\eqref{eq_app_rayleigh} on both sides gives, for the lower bound,
\begin{equation}
\|\theta\mathbf{x}\|^{2}
     \;\geq\;
     \frac{\|\theta\mathbf{x}\|_{\oplus}^{2}}{\lambda_{\max}(\widetilde{\mathbf{G}})}
     \;\geq\;
     \frac{\eta(\theta)\,\|\mathbf{x}\|_{\oplus}^{2}}{\lambda_{\max}(\widetilde{\mathbf{G}})}
     \;\geq\;
     \frac{\lambda_{\min}(\mathbf{G})}{\lambda_{\max}(\widetilde{\mathbf{G}})}\,
     \eta(\theta)\,\|\mathbf{x}\|^{2}
     ,
\end{equation}
and, for the upper bound,
\begin{equation}
\|\theta\mathbf{x}\|^{2}
     \;\leq\;
     \frac{\|\theta\mathbf{x}\|_{\oplus}^{2}}{\lambda_{\min}(\widetilde{\mathbf{G}})}
     \;\leq\;
     \frac{\|\mathbf{x}\|_{\oplus}^{2}}{\lambda_{\min}(\widetilde{\mathbf{G}})}
     \;\leq\;
     \frac{\lambda_{\max}(\mathbf{G})}{\lambda_{\min}(\widetilde{\mathbf{G}})}\,
     \|\mathbf{x}\|^{2}
     ,
\end{equation}
which together are~\eqref{eq:ricdiag}. The ratio of the two frame constants
so obtained is
\begin{equation}
\frac{\alpha}{\beta}
     =
     \frac{\lambda_{\min}(\mathbf{G})\,\lambda_{\min}(\widetilde{\mathbf{G}})}
          {\lambda_{\max}(\mathbf{G})\,\lambda_{\max}(\widetilde{\mathbf{G}})}\,
     \eta(\theta)
     =
     \frac{\eta(\theta)}
          {\kappa(\mathbf{G})\,\kappa(\widetilde{\mathbf{G}})}
     ,
\end{equation}
and centering the frame bounds by the global rescaling
$\theta\mapsto\sqrt{2/(\alpha+\beta)}\,\theta$ exactly as in the proof of
Proposition~\ref{prop:ric} yields
$\delta_{s}\leq(\beta-\alpha)/(\beta+\alpha)$, which
is~\eqref{eq:ricdiag2}. Finally, when both shifts are normal the
projections are orthogonal, so $\mathbf{\Pi}_{c}^{*}\mathbf{\Pi}_{c}=\mathbf{\Pi}_{c}$
and $\mathbf{G}=\sum_{c}\mathbf{\Pi}_{c}=\mathbf{I}$, and likewise
$\widetilde{\mathbf{G}}=\mathbf{I}$, so
$\kappa(\mathbf{G})\kappa(\widetilde{\mathbf{G}})=1$
and~\eqref{eq:ricdiag2} reduces to~\eqref{eq:ricbound}.
\end{proof}

%%%%%%%%%%%%%%%%%%%%%%%%%%%%%%%%%%%%%%%%%%%%%%
%%%%%%%%%%%%%%%%%%%%%%%%%%%%%%%%%%%%%%%%%%%%%%
\subsection{Proof of Proposition~\ref{prop_equivariance}}
\label{proof_prop_equivariance}

\begin{proof}
(i) Fix $\lambda$ as in the hypothesis and set $c:=p(\lambda)\in\ccalC_p$,
so that $\ccalE_{\lambda}$ is one of the summands of
$\ccalE_{c}=\bigoplus_{\lambda':p(\lambda')=c}\ccalE_{\lambda'}$. We show
that a single failure at the level of $\lambda$ propagates, in two more
steps, all the way up to $\eta(\theta)$.
\emph{Step 1: $\theta\vert_{\ccalE_{\lambda}}$ has a nontrivial kernel.}
If $\lambda\notin\sigma(\widetilde{\mathbf{S}})$, then
$\widetilde{\ccalE}_{\lambda}=\{0\}$, and since
$\theta(\ccalE_{\lambda})\subseteq\widetilde{\ccalE}_{\lambda}$ for every
$\theta\in\Theta_{t}$, the map $\theta\vert_{\ccalE_{\lambda}}$ vanishes
identically -- its kernel is all of $\ccalE_{\lambda}$. If instead
$m_{\lambda}>\widetilde{m}_{\lambda}$, then
$\theta\vert_{\ccalE_{\lambda}}:\ccalE_{\lambda}\to\widetilde{\ccalE}_{\lambda}$
maps a higher-dimensional space into a lower-dimensional one, so
rank-nullity forbids it from being injective. Either way, there is a
nonzero $\mathbf{v}\in\ccalE_{\lambda}$ with $\theta\mathbf{v}=\mathbf{0}$.
\emph{Step 2: this forces $\eta_{c}(\theta)=0$ at the class level.}
Since $\ccalE_{\lambda}\subseteq\ccalE_{c}$, the vector $\mathbf{v}$ is
also a vector of $\ccalE_{c}$, and
$\theta_{c}\mathbf{v}=\theta\mathbf{v}=\mathbf{0}$ because $\theta_{c}$
is simply $\theta$ restricted to $\ccalE_{c}$. Now,
$\sigma_{\min}(\theta_{c})$ is defined as the \emph{smallest} value of
$\|\theta_{c}\mathbf{x}\|$ over \emph{all} unit vectors
$\mathbf{x}\in\ccalE_{c}$, not only those coming from $\ccalE_{\lambda}$;
having found one unit vector (after rescaling $\mathbf{v}$) on which
$\theta_{c}$ vanishes is therefore already enough to force
$\sigma_{\min}(\theta_{c})=0$, and hence $\eta_{c}(\theta)=0$.
\emph{Step 3: one bad class zeroes out the global richness.} By
definition $\eta(\theta)=\min_{c'\in\ccalC_p}\eta_{c'}(\theta)$, a
minimum taken over \emph{every} coincidence class. Since $\eta_{c}(\theta)=0$
for the class found in Step~2, this minimum can be no larger than $0$;
as $\eta(\theta)\geq0$ always, $\eta(\theta)=0$. The eigenvalue
$\lambda$ was arbitrary among those satisfying the hypothesis, and
$\theta\in\Theta_{t}$ was arbitrary, so $\eta(\theta)=0$ for every
$\theta\in\Theta_{t}$ -- even though $\Theta_{p}$, which only has to
respect the coarser class-level routing rather than this finer
$\lambda$-level routing, may still contain a map with $\eta(\theta)=1$.
(ii) \emph{Necessity} is exactly the contrapositive of (i): if some
$\theta\in\Theta_{t}$ has $\eta(\theta)\neq0$, then no $\lambda$ can
satisfy (i)'s hypothesis, which is precisely the stated condition that
every $\lambda\in\sigma(\mathbf{S})$ with $p(\lambda)\in\ccalC_{p}$ has
$\lambda\in\sigma(\widetilde{\mathbf{S}})$ and $m_{\lambda}\leq\widetilde{m}_{\lambda}$.
\emph{Sufficiency} is proved by an explicit construction, one
coincidence class at a time. Fix $c\in\ccalC_{p}$ and, for every raw
eigenvalue $\lambda$ with $p(\lambda)=c$, let
$\theta\vert_{\ccalE_{\lambda}}$ be a scaled isometry into
$\widetilde{\ccalE}_{\lambda}$ (possible since $m_{\lambda}\leq
\widetilde{m}_{\lambda}$), using \emph{one common scale factor}
$\sigma_{c}$ for every such $\lambda$ in the class. The common scale is
the crux of the argument: $\eta_{c}(\theta)$ is computed from
$\sigma_{\min}(\theta_{c})$ and $\sigma_{\max}(\theta_{c})$ over the
\emph{whole} space $\ccalE_{c}=\bigoplus_{\lambda:p(\lambda)=c}\ccalE_{\lambda}$,
not eigenvalue by eigenvalue -- so if two eigenvalues in the same class
were assigned different scales, a vector mixing directions from both
would be stretched unevenly by $\theta_{c}$, giving
$\sigma_{\min}(\theta_{c})<\sigma_{\max}(\theta_{c})$ and pulling
$\eta_{c}(\theta)$ below $1$. With a single scale $\sigma_{c}$, however,
$\theta_{c}$ acts as the same multiple of an isometry in every direction
of $\ccalE_{c}$, so $\sigma_{\min}(\theta_{c})=\sigma_{\max}(\theta_{c})$
and $\eta_{c}(\theta)=1$. Repeating this construction independently for
every $c\in\ccalC_{p}$ gives $\eta_{c}(\theta)=1$ for every class, and
therefore $\eta(\theta)=\min_{c\in\ccalC_{p}}\eta_{c}(\theta)=1$.
\end{proof}